\documentclass[times,review,preprint,authoryear]{elsarticle}
\usepackage{medima}

\usepackage{framed,multirow}

\usepackage{amssymb}
\usepackage{latexsym}
\usepackage{booktabs}
\usepackage{multirow}
\usepackage{tabularx}
\usepackage{amsmath}

\DeclareMathOperator*{\argmin}{arg\,min}
\usepackage{float}
\usepackage{amsfonts}
\usepackage{color,soul}
\usepackage[ruled]{algorithm2e}
\usepackage{url}
\usepackage{xcolor}
\usepackage{qting}
\usepackage{hyperref}

\definecolor{newcolor}{rgb}{.8,.349,.1}

\journal{Medical Image Analysis}

\begin{document}

\verso{Fanwei Kong \textit{et~al.}}

\begin{frontmatter}
\title{A Deep-Learning Approach For Direct Whole-Heart Mesh Reconstruction}%

\author[1]{Fanwei \snm{Kong}\corref{cor1}}
\ead{fanwei\_kong@berkeley.edu}
\cortext[cor1]{Corresponding author}
\author[2]{Nathan \snm{Wilson}}
\ead{nwilson@osmsc.com}
\author[1]{Shawn \snm{Shadden}\corref{cor1}}
\ead{shadden@berkeley.edu}
\address[1]{Mechanical Engineering Department, University of California, Berkeley, Berkeley, CA 94709}
\address[2]{Open Source Medical Software Corporation, Santa Monica, CA}

\begin{abstract}
Automated construction of surface geometries of cardiac structures from volumetric medical images is important for a number of clinical applications. While deep-learning-based approaches have demonstrated promising reconstruction precision, these approaches have mostly focused on voxel-wise segmentation followed by surface reconstruction and post-processing techniques. However, such approaches suffer from a number of limitations including disconnected regions or incorrect surface topology due to erroneous segmentation and stair-case artifacts due to limited segmentation resolution. We propose a novel deep-learning-based approach that directly predicts whole heart surface meshes from volumetric CT and MR image data. Our approach leverages a graph convolutional neural network to predict deformation on mesh vertices from a pre-defined mesh template to reconstruct multiple anatomical structures in a 3D image volume. \textcolor{black}{Our method demonstrated promising performance of generating} \q{abstract}{whole heart reconstructions with as good or better accuracy than prior deep-learning-based methods on both CT and MR data. Furthermore, by deforming a template mesh, our method can generate whole heart geometries with better anatomical consistency and produce high-resolution geometries from lower resolution input image data.} \textcolor{black}{Our method was also able to produce temporally-consistent} surface mesh predictions for heart motion from CT or MR cine sequences, and therefore can potentially be applied for efficiently constructing 4D whole heart dynamics. \q{code}{Our code and pre-trained networks are available at \url{https://github.com/fkong7/MeshDeformNet}
}
\end{abstract}


\end{frontmatter}

\section{Introduction}

Three-dimensional (3D) geometries of anatomical structures reconstructed from volumetric medical images are increasingly used for a number of clinical applications, such as patient-specific visualization \citep{GonzalezIzard2020}, physics-based simulation, virtual surgery planning and morphology assessment \citep{Prakosa2018, Bucioli2017}. As cardiovascular diseases are the leading causes of mortality, one area of research that currently receives considerable attention is computational modeling and visualization of the heart from patient-specific image data \citep{Prakosa2018, Chnafa2016, Khalafvand2012}. 
Creating accurate patient-specific models of the whole heart from image data has traditionally required significant time and human effort, limiting clinical applications and high-throughput, large-cohort analyses of patient-specific cardiac functions \citep{Mittal}. While surface representation of the whole heart is important for the aforementioned applications, most studies have focused on \textcolor{black}{image segmentation} rather than direct surface reconstruction \citep{Payer2018, BAI201598, Ye2019}. Nevertheless, accurate and reliable automatic whole heart segmentation remains an ongoing challenge and an active research direction \citep{ZHUANG2019, Zhuang2013,Peng2016}. Much of the challenge is related to the complex geometries of the heart, large structural deformation over the cardiac cycle, difficulties in differentiating individual cardiac structures from each other and the surrounding tissue, \textcolor{black}{as well as variations across individuals, different imaging modalities, systems, and centers.}

Prior cardiac model construction efforts have typically adopted a multistage approach whereby 3D segmentations of cardiac structures are first obtained from image volumes, meshes of the segmented regions are then generated using marching cube algorithms, and finally manual surface post-processing or editing is performed \citep{Lorensen1987, KONG2020, Maher2019, Augustin2016EM}. The quality of reconstructed surfaces highly depends on the quality of segmentation and the complexity of the anatomical structures. Automatic heart segmentation has been a popular research topic and previously published algorithms have been summarized in detail \citep{Zhuang2013, ZHUANG2019, Peng2016, Habijan2020}. Generally, there are two common approaches to whole heart segmentation: multi-atlas segmentation (MAS) \citep{BAI201598, Zhuang2015MAS, ZHUANG2016MAS} and \textcolor{black}{deep-learning-based segmentation} \citep{Ronneberger2015, cicek3DUNET}. Compared with MAS, \textcolor{black}{deep-learning-based} approaches have become more popular as they have demonstrated higher segmentation precision \citep{ZHUANG2019, Payer2018} and are much faster in practice. While a couple of recent studies have reduced the processing time of MAS approaches down to a couple of minutes \citep{Bui2020, BUI2020104019}, \textcolor{black}{deep-learning-based} approach can generally process a whole heart segmentation within a couple of seconds. However, while \textcolor{black}{deep-learning-based} methods may produce segmentations that achieve high average voxel-wise accuracy, they can contain extraneous regions and other nonphysical artifacts. Correcting such artifacts would require a number of carefully designed post-processing steps and sometimes manual efforts \citep{KONG2020}. Indeed, since the CNN-based segmentation methods are based on classification of each image voxel to a particular tissue class, the neural networks are often trained to reduce voxel-wise discrepancy between the predicted segmentation and the ground truth and therefore lack awareness of the overall anatomy and topology of the target organs. Moreover, CNN-based 3D segmentation methods are memory intensive and therefore require downsampling of the data to fit within memory and thus can only generate segmentation with limited resolution. However, high-resolution geometries are often required for down-stream applications such as computational simulations, and direct or low-resolution segmentation will often produce surfaces with staircase artifacts that require additional post-processing \citep{Wei2018, WEI20131223, UPDEGROVE201616}. 

Compared with representing whole heart geometries as segmentations on a dense voxel grid, representing the geometries as meshes is a more compact representation, as only point coordinates on the organ boundaries need to be stored. This advantage may enable efficient reconstruction of high-resolution surface meshes on a limited memory budget and avoid the stair-case artifacts of surfaces constructed from low-resolution 3D segmentation. Moreover, for low-resolution input images, voxel-wise segmentation would be a coarse representation of the underlying cardiac structures, but a surface mesh representation can still function as a smoother and more realistic representation of the shapes as the mesh vertices are defined in a continuous coordinate space and do not have to align with the input voxel grid. 

Some studies have adopted a model-based approach to directly fit surfaces meshes of the heart to target images \citep{Ecabert2008, ECABERT2011863, PETERS201070}. Such approaches deform a template mesh using local optimization to match with tissue boundaries on input images. However, they are often sensitive to initialization and require complicated steps and manual efforts to construct a mean template of the heart. \q{intro_zhang}{A recent study by \cite{ZhangMo2020} proposed deep learning to learn the initialization of the active contour method--a model-based approach--to help solve for the contours of the target tissues. Alternatively, others have turned to pure deep learning methods that do not require test-time optimization.} 
\q{intro_ye}{ \cite{YeMeng2020} proposed a deep learning approach to jointly predict the segmentation and the geometry of the left ventricle in the form of a point cloud from the image data. 
 }

Recent progress on geometric deep learning has extended the concepts of convolutional neural network on irregular graphs \citep{Defferrard2016, Bronstein2017}. Recent deep-learning-based approaches have shown promise in reconstructing shapes as surface meshes from image data using graph convolutional neural networks \citep{Pixel2Mesh, Wen2019Pixel2MeshM3, Image2mesh}. However, these approaches have focused on reconstructing a single shape from a 2D camera image and thus cannot be directly applied to reconstructing multiple anatomical structures from volumetric medical image data. A recent study from \cite{Voxel2Mesh} extended the work of \cite{Pixel2Mesh} to 3D volumetric medical image data and demonstrated improved segmentation results. However, their method demonstrated success only on simple geometries such the liver, hippocampus and synaptic junction but not on the whole heart that involves multiple cardiac structures with widely varying shapes. 

To overcome these shortcomings, we explore the problem of using a deep-learning-based approach to directly predict surface meshes of multiple cardiac structures from volumetric image data. Our approach leverages a graph convolutional neural network to predict deformation on mesh vertices from a pre-defined mesh template to fit multiple anatomical structures in a 3D image volume. The mesh deformation is conditioned on image features extracted by a CNN-based image encoder. Since cardiac structures such as heart chambers are \textcolor{black}{homeomorphic} to a sphere, we use spheres as our initial mesh templates, which can be considered as a topological prior of the cardiac structures. Compared with classification-based approaches, our approach can reduce extraneous regions that are anatomically inconsistent. Using a generic initial mesh also enables our approach to be easily adapted to other anatomical structures.  

The key contributions of our work are as follows:  
\begin{enumerate}
    \item We propose the first end-to-end deep-learning-based approach of predicting multiple anatomical structures in the form of surfaces meshes from 3D image data. We show that our method was able to better produce whole-heart geometries from both CT and MR images compared to classification-based approaches. 
    \item We investigate and compare the impact of dataset size and variability on whole-heart reconstruction performance to different methods. When having trained on both small and larger training datasets, our method demonstrated better Dice scores for most of the cardiac structures reconstructed than prior approaches. 
    \item As cardiac MR image data often have large variation across different data sources, we compare different methods and demonstrate the advantage of our approach on MR images with varying through-plane resolution as well as on low-resolution MR images that differ significantly from our training datasets.
    \item Since our approach predicts deformation from a template mesh, we show that our reconstructions generally have point correspondence across different time frames and different patients by consistently mapping mesh vertices on the templates to similar structural regions of the heart. We demonstrate the potential application of our method on efficiently constructing 4D whole heart dynamics that captures the motion of a beating heart from a time-series of images. 
\end{enumerate}

\section{Methods}
\subsection{Dataset Information}
Since cardiac medical image data is sensitive to a number of factors, including differences in vendors, modalities and acquisition protocols across clinical centers, deep-learning-based methods can be easily biased to these factors. Therefore, we aimed to develop our models using whole heart image data collected from different sources, vendors and imaging modalities. We included data from four existing public datasets that contain contrast-enhanced CT images or MR images that cover the whole heart. These four datasets are from the multi-modality whole heart segmentation challenge (MMWHS) \cite{ZHUANG2019}, orCalScore challenge \citep{orCaScore}, left atrial wall thickness challenge (SLAWT) \citep{SLAWT} and left atrial segmentation challenge (LASC) \citep{LASC}. The use of such diverse data enables us to not only better evaluate the reconstruction accuracy of our trained model but also evaluate the impact of dataset size and variability on model performance.

Additional time-series CT and MR images were collected to evaluate the performance of our trained neural network models on time-series image data acquired from different data sources from the training data. The time-series CT data were from 10 patients with left ventricular diastolic dysfunction. The 9 sets of cine cardic MR data were from 5 healthy subjects and 4 patients with cardiac diseases.  All data was de-identified and previously collected for other purposes. The details of the datasets used and collected are described in the following sub-sections and summarized in Table \ref{table:data}. We followed the same method of \cite{ZHUANG2019} to manually delineate seven cardiac structures: LV, LA, RA, RV, myocardium, aorta and pulmonary artery for the collected image data that did not have ground truth annotations of the whole heart.

\begin{table}[]
\caption{Summary of data characteristics for whole heart CT and MR data included.}\label{table:data}
\resizebox{\textwidth}{!}{%

\begin{tabularx}{1.5\linewidth}{*{5}{>{\raggedright\arraybackslash}X}|*{3}{>{\raggedright\arraybackslash}X}}
\hline
                              & \multicolumn{4}{c|}{CT data}                                                             & \multicolumn{3}{c}{MR data}                                                                   \\ \cline{2-8} 
                              & MMWHS \citep{ZHUANG2019}       & OrCaScore \citep{orCaScore}                        & SLAWT  \citep{SLAWT}                 & time-series CT    & MMWHS \citep{ZHUANG2019}                                        & LASC  \citep{LASC}                & cine MR              \\ \hline
Vendor                        & Philips      & GE, Philips, Siemens and Toshiba & Philips Achieva 256 iCT & GE           & 1.5T Philips and 1.5T Siemens Magnetom Avanto & 1.5 T Philips Achieva & 1.5 T Philips          \\ \hline
\# of clinical sites involved & 2     & 4                                & 1                       & 1            & 2                                             & 1                     & 1                      \\ \hline
\# of 3D image volumes        & 60           & 72                               & 10                      & 100          & 60                                            & 27                    & 200                    \\ \hline
\# of patients involved        & 60           & 72                               & 4                       & 10           & 60                                            & 27                    & 10                   \\ \hline
In-place resolution (mm)      & 0.78 by 0.78 & 0.4-0.5 by 0.4-0.5               & 0.4 by 0.4              & 0.44 by 0.44 & 1.6-2.0 by 1.6-2.0                            & 1.25 by 1.25          & 0.65-1.75 by 0.65-1.75 \\ \hline
Slice thickness (mm)          & 1.6          & 0.5-0.625                        & 0.8-1.0                 & 0.625        & 2.0-3.2                                       & 2.7                   & 8-10                     \\ \hline
Temporal resolution (ms)      & N/A          & N/A                              & N/A                     & 100          & N/A                                           & N/A                   & 50                     \\ \hline
Public or private             & public       & public                           & public                  & private      & public                                        & public                & \textcolor{black}{private}                 \\ \hline
\end{tabularx}
}
\end{table}

\subsection{Geometry Reconstruction From Volumetric Images}
\begin{figure}[h]
\centering
\includegraphics[width=\textwidth]{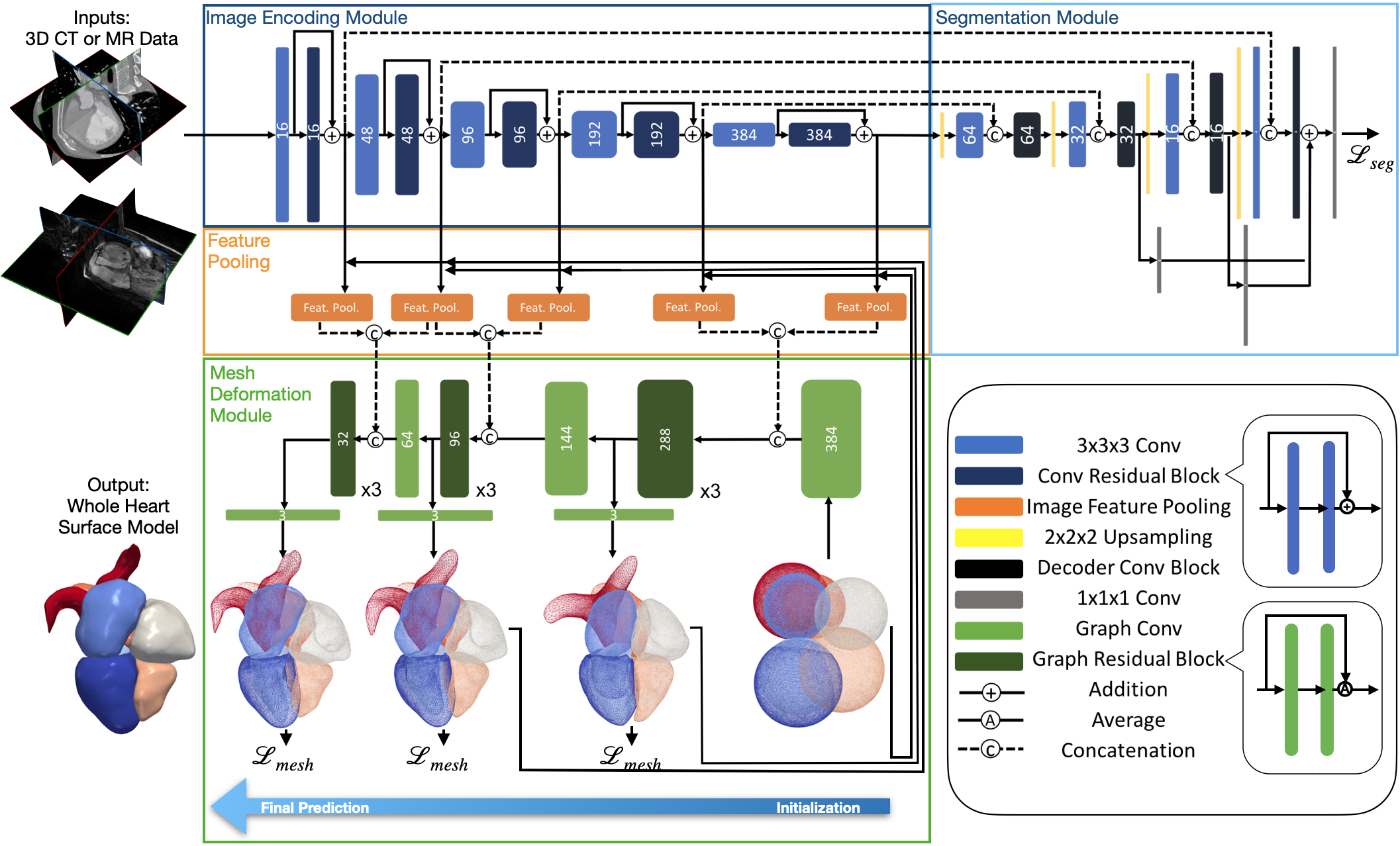}
\caption{Diagram of the proposed automatic whole heart reconstruction approach. The framework uses 3D convolutional layers (shown in blue) to encode image features and predict a binary segmentation map from an input image volume. The corresponding image features are sampled by pooling layers (shown in orange) based on the vertex coordinates of the template mesh. From the combined image and mesh features, graph convolutional layers (shown in green) are then used to predict the deformation of mesh vertices to generate the final mesh predictions} 
\label{figure:network}
\end{figure}

Our framework consists of three components to predict the whole-heart meshes from a volumetric input image: (1) an image encoding module that extracts and encodes image features, (2) a mesh deformation module that combines features from images and meshes to predict deformation of mesh vertices, and (3) a segmentation module that predicts a binary segmentation map to allow additional supervision using ground truth annotations. Figure \ref{figure:network} shows the overall architecture.
\subsubsection{Image Encoding Module}
For an input image data, the image encoding module uses a series of 3D convolutional layers to extract volumetric image feature maps at multiple resolutions. These feature maps are required by the following mesh deformation module to predict whole-heart geometries. Therefore, the image encoder should both be effective for better geometric reconstruction and be memory-efficient to process a $128\times128\times128$ volumetric input image in a single pass. Our image feature encoder is based on an improved 3D UNet architecture that was designed to work effectively for large volumetric image data \citep{Isensee2017}. Briefly, the feature encoder architecture consists of multiple levels of residual blocks that encode increasingly abstract representations of the input. Residual connections are known to facilitate gradient propagation during training and improve generalization \citep{He2016}. Each residual block contains two $3\times3\times3$ convolutional layers and a dropout layer before the last convolutional layer. The input to the first convolutional layer is then added to the output of the last one. After each residual block, we use a $3\times3\times3$ convolutional layer with input stride 2 to reduce the resolution of the feature maps. 

\subsubsection{Segmentation Module}
While our purpose is to reconstruct surface meshes directly from image data, the ground truth segmentation can function as an additional supervision to the network to further facilitate training. From our experiments, including the segmentation module helped avoid non-manifold geometries due to local minimums and thus improve reconstruction accuracy. Since the ground truth mesh is a sparse representation of the cardiac structures compared with a volumetric segmentation, including the segmentation as a dense supervision with skip connections to the image feature encoder can improve gradient propagation to the image encoding module to better interpret the full volumetric input data. However, since we are only interested in reconstructing meshes, rather than predicting segmentations for all cardiac structure, our segmentation module is trained to predict only a binary segmentation representing the occupancy of the heart in the input image. The adopted network architecture is simplified from the decoder architecture of \cite{Isensee2017} with only a small number of filters in the convolutional layers. Briefly, the segmentation module contains multiple levels of decoder convolutional blocks that correspond to the residual blocks from the image encoding module to reconstruct segmentation from extracted features. Following a $3\times3\times3$ convolution of the up-sampled intermediate output, a decoder convolutional block concatenates the current output with the corresponding output from the residual blocks of the image encoding module and then uses a $1\times1\times1$ convolutional layer to process the concatenated features. Binary segmentation predictions were generated from three different levels of the segmentation module and added together to form the final prediction. 

\subsubsection{Graph Convolution on  Mesh}
Our neural network uses graph convolutions on a template mesh to predict deformation vectors on its vertices. Unlike for structured data such as images, convolution in the spatial domain is not well defined for manifold structures such as meshes. Therefore, we apply graph convolution in the frequency domain following recent process in graph convolutional neural networks \citep{Bronstein2017, Defferrard2016}. Briefly, our template mesh is represented by a graph $\mathcal{M} = (\mathcal{V}, \mathcal{E})$, where $\mathcal{V}=\{v_i\}_{i=1}^N$ is the set of $N$ vertices and $\mathcal{E}=\{e_i\}_{i=1}^E$ is the set of $E$ edges that define the connections among mesh vertices. The graph adjacency matrix $A \in \{0,1\}^{N\times N}$ is a sparse matrix that defines the connection between each pair of vertices, with $A_{ij}=0$ if \textcolor{black}{vertices $v_i$ and $v_j$} are not connected and $A_{ij}=1$ if the two vertices are connected. The degree matrix $D$ is a diagonal matrix that represents the degree of each vertex, with $D_{ii} = \sum_j A_{ji}$. Therefore, the graph \textcolor{black}{Laplacian} matrix is a real and symmetric matrix defined as $L = D-A$, which can then be normalized as $L_{norm} = I - D^{-1/2} A D^{-1/2}$. The \textcolor{black}{normalized Laplacian} matrix can be diagonalized by the Fourier basis on graph $U\in\mathbb{R}^{N\times N}$ as $L_{norm} = U\Lambda U^T$. The columns of $U$ are the orthogonal eigenvectors of $L$ and $\Lambda$ is a diagonal matrix containing the corresponding eigenvalues. The Fourier transform of a function defined on mesh vertices, $f \in L^2(\mathcal{V})$, is thus described by $\hat{f} = U^T f$ and the inverse \textcolor{black}{Fourier} transform is $f = U \hat{f}$. Therefore,  convolution between $f$ and $g \in L^2(\mathcal{V})$ is described as $f * g = U (( U^T f) \odot (U^T g))$. If we parameterize $g$ with learnable weights, a graph convolution layer can then be defined as $f_{out} = \sigma(U g_\theta(\Lambda)U^Tf_{in})$, where $f_{in}$ and $f_{out}$ are the input and output and $\sigma$ is the ReLU activation function. 

The above expression is computationally expensive for meshes with a large number of vertices, since $U$ is not sparse and the number of parameters required can be as many as the number of vertices. Therefore, we followed \cite{Defferrard2016} to approximate $g_\theta(\Lambda)$ using Chebyshev polynomials \textcolor{black}{so that} $Ug_\theta(\Lambda)U^T = \sum_{k=0}^K \theta_k T_k(\Tilde{L})$, \textcolor{black}{where $\Tilde{L}$ is the scaled sparse Laplacian matrix $\Tilde{L} = 2L_{norm}/\lambda_{max} - I$, where $\lambda_{max}$ is the maximum eigenvalue of $L_{norm}$.} \textcolor{black}{$\theta_k$} is the parameter for the $k$th order Chebyshev polynomial and $T_k$ is the $k$th order polynomial that can be computed recursively as $T_0 = I$, $T_1 = \Tilde{L}$ and $T_k(\Tilde{L}) = 2 \Tilde{L} T_{k-1}(\Tilde{L}) - T_{k-2}(\Tilde{L})$. We chose $K=1$ since a lower order polynomial can effectively avoid fitting the noise on our ground truth surfaces and reduce the amount of parameters to learn.  \q{graph_conv}{Therefore, the graph convolution on the mesh using a first-order Chebyshev polynomial approximation is described as $f_{out} = \sigma(\theta_0 f_{in} + \theta_1 f_{in} \Tilde{L} )$, where $\theta_0, \theta_1 \in \mathbb{R}^{d_{out}\times d_{in}}$ are trainable weights. $f_{out} \in \mathbb{R}^{d_{out}\times N}$ and  $f_{in} \in \mathbb{R}^{d_{in}\times N}$ are, respectively, the input and output feature matrices, where $d_{in}$ and $d_{out}$ are, respectively, the input and output dimensions of the mesh features. }

\subsubsection{Mesh Initialization}
Our method uses a single network to simultaneously deform multiple sphere templates to corresponding cardiac structures on the input image. Since the relative locations and scales of different cardiac structures of the heart are generally consistent across a population, we leverage this prior knowledge into our neural network by scaling and positioning the corresponding initial sphere mesh template based on the relative sizes and locations of the cardiac structures. We then used a graph convolution layer to augment the coordinates of the initial meshes such that they have comparable contribution as the image features, in terms of the length of feature vectors, to the following deformation block. \q{mesh_init}{Namely, after pre-processing the volumetric training data and obtaining the corresponding ground truth meshes as described in detail in Appendix \ref{appendix:pre-processing}, we computed the corresponding image coordinates of the vertices of the surface meshes in the volumetric training image data. For each cardiac structure, we then computed the average centroid location and the average length between surface and centroid, across all the ground truth meshes in the training data. For each input image, we then used this approximated center and radius to initialize each sphere.} By having a closer initialization compared with using centered unit spheres as in prior approaches \citep{Voxel2Mesh, Pixel2Mesh}, our network can have reduced distance between predictions and ground truths and thus avoid large deformation during the early phase of training. From our experiments, this is an important and effective technique to avoid getting stuck in local minimums and achieve faster convergence. 

\subsubsection{Mesh Deformation Module}
Our proposed mesh deformation module consists of three deformation blocks with graph convolutional layers that progressively deform our initial template meshes based on both existing mesh vertex features and image features extracted from the image encoding module. \q{shared_decoder}{Meshes of all different cardiac structures are deformed simultaneously by these shared mesh deformation blocks.} The volumetric feature maps have increasing level of abstraction but decreasing spatial resolution as we progress deeper in the image encoding module. Therefore, as shown in Figure \ref{figure:network}, we used more abstracted, high-level image feature maps for the initial mesh deformation blocks to learn the general shapes of cardiac structures while using low-level, high-resolution feature maps for the later mesh deformation blocks to produce more accurate predictions with detailed features. For each mesh deformation block, we project image features from the image encoding module to the mesh vertices and then concatenate the extracted image feature vector with the existing vertex feature vector. As we deform the mesh through multiple deformation blocks, we decrease the size of the graph convolutional filters to reduce the dimension of mesh feature vectors to match with the reduced number of filters used in upper levels of the image encoding module. Within each mesh deformation block, the concatenated feature vectors are processed by three graph residual blocks, which contains two graph convolutional layers with residual connections. We then use an additional graph convolutional layer to predict deformation as 3D feature vectors on mesh vertices and add those with the vertex coordinates of the initial mesh or the mesh from the previous deformation block to obtain the current predicted vertex coordinates. To project corresponding image features onto mesh vertices, from the vertex locations of the initial or previously deformed mesh, we compute the corresponding image coordinates in the volumetric image feature maps. We then tri-linearly interpolate the feature vectors that correspond to the 8 neighboring voxels of the computed image coordinates in the volumetric feature maps. 

\subsection{Loss functions}
The training of our networks was supervised by 3D ground truth meshes of the whole heart as well as a binary segmentation indicating occupancy of the heart on the voxel grid that corresponds to the input image volume. The whole heart meshes were extracted from segmentation of cardiac structures using the marching cube algorithm and the binary segmentation was also obtained from segmentation by setting all non-background voxels to 1 and the rest to 0. We used two categories of loss functions, geometry consistency losses and regularization losses in the training process. The geometry consistency losses include point and normal consistency losses while the regularization losses include edge length and \textcolor{black}{Laplacian} losses. 

\subsubsection{Segmentation loss} 
We used a hybrid loss function that contained both cross-entropy and dice-score losses. \textcolor{black}{This loss has been used in training UNets and has demonstrated promising results on various medical image segmentation tasks \citep{Isensee2021}.} Namely, let $L_{occupancy}(I_p, I_g)$ denote the loss of between the predicted occupancy probability map $I_P$ and the ground truth binary segmentation of the whole heart $I_G$. The hybrid loss function was 
\textcolor{black}{\begin{equation}
    \mathcal{L}_{occupancy}(I_P, I_G) = -\sum_{x\in I_G} \left(I_G(x) \log(I_P(x)) + (1-I_G(x)) \log(1-I_P(x)) \right)
      - \frac{2\sum_{x\in I} I_G(x) I_P(x)}{\sum_{x\in I} I_G(x)+\sum_{x\in I} I_P(x)}
    \label{loss}
\end{equation}}
where $x$ denotes the pixel in the input image $I$. 
\subsubsection{Point loss}
We used Chamfer loss to regulate the accuracy of the vertex locations on predicted meshes. For a point from the predicted mesh or the ground truth mesh, Chamfer loss finds the nearest vertex in the other point set and adds up all pair-wise distances.  The point loss is defined by, 
\begin{equation}
    \mathcal{L}_{point} (\mathbf{P}_i, \mathbf{G}_i) =\sum_{\substack{\mathbf{p}\in \mathbf{P}_i}} \min_{\substack{\mathbf{g}\in \mathbf{G}_i}} ||\mathbf{p}-\mathbf{g}||_2^2 + \sum_{\substack{\mathbf{g}\in \mathbf{G}_i}} \min_{\substack{\mathbf{p}\in \mathbf{P}_i}} ||\mathbf{p}-\mathbf{g}||_2^2 
\end{equation}
where $\mathbf{p}$ and $\mathbf{g}$ are, respectively, points from the vertex sets of the predicted mesh $\mathbf{P}_i$ and the ground truth mesh $\mathbf{G}_i$ of cardiac structure $i$.

\subsubsection{Normal loss} We used a normal consistency loss to regulate the accuracy of the surface normal on the predicted meshes. For each point, the surface normal is estimated by the cross product between two edges of a face connected to the point. The predicted surface normal is then compared with the ground truth surface normal at the nearest vertex. Namely, 
\begin{equation}
    \mathcal{L}_{normal} (\mathbf{P}_i, \mathbf{G}_i) = \textcolor{black}{ \sum_{\substack{\mathbf{p}\in \mathbf{P}_i}; \mathbf{g}=\argmin \limits_{\substack{\mathbf{g} \in \mathbf{G}_i}}||\mathbf{p}-\mathbf{g}||_2^2 } } \left\|(\mathbf{p}_1 - \mathbf{p}) \times (\mathbf{p}_2 - \mathbf{p}) - \mathbf{n_g} \right\|_2^2
\end{equation}
where $\mathbf{p}_1$ and $\mathbf{p}_2$ are the two vertices sharing the same face with vertex $\mathbf{p}$. 

\subsubsection{Edge length loss} 
We used an edge length loss to encourage a more uniform mesh density on the predictions. That is, we regularize the difference between each edge length and an estimated average edge length $\mathbf{\mu}_i$ of the corresponding cardiac structure $\mathbf{G}_i$. Namely, we compute the average surface area of our ground truth mesh for each cardiac structure and estimate the average edge length based on the surface area ratio between the template and ground truth meshes, leading to 
\begin{equation}
    \label{eq:edge_loss}
    \textcolor{black}{\mathcal{L}_{edge}(\mathbf{P}_i)} =\sum_{\mathbf{p} \in \mathbf{P}_i} \sum_{\mathbf{k}_p\in \mathcal{N}(\mathbf{p})} \left| \left\|\mathbf{p}-\textcolor{black}{\mathbf{k_p}}\right\|_2^2 - \mu_i^2 \right| \;,
\end{equation}
\textcolor{black}{where $\mathcal{N(\mathbf{p})}$ represents the neighborhood of vertex $\mathbf{p}$.} 
\subsubsection{Laplacian loss} To encourage a smoother mesh prediction, we used a \textcolor{black}{Laplacian} loss to regularize the difference between a vertex location $\mathbf{p}$ and the mean location of its neighboring vertices $\mathbf{k_p}$ as
\begin{equation}
    \mathcal{L}_{lap} (\mathbf{P}_i) = \sum_{\mathbf{p}\in \mathbf{P}_i} \left\|\mathbf{p}-\sum_{\mathbf{k_p}\in \mathcal{N}(\mathbf{p})} \frac{1}{||\mathcal{N}(\mathbf{p})||}\mathbf{k_p} \right\|_2^2 \;.
\end{equation}

\subsubsection{Total loss}
Prior approaches of mesh reconstruction from images commonly formulated the total loss function as a weighted sum of multiple loss functions \citep{Pixel2Mesh, Voxel2Mesh}. However, for multi-loss regression problems, different loss functions are different in scales. Manually tuning the weight assigned to each loss function is difficult and expensive since losses can differ by orders of magnitude. Therefore, we express the total loss on predicted meshes as a weighted geometric mean of the individual losses so that the gradient for an individual loss function can be invariant to its scale relative to other loss functions \textcolor{black}{ \citep{Chennupati_2019_CVPR_Workshops}}. Thus, for predicted meshes $\mathbf{G}$ and ground truth meshes $\mathbf{P}$ with $N$ cardiac structures, the total mesh loss is expressed as, 
\begin{equation}
    \label{eq:mesh_loss}
    \mathcal{L}_{mesh}(\mathbf{P}, \mathbf{G}) = \sum_i^N \mathcal{L}_{point} (\mathbf{P}_i, \mathbf{G}_i)^{\lambda_1}  \mathcal{L}_{normal} (\mathbf{P}_i, \mathbf{G}_i)^{\lambda_2} \textcolor{black}{\mathcal{L}_{edge}(\mathbf{P}_i)^{\lambda_3} }\mathcal{L}_{lap} (\mathbf{P}_i)^{\lambda_4} \;,
\end{equation}
where each $\lambda$ is a hyperparameter to weight each individual loss based on its importance without being affected by its scale. We can thus choose hyperparameters from a consistent range for all the losses. \q{8set}{We generated 8 sets of random numbers ranging from 0 to 1 and chose the best out of the 8 sets of hyperparameters that produced the smallest point loss on the validation data. The chosen hyperparameters are $\lambda_1 = 0.3$, $\lambda_2=0.46$, $\lambda_3 = 0.16$ and $\lambda_4 = 0.05$}. For total loss, we added up losses from all three deformation blocks as well as the binary segmentation loss: 
\textcolor{black}{\begin{equation}
    \mathcal{L}_{total} = \mathcal{L}_{mesh}(\mathbf{P}^{B1}, \mathbf{G}) + \mathcal{L}_{mesh}(\mathbf{P}^{B2}, \mathbf{G}) + \mathcal{L}_{mesh}(\mathbf{P}^{B3}, \mathbf{G}) + \mathcal{L}_{occupancy}(I_p, I_g) \; .
\end{equation}}
\textcolor{black}{The network parameters were computed by minimizing the total loss function using the Adam stochastic gradient descent algorithm \citep{adam}.}

\section{Experiments and Results}
\subsection{Baselines}
We considered \textcolor{black}{the following} three baselines to compare our method against: 2D UNet \citep{Ronneberger2015}, a residual 3D UNet \citep{Isensee2017} and Voxel2Mesh \citep{Voxel2Mesh}. The UNets are arguably the most successful architecture for medical image segmentation and thus can function as strong baselines. In particular, the 2D UNet is a part of the whole-heart segmentation framework implemented in \cite{KONG2020} that recently demonstrated state-of-the-art performance on the MMWHS challenge dataset. The residual 3D UNet has demonstrated improved performance than a regular 3D UNet and won the KiTS2019 Challenge \citep{Isensee2019AnAA, HELLER2021101821}. To ensure a fair comparison, the same network architecture and convolutional filter numbers were used for the image encoding module between our method and the residual 3D UNet and the same image pre-processing and augmentation methods were applied during the training of all methods. For Voxel2Mesh, we reduced the resolution of the template mesh such that the total memory consumption during training can fit within the memory available on our Nvidia GeForce GTX 1080 Ti GPU (11 GB). The final mesh resolution is thus halved compared to the original implementation \citep{Voxel2Mesh} and contains 3663 vertices for each cardiac structures. In contrast, our method can process a template mesh with 11494 mesh vertices for each cardiac structures within the available GPU memory.

\subsection{Whole Heart Reconstruction for CT and MR images}

We first compare the performance of whole-heart reconstruction from our method against our baselines. In this experiment, we trained and validated our method using both CT and MR images collected from existing public datasets except for the held-out test dataset of the MMWHS challenge, which we used for test-time evaluation. Our training set thus contained 87 CT images and 41 MR images and the validation set contained 15 CT images and 6 MR images. The MMWHS held-out test dataset contained 40 CT images and 40 MR images. We analyzed the performance of our method against baselines in terms of both the accuracy and the quality of the surface reconstructions. \q{surf2im}{We converted the surface predictions of our method and those of Voxel2Mesh to segmentations at the spatial resolution of the input image data, which is the same as the resolution of the segmentations produced by 2D UNet and 3D UNet. This allowed us to evaluate the accuracy of different methods at the same resolution against the ground truth segmentation using the executable provided by the MMWHS challenge organizers. }We also manually labeled the testing images and compared this with the ground truth segmentation of the MMWHS challenge to provide a comparison between the evaluated reconstruction accuracy of our deep-learning-based method and the inter-observer variability in manual delineations. The surface quality was evaluated in terms of surface smoothness, normal consistency and topological correctness.

\begin{table}
\centering
\caption{A comparison of prediction accuracy on MMWHS MR and CT test datasets from different methods. }\label{table:data_large}
\resizebox{\textwidth}{!}{%
\begin{tabular}{lllrrrrrrrr}
\toprule
  &         &    &     Epi &      LA &      LV &      RA &      RV &      Ao &      PA &      WH \\
\midrule
\multirow{20}{*}{CT} & \multirow{5}{*}{Dice ($\uparrow$)} & Ours &   \textbf{0.899} &  \textbf{0.932} &   \textbf{0.940} &   \textbf{0.892} &   \textbf{0.910} &   \textbf{0.950} &   \textbf{0.852} &   \textbf{0.918} \\
  &         & 2DUNet &   \textbf{0.899} &   0.931 &   0.931 &   0.877 &   0.905 &   0.934 &   0.832 &   0.911 \\
  &         & 3DUNet &   0.863 &   0.902 &   0.923 &   0.868 &   0.876 &   0.923 &   0.813 &   0.888 \\
  &         & Voxel2Mesh &   0.775 &   0.888 &   0.910 &   0.857 &   0.885 &   0.874 &   0.758 &   0.865 \\
\cline{3-11}
  &         & Manual &   0.919 &   0.938 &   0.941 &   0.894 &   0.917 &   0.955 &   0.854 &   0.925 \\
\cline{2-11}
  & \multirow{5}{*}{Jaccard ($\uparrow$)} & Ours &   \textbf{0.819} &   \textbf{0.875} &   \textbf{0.888} &   \textbf{0.809} &   \textbf{0.837} &   \textbf{0.905} &   \textbf{0.755} &   \textbf{0.849} \\
  &         & 2DUNet &   0.817 &   0.872 &   0.873 &   0.787 &   0.828 &   0.879 &   0.726 &   0.837 \\
  &         & 3DUNet &   0.762 &   0.825 &   0.861 &   0.769 &   0.783 &   0.860 &   0.695 &   0.799 \\
  &         & Voxel2Mesh &   0.638 &   0.801 &   0.839 &   0.754 &   0.795 &   0.778 &   0.619 &   0.763 \\
\cline{3-11}
  &         & Manual &   0.852 &   0.884 &   0.890 &   0.814 &   0.848 &   0.914 &   0.759 &   0.860 \\
\cline{2-11}
  & \multirow{5}{*}{ASSD (mm) ($\downarrow$)} & Ours &   1.335 &   \textbf{1.042} &   \textbf{0.842} &   \textbf{1.583} &   1.176 &   \textbf{0.531} &   1.904 &   1.213 \\
  &         & 2DUNet &   \textbf{0.808} &   1.049 &   0.905 &   1.719 &   \textbf{1.064} &   0.645 &   \textbf{1.551} &   \textbf{1.088} \\
  &         & 3DUNet &   1.443 &   1.528 &   1.024 &   1.943 &   1.663 &   0.814 &   2.194 &   1.552 \\
  &         & Voxel2Mesh &   1.714 &   1.696 &   1.266 &   2.020 &   1.492 &   1.341 &   3.398 &   1.848 \\
\cline{3-11}
  &         & Manual &   1.437 &   0.936 &   0.815 &   1.541 &   0.983 &   0.480 &   1.455 &   1.106 \\
\cline{2-11}
  & \multirow{5}{*}{HD (mm) ($\downarrow$)} & Ours &  14.393 &  10.407 &  10.325 &  13.639 &  13.360 &   \textbf{9.407} &  \textbf{26.616} &  28.035 \\
  &         & 2DUNet &   \textbf{9.980} &   8.773 &   \textbf{6.098} &  13.624 &  10.016 &  10.013 &  27.834 &  28.727 \\
  &         & 3DUNet &  13.635 &  10.814 &   9.580 &  16.031 &  15.635 &  13.326 &  26.941 &  31.088 \\
  &         & Voxel2Mesh &  13.564 &   \textbf{8.743} &   6.248 &  \textbf{12.116} &   \textbf{9.601} &  12.080 &  26.252 &  \textbf{27.459} \\
\cline{3-11}
  &         & Manual &  14.446 &  12.677 &  12.619 &  15.313 &  13.496 &  11.189 &  25.449 &  27.181 \\
\cline{1-11}
\cline{2-11}
\multirow{20}{*}{MR} & \multirow{5}{*}{Dice ($\uparrow$)} & Ours &   \textbf{0.797} &   \textbf{0.881} &   \textbf{0.922} &   \textbf{0.888} &   \textbf{0.892} &   \textbf{0.890} &   \textbf{0.816} &   \textbf{0.882} \\
  &         & 2DUNet &   0.795 &   0.864 &   0.896 &   0.852 &   0.865 &   0.869 &   0.772 &   0.859 \\
  &         & 3DUNet &   0.761 &   0.852 &   0.879 &   0.866 &   0.828 &   0.742 &   0.764 &   0.840 \\
  &         & Voxel2Mesh &   0.602 &   0.734 &   0.852 &   0.774 &   0.830 &   0.700 &   0.506 &   0.766 \\
\cline{3-11}
  &         & Manual &   0.830 &   0.885 &   0.925 &   0.887 &   0.894 &   0.885 &   0.807 &   0.887 \\
\cline{2-11}
  & \multirow{5}{*}{Jaccard ($\uparrow$)} & Ours &   \textbf{0.671} &   \textbf{0.791} &   \textbf{0.85}8 &   \textbf{0.801} &  \textbf{ 0.812} &   \textbf{0.805} &   \textbf{0.697} &   \textbf{0.790} \\
  &         & 2DUNet &   0.668 &   0.765 &   0.817 &   0.752 &   0.771 &   0.774 &   0.641 &   0.757 \\
  &         & 3DUNet &   0.626 &   0.756 &   0.802 &   0.766 &   0.728 &   0.650 &   0.639 &   0.732 \\
  &         & Voxel2Mesh &   0.443 &   0.584 &   0.752 &   0.635 &   0.721 &   0.552 &   0.352 &   0.626 \\
\cline{3-11}
  &         & Manual &   0.713 &   0.797 &   0.862 &   0.799 &   0.812 &   0.798 &   0.681 &   0.798 \\
\cline{2-11}
  & \multirow{5}{*}{ASSD (mm) ($\downarrow$)} & Ours &   2.198 &   \textbf{1.401} &   \textbf{1.183} &   \textbf{1.611} &   \textbf{1.333} &   2.648 &   2.689 &   1.775 \\
  &         & 2DUNet &   \textbf{1.830} &   1.488 &   1.455 &   1.715 &   1.483 &   \textbf{2.447} &   \textbf{1.820} &   \textbf{1.690} \\
  &         & 3DUNet &   2.175 &   2.503 &   1.836 &   1.890 &   2.871 &   4.092 &   1.952 &   2.037 \\
  &         & Voxel2Mesh &   2.505 &   3.365 &   2.506 &   3.475 &   2.233 &   4.614 &   6.078 &   3.359 \\
\cline{3-11}
  &         & Manual &   1.837 &   1.301 &   1.070 &   1.463 &   1.218 &   2.159 &   1.581 &   1.485 \\
\cline{2-11}
  & \multirow{5}{*}{HD (mm) ($\downarrow$)} & Ours &  \textbf{16.923} &  11.723 &  10.891 &  14.810 &  13.463 &  \textbf{22.219} &  19.345 &  \textbf{27.701} \\
  &         & 2DUNet &  19.139 &  \textbf{10.781} &   \textbf{9.958} &  \textbf{14.530} &  \textbf{13.082} &  22.567 &  \textbf{16.721} &  28.350 \\
  &         & 3DUNet &  28.159 &  23.640 &  21.494 &  18.949 &  21.095 &  37.937 &  17.055 &  43.022 \\
  &         & Voxel2Mesh &  20.156 &  13.416 &  10.301 &  15.796 &  11.672 &  27.806 &  26.464 &  33.020 \\
\cline{3-11}
  &         & Manual &  15.854 &  12.444 &  12.125 &  14.376 &  13.145 &  21.783 &  13.754 &  25.336 \\
\bottomrule
\end{tabular}
}
\end{table}

Table \ref{table:data_large} shows the average Dice and Jaccard scores, \textcolor{black}{average symmetric surface distance (ASSD) and Hausdorff distance (HD)} of the reconstruction results of both the whole heart and individual cardiac structures for the MMWHS test dataset. For both CT and MR data, our method consistently outperformed our baselines in terms of Dice and Jaccard scores for both whole heart and all individual cardiac structures. In terms of surface ASSD and HD measures for the whole heart or individual cardiac structures, our method was the best or the second among the four deep-learning-based methods compared. To provide further details on segmentation accuracy, Figure \ref{figure:boxplot-large} gives the distribution of different segmentation accuracy metrics for whole heart and individual cardiac structures. Overall, our method demonstrated advantages of whole heart reconstruction for both CT and MR images, and 2D UNet was the closest to ours compared with 3D UNet or Voxel2Mesh. All methods produced better reconstruction for CT images than for MR images. Furthermore, there are no significant differences between the evaluated Dice scores of our methods and those of our manual labeling, except for left ventricle epicardium (p$<$0.05). That is, the discrepancy between our predicted whole-heart reconstruction and the ground truths provided by the MMWHS challenge is comparable to the inter-observer variability of manual whole-heart segmentation.

\begin{figure}[H]
\centering
\includegraphics[width=\textwidth]{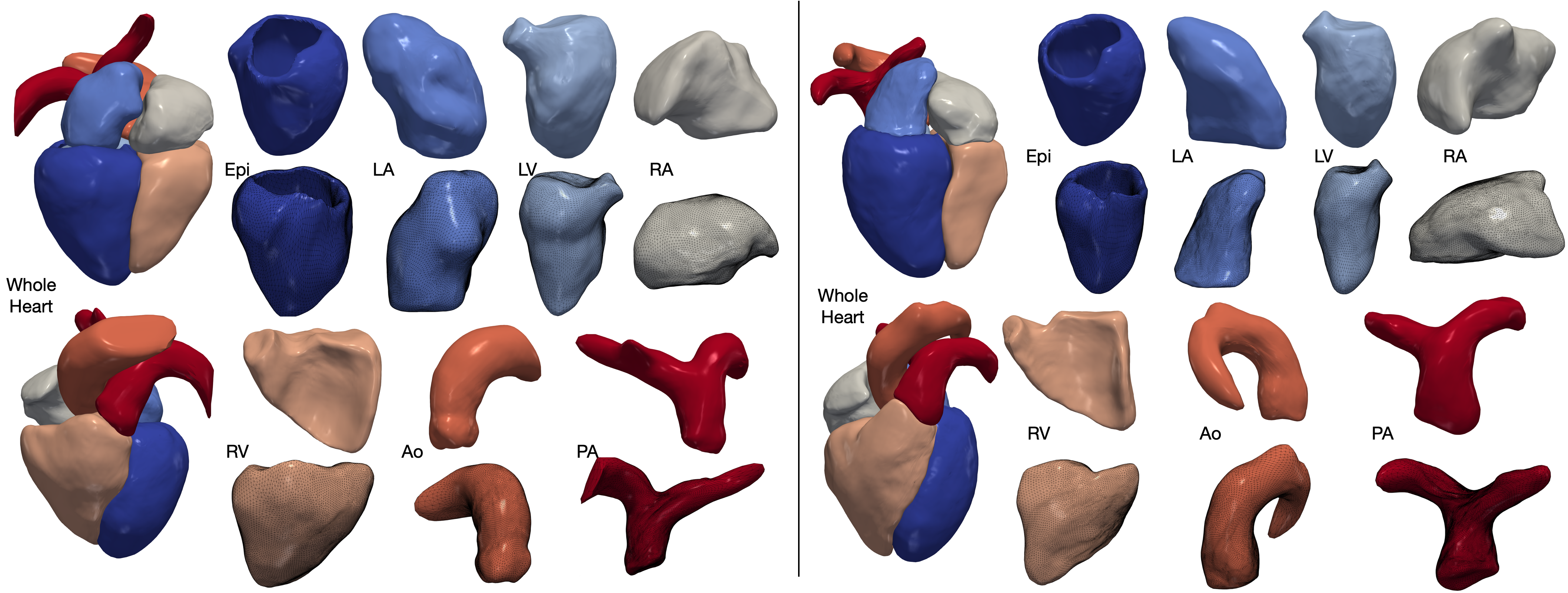}   
\caption{Example reconstructions from our method for CT (left) and MR (right) data selected from MMWHS test dataset. Our method reconstructs the whole heart consisting of seven cardiac structures, including the four heart chambers, left ventricle epicardium, aorta and pulmonary arteries. Geometry of each reconstructed cardiac structure is demonstrated in two different views, with the bottom view also displaying the meshes.} 
\label{figure:structures}
\end{figure}

Figure \ref{figure:structures} displays two examples of the reconstruction results for CT and MR from the MMWHS test dataset, including the surface meshes of individual cardiac structures. Despite starting from a generic template, our method is able to accurately map a template sphere to various cardiac structures with disparate shapes such as the left ventricle epicardium and the pulmonary artery. Moreover, we are able to generate smooth surface reconstruction with consistent normal while capturing the details of individual cardiac structures such as mitral annulus on the left ventricle epicardium, aortic outlet of the left ventricle and the aortic sinus. 

\begin{figure}[H]
\centering
\includegraphics[width=\textwidth]{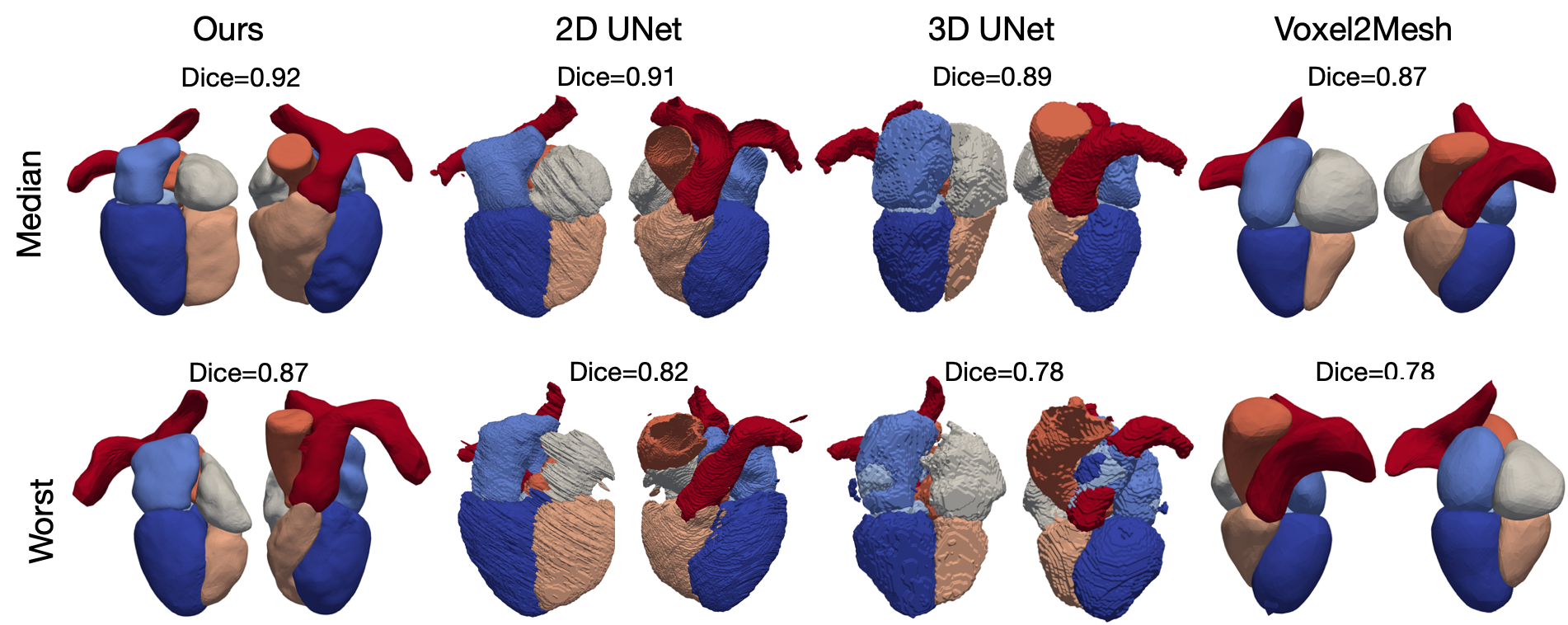}
\caption{Visualizations of the median and worst reconstruction results among the MMWHS CT test dataset in terms of whole-heart Dice scores for all compared methods.} 
\label{figure:ct_median_worst}
\end{figure}

\begin{figure}[H]
\centering
\includegraphics[width=\textwidth]{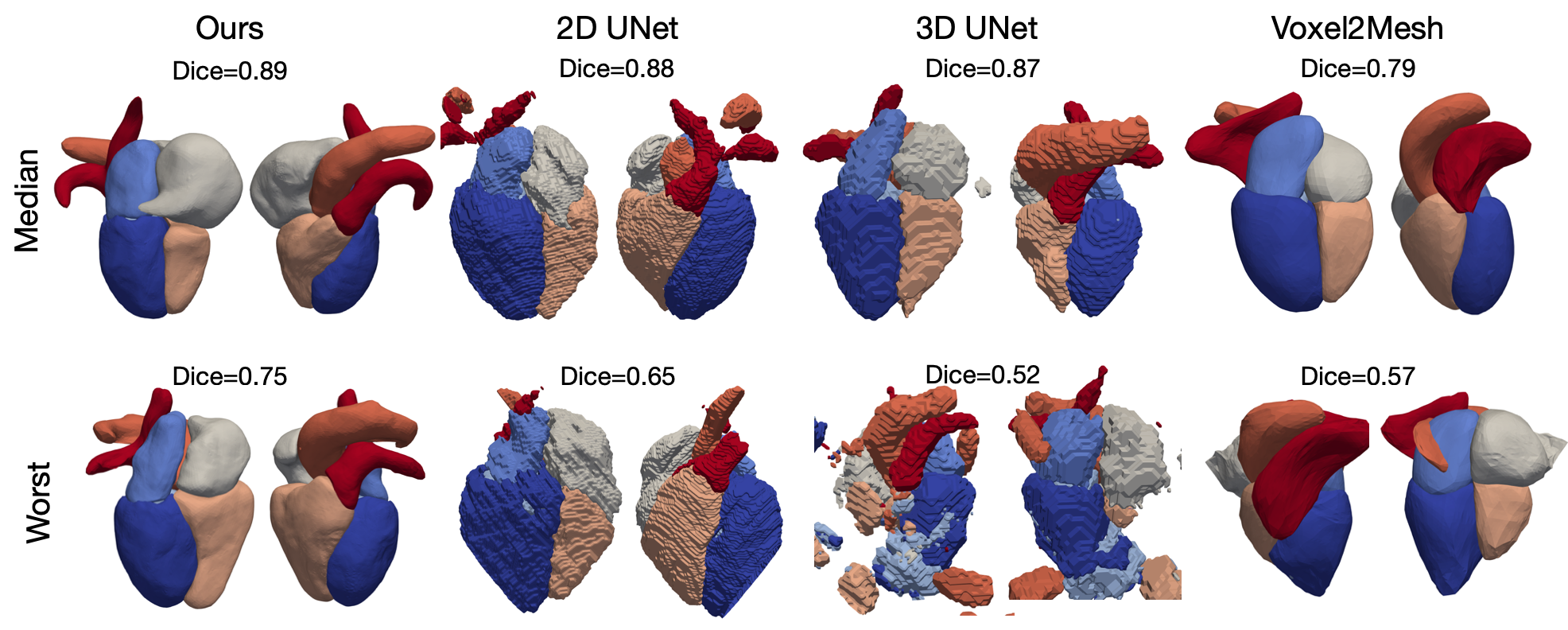}
\caption{Visualizations of the median and worst reconstruction results among the MMWHS MR test dataset in terms of whole-heart Dice scores for all compared methods.} 
\label{figure:mr_median_worst}
\end{figure}

Figure \ref{figure:ct_median_worst} and \ref{figure:mr_median_worst} visualize the median and worst results from the different methods for CT and MR images, respectively, from the MMWHS test dataset. The surface meshes of 2D UNet and 3D UNet were extracted from the segmentation results using the marching cube algorithm. As shown, our method is able to construct smooth geometries while segmentation based methods, such as 2D UNet or 3D UNet, produced surfaces with staircase artifacts. Such artifacts require surface post-processing techniques such as Laplacian smoothing that often also degrade true features. Generally, all four methods are able to produce reasonable median cases from CT data. For MR data, our method produced reasonable reconstructions, while the 2D UNet and 3D UNet produced reconstructions with disconnected regions that would require post-processing to remove or connect. Voxel2Mesh was unable to capture detailed shapes of some structures such as the bifurcation of the pulmonary artery branches. In the worst cases for both CT and MR, our method nonetheless produced realistic shapes. However, 2D UNet and 3D UNet predicted geometries with missing parts, noisy surfaces, incorrect classifications and/or disconnected regions that would require significant post-processing. Voxel2Mesh predicted worst-case geometries that deviated largely from ground truths and had major surface artifacts. \q{surfacequality}{To provide quantitative comparison on the surface quality produced by different methods, Table \ref{table:surface} displays average normal error (ANE), average normalized Laplacian distance (ANLD), and percentage mesh self-intersection of the reconstruction results. The average normal error measures the discrepancy between the point normals on the reconstruction and the ground truth. The ANLD measures the local smoothness of the meshes. The percentage self-intersection measures the local topological correctness of the meshes. Detailed definitions of these metrics can be found in Appendix \ref{appendix:metrics}. Overall, our method demonstrated the best surface smoothness and normal consistency for all cardiac structures for CT data and for most cardiac structures for MR data. For topology correctness, our method produced meshes with a small number of self-intersections. In contrast, the segmentation-based approaches apply the Marching Cube algorithm to generate uniform and watertight surface meshes without self-intersection.} 

\begin{table}
\centering{\color{black}
\caption{A comparison of the quality of the whole heart surfaces from different methods on MMWHS MR and CT test datasets.}\label{table:surface}
\resizebox{\textwidth}{!}{%
\begin{tabular}{lllrrrrrrr}
\toprule
   &     &            &    Epi &     LA &     LV &     RA &     RV &     Ao &     PA \\
\midrule
\multirow{12}{*}{CT} & \multirow{4}{*}{ANE ($\downarrow$)} & Ours &  \textbf{0.004} &  \textbf{0.003} &  \textbf{0.003} &  \textbf{0.006} &  \textbf{0.006} &  \textbf{0.002} &  \textbf{0.008} \\
   &     & 2DUNet &  0.036 &  0.012 &  0.014 &  0.022 &  0.030 &  0.010 &  0.023 \\
   &     & 3DUNet &  0.033 &  0.015 &  0.014 &  0.017 &  0.021 &  0.010 &  0.016 \\
   &     & Voxel2Mesh &  0.136 &  0.171 &  0.129 &  0.150 &  0.136 &  0.143 &  0.105 \\
\cline{2-10}
   & \multirow{4}{*}{ANLD ($\downarrow$)} & Ours &  \textbf{0.091} &  \textbf{0.078} &  \textbf{0.085} &  \textbf{0.085} &  \textbf{0.080} &  \textbf{0.076} &  \textbf{0.090} \\
   &     & 2DUNet &  0.287 &  0.280 &  0.287 &  0.282 &  0.286 &  0.265 &  0.278 \\
   &     & 3DUNet &  0.292 &  0.284 &  0.295 &  0.295 &  0.290 &  0.273 &  0.291 \\
   &     & Voxel2Mesh &  0.113 &  0.119 &  0.129 &  0.134 &  0.126 &  0.140 &  0.160 \\
\cline{2-10}
   & \multirow{4}{*}{Intersection (\%) ($\downarrow$)} & Ours &  0.014 &  0.006 &  0.017 &  0.007 &  0.024 &  0.005 &  0.049 \\
   &     & 2DUNet &  \textbf{0.000} &  \textbf{0.000} &  \textbf{0.000} &  \textbf{0.000} &  \textbf{0.000} &  \textbf{0.000} &  \textbf{0.000} \\
   &     & 3DUNet &  \textbf{\textbf{0.000}} &  \textbf{0.000} &  \textbf{0.000} &  \textbf{0.000} &  \textbf{0.000} &  \textbf{0.000} &  \textbf{0.000} \\
   &     & Voxel2Mesh &  0.269 &  \textbf{0.000} &  \textbf{0.000} &  0.003 &  \textbf{0.000} &  \textbf{0.000} &  0.020 \\
\cline{1-10}
\cline{2-10}
\multirow{12}{*}{MR} & \multirow{4}{*}{ANE ($\downarrow$)} & Ours &  \textbf{0.015} &  \textbf{0.012} &  \textbf{0.007} &  \textbf{0.010} &  \textbf{0.013} &  \textbf{0.015} &  0.017 \\
   &     & 2DUNet &  0.057 &  0.016 &  0.022 &  0.024 &  0.033 &  0.026 &  0.018 \\
   &     & 3DUNet &  0.056 &  0.017 &  0.035 &  0.020 &  0.034 &  0.037 &  \textbf{0.014} \\
   &     & Voxel2Mesh &  0.104 &  0.189 &  0.130 &  0.136 &  0.150 &  0.123 &  0.160 \\
\cline{2-10}
   & \multirow{4}{*}{ANLD ($\downarrow$)} & Ours &  \textbf{0.103} &  \textbf{0.093} &  \textbf{0.088} &  \textbf{0.101} &  \textbf{0.092} &  \textbf{0.088} &  \textbf{0.103} \\
   &     & 2DUNet &  0.287 &  0.274 &  0.285 &  0.282 &  0.276 &  0.275 &  0.289 \\
   &     & 3DUNet &  0.296 &  0.283 &  0.299 &  0.297 &  0.288 &  0.296 &  0.304 \\
   &     & Voxel2Mesh &  0.130 &  0.132 &  0.129 &  0.144 &  0.139 &  0.155 &  0.159 \\
\cline{2-10}
   & \multirow{4}{*}{Intersection (\%) ($\downarrow$)} & Ours &  0.069 &  0.018 &  0.023 &  0.069 &  0.069 &  0.108 &  0.134 \\
   &     & 2DUNet &  \textbf{0.000} & \textbf{0.000} &  \textbf{0.000} &  \textbf{0.000} &  \textbf{0.000} &  \textbf{0.000} &  \textbf{0.000} \\
   &     & 3DUNet &  \textbf{0.000} &  \textbf{0.000} &  \textbf{0.000} &  \textbf{0.000} &  \textbf{0.000} &  \textbf{0.000} &  \textbf{0.000} \\
   &     & Voxel2Mesh &  0.189 &  \textbf{0.000} &  \textbf{0.000} &  0.070 &  \textbf{0.000} &  0.059 &  0.020 \\
\bottomrule 
\end{tabular}
}}
\end{table}

\begin{figure}[H]
\centering
\includegraphics[width=0.9\textwidth]{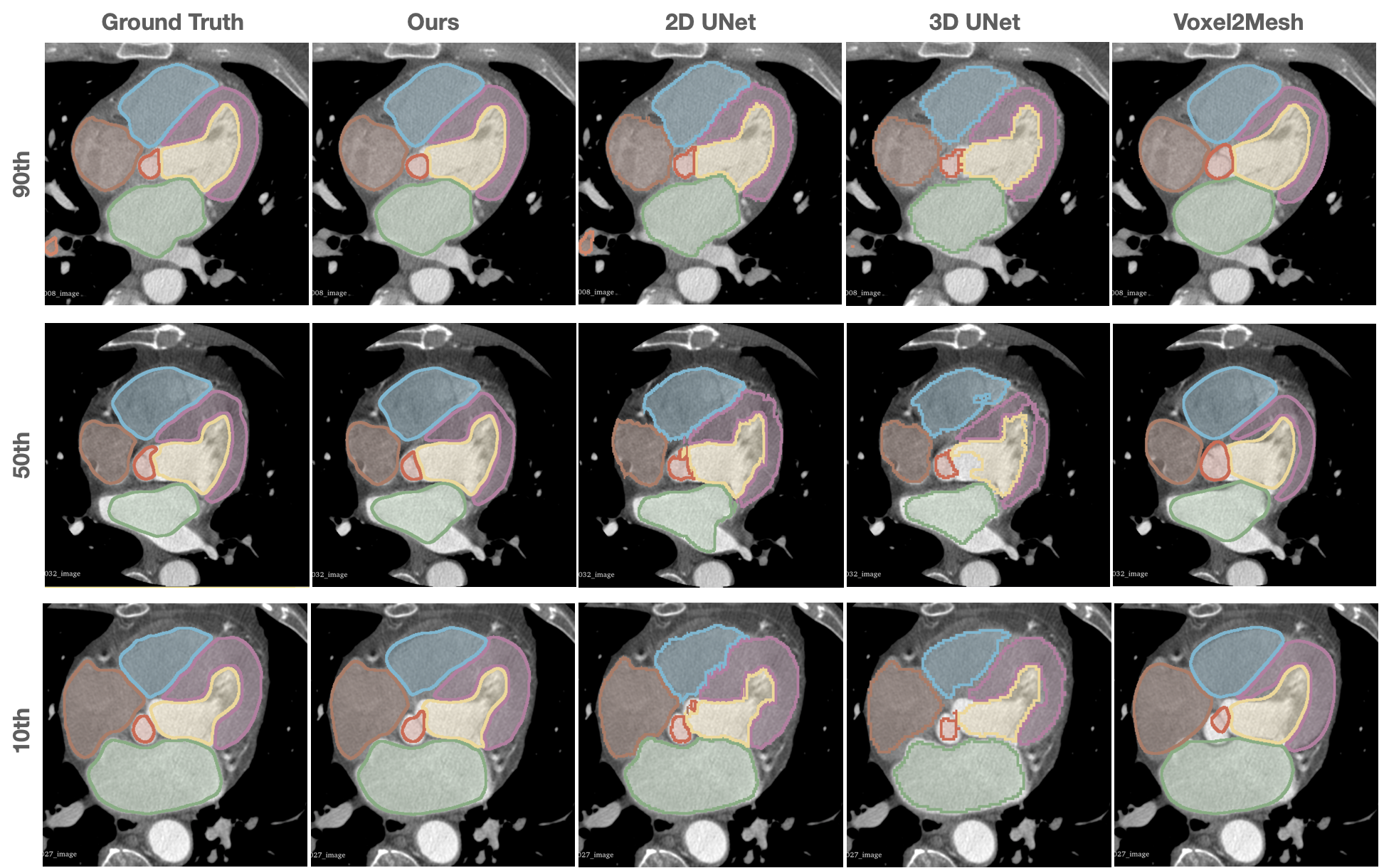}
\caption{Comparison of the predicted whole heart surfaces from different methods for CT test cases. Different rows demonstrated the \textcolor{black}{zoomed-in axial view of the} images and predictions from different test cases with the 10th, 50th, 90th percentiles of Dice scores based on our method.} 
\label{figure:ct_seg_compare}
\end{figure}

\begin{figure}[H]
\centering
\includegraphics[width=0.9\textwidth]{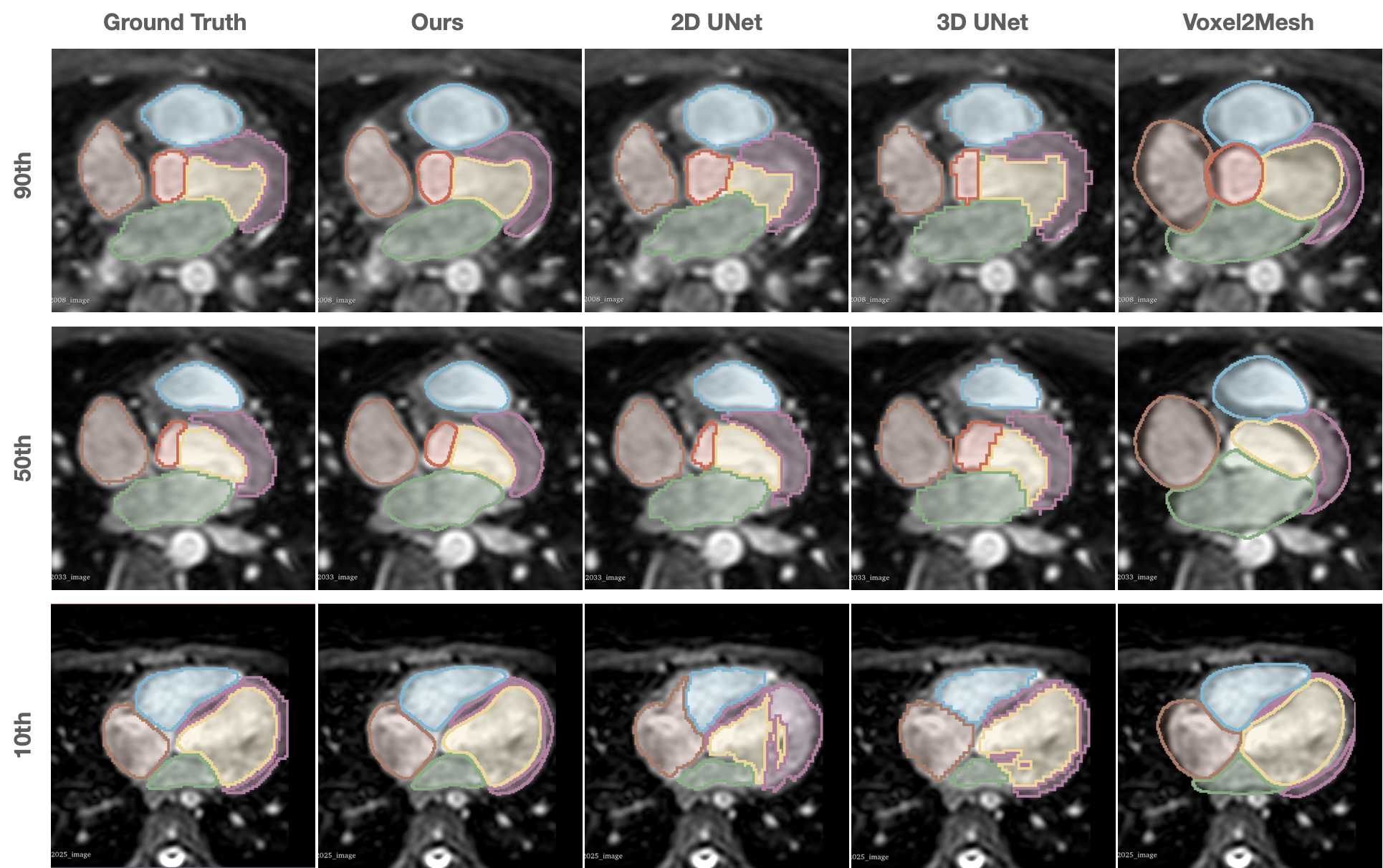}
\caption{Comparison of the predicted whole heart surfaces from different methods for MR test cases. Different rows demonstrated the \textcolor{black}{zoomed-in axial view of the} images and predictions from different test cases with the 10th, 50th, 90th percentiles of Dice scores based on our method.} 
\label{figure:mr_seg_compare}
\end{figure}

Figures \ref{figure:ct_seg_compare} and \ref{figure:mr_seg_compare} provide further qualitative comparisons of the results from the different methods. As shown in Fig.~\ref{figure:mr_seg_compare}, our method was able to generate smoother reconstruction than the ground truth segmentation on MR images that have relatively large voxel spacing. In contrast, 2D UNet that produces segmentation on a slice-by-slice manner along the sagittal view, may suffer from inconsistency between adjacent slices, leading to coarse segmentation when looking from the axial view that the 2D UNet was not trained on. 3D UNet, limited by the memory constrain of GPU, can only produce coarse segmentation on a down-sampled voxel grid of $128\times 128 \times 128$ for high-resolution CT image data. Although Voxel2Mesh can also produce smooth surface meshes, it tends to predict surfaces that lack shape details and do not match well with the true boundary of many cardiac structures. 

\begin{figure}[H]
\centering
\includegraphics[width=0.9\textwidth]{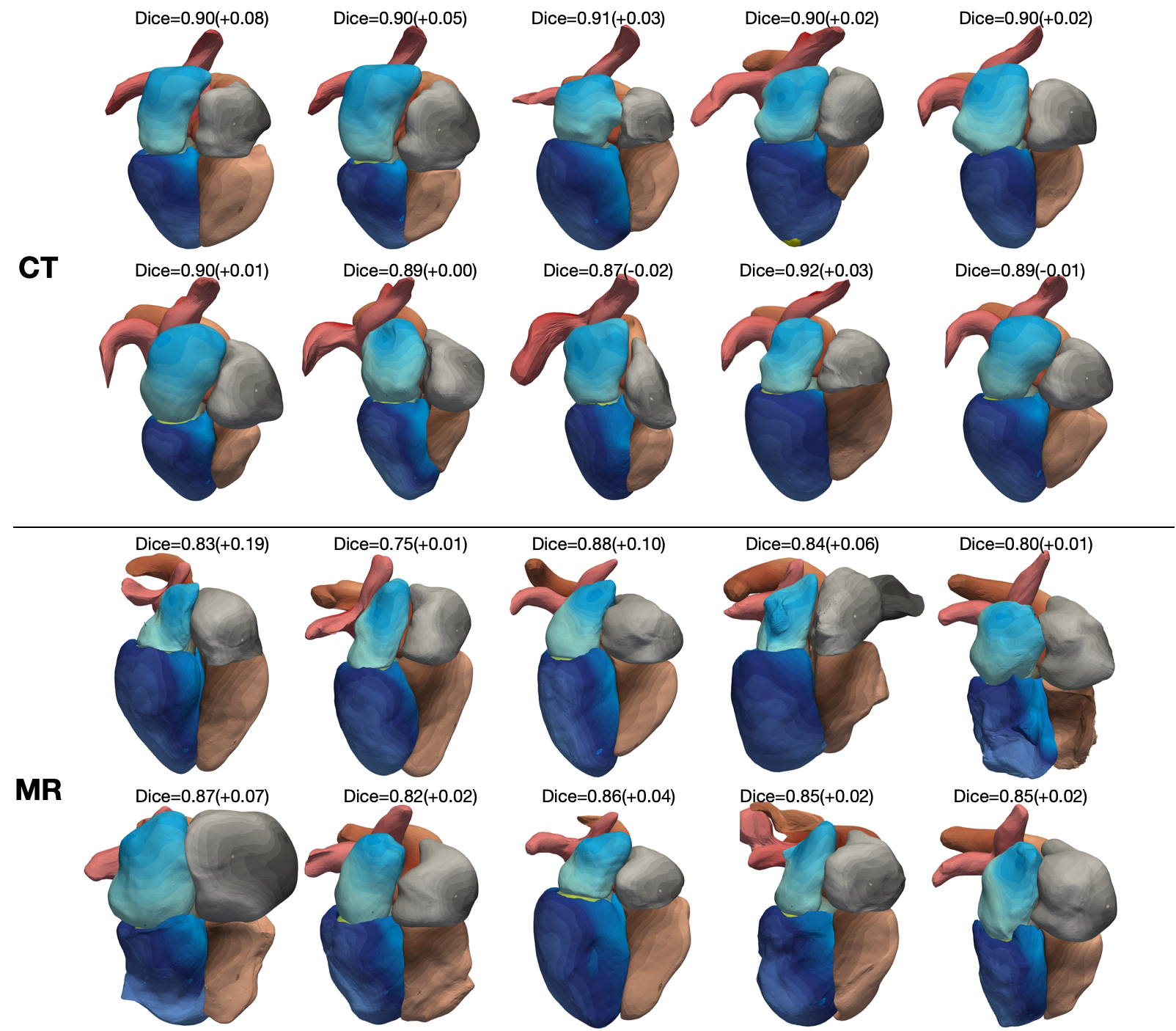}
\caption{Whole heart reconstruction results from the 10 most challenging CT and MR images for which 2D UNet predicted less accurate segmentations in terms of Dice scores compared with the rest images in the MMWHS test datasets. On top of each case is the whole-heart Dice score of our result and the difference in whole-heart Dice score compared with 2D UNet reconstruction. The color map denotes the indices of mesh vertices and demonstrates the correspondence of mesh vertices across reconstructed meshes from different images. } 
\label{figure:example_meshes}
\end{figure}

Figure \ref{figure:example_meshes} shows reconstruction results for the 10 most challenging CT and MR images for which 2D UNet (the method that demonstrated closest performance to our method) predicted less accurate segmentations in terms of Dice scores compared with the rest images in the test datasets. For all the 10 MR images and 8 out of the 10 CT images, our method produced whole-heart reconstructions with improved Dice scores. For all these CT cases, we were able to generate accurate reconstruction with Dice scores above 0.87 and smooth surfaces without obvious artifacts. However, for the 10 MR cases, although we demonstrated improvement against 2D UNet predictions, we observed buckling and bumpiness on mesh surfaces of one or more cardiac structures for 5 out of 10 cases. 

\begin{figure}[H]
\centering
\includegraphics[width=0.8\textwidth]{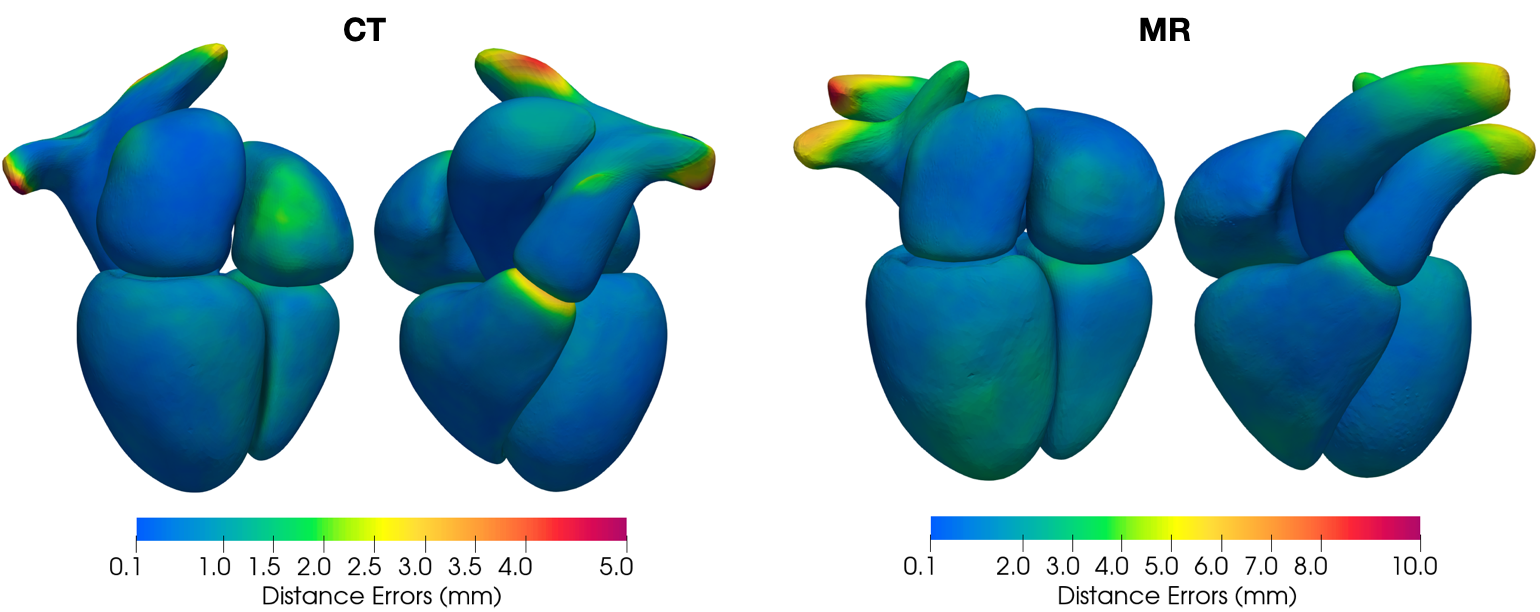}
\caption{Distribution of the average surface distance errors on mean whole heart shapes from the CT and MR data in MMWHS test dataset.} 
\label{figure:distance_error}
\end{figure}

Interestingly, as indicated by the point-correspondence color maps in Figure \ref{figure:example_meshes}, although we did not explicitly train our method to generate feature-corresponding meshes across different input images, our method was generally able to consistently deform template meshes to map mesh vertices to similar structural features of the heart for different images. This behavior allowed convenient generation of the mean whole heart shapes from the test dataset by computing the average coordinates of each vertex. Figure \ref{figure:distance_error} demonstrates the mean whole heart shapes for MR and CT images from the MMWHS test dataset, respectively, and the distribution of the average surface distance errors on the whole heart compared with manual ground truths. For both CT and MR data, locations that suffer from higher surface errors include the ends of the aorta and pulmonary arteries, boundaries between the right ventricle and the pulmonary artery, boundaries between the right atrium and the ventricle, and the inferior vena cava region on the right atrium. We note that several of the locations of largest error are artificial boundaries, or arbitrary truncations of vessels extending away form the heart. 

\subsection{Generalization to Low-Resolution MR Images}
Cardiac MR image data are often acquired in a slice-by-slice manner and thus often vary in through-plane resolution due to the use of different acquisition protocols and vendors. For MR images with low through-plane resolution, accurately constructing smooth surface geometries is challenging since a method would need to complete the cardiac structures that are not captured between the slices. Therefore, having trained our method on MR images with high through-plane resolution to produce detailed whole heart geometries, we evaluate the performance of our method on MR images with lower through-plane resolution and compare it with our baselines. To disentangle the effect of through-plane resolution from the effect of other variations of MR images, we first generate low-resolution MR images from our validation data by down-sampling the images to various slice thicknesses. We then evaluate the robustness of different methods to challenging real low-resolution MR images that significantly differ from our training datasets. Namely, we used data from our cine MR images, which were acquired with large slice thicknesses (8-10 mm), different acquisition planes, and from a different clinical center.  

\begin{figure}[H]
\centering
\includegraphics[width=0.9\textwidth]{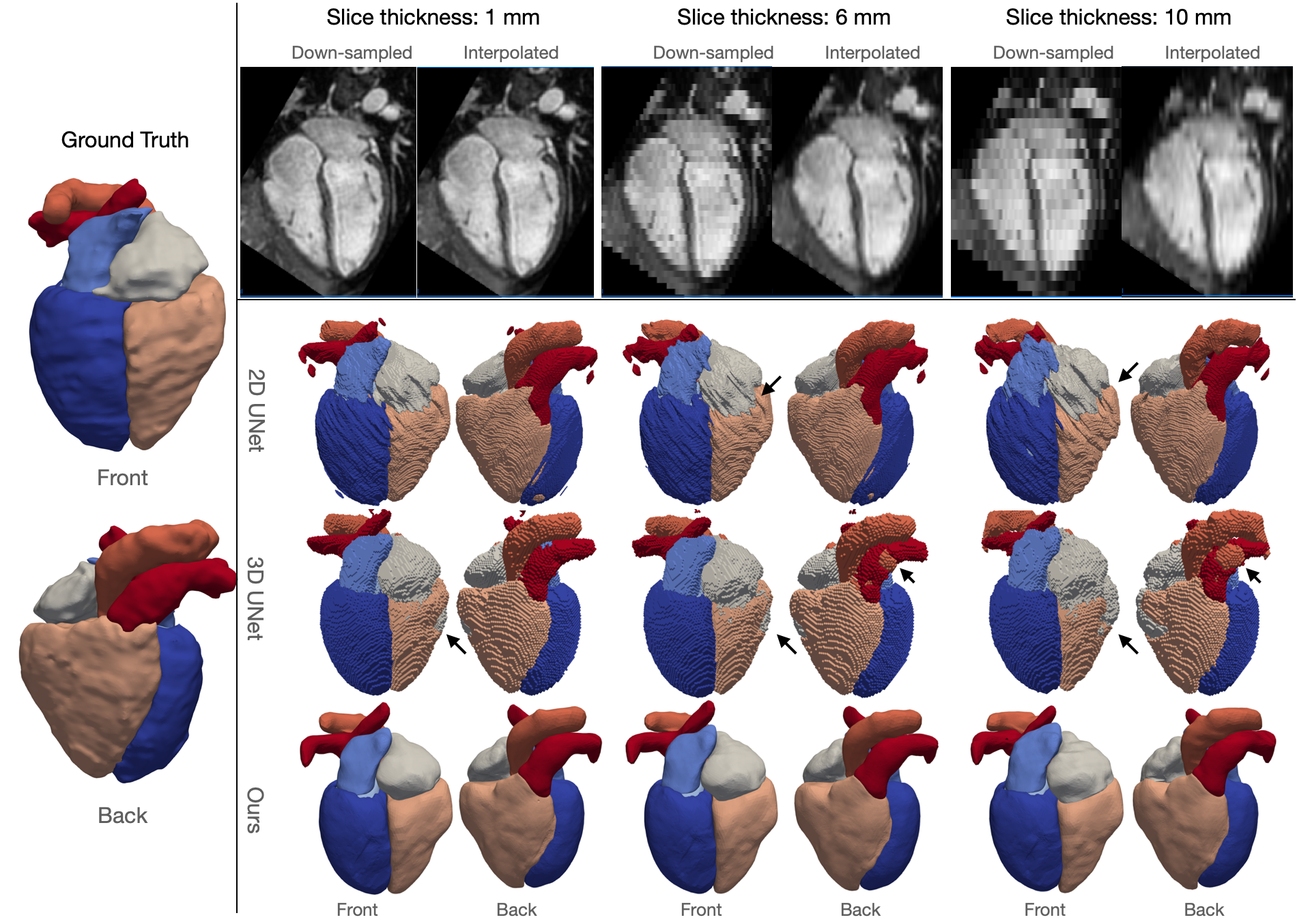}
\caption{Robustness of different methods to through-plane resolution changes of MR images. Left panel shows the front and back views of the ground truth surfaces; top panel shows example slices along the down-sampling axis of images down-sampled to varying slice thicknesses, and bottom panel shows front and back views of predicted whole-heart surfaces from different methods corresponding to different slice thickness values.} 
\label{figure:resolution}
\end{figure}

\subsubsection{Synthetic Low-resolution MR data} 
\begin{figure}[H]
\centering
\includegraphics[width=0.8\textwidth]{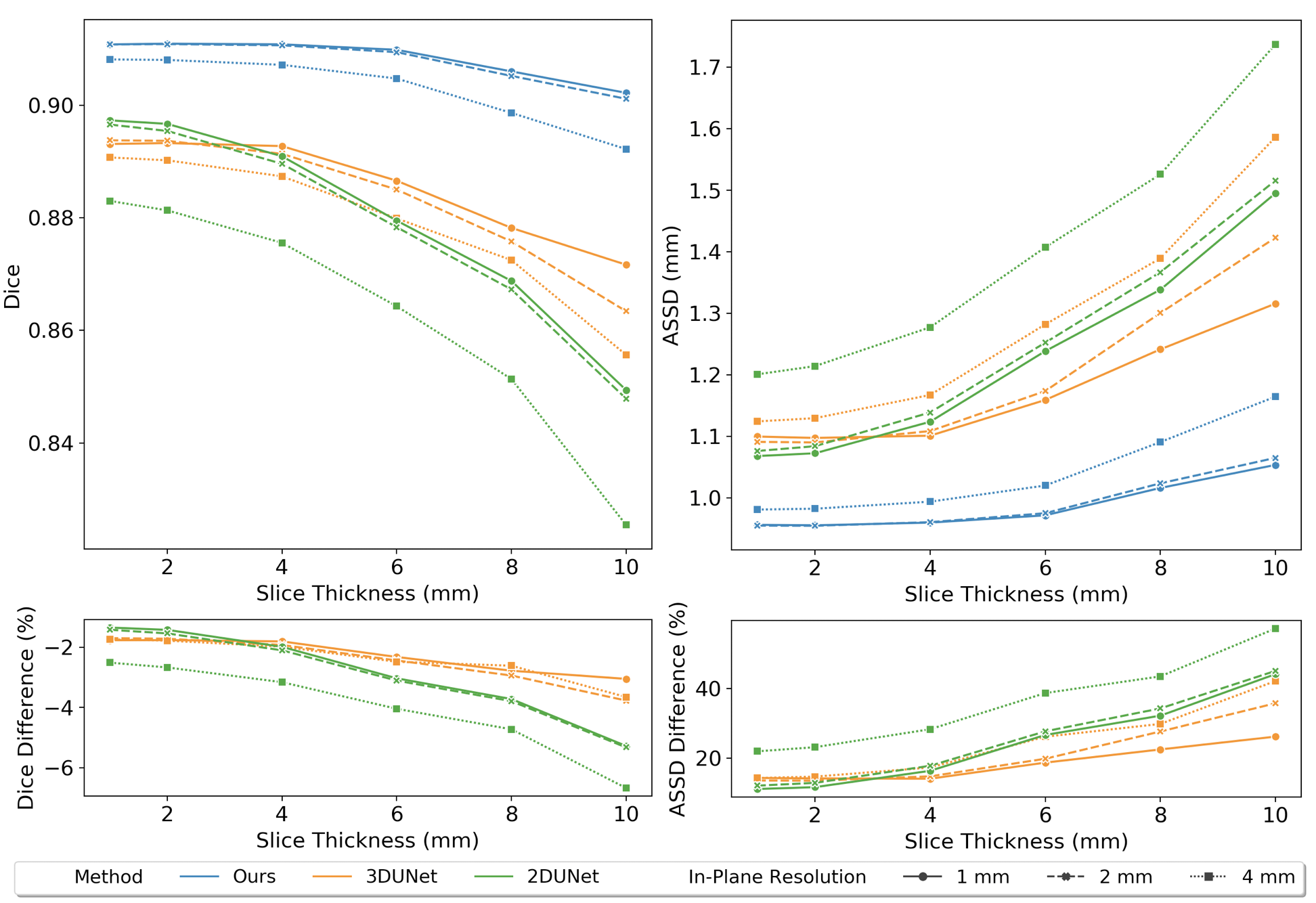}
\caption{Relation of Dice and ASSD values of whole-heart surfaces to through-plane resolution of MR images. Comparison between different methods and different in-plane resolutions are indicated by lines with different color and different styles, respectively. The bottom panel shows the average percentage differences of Dice or ASSD values between our method and 2D UNet or 3D UNet across all validation images. } 
\label{figure:resolution_line}
\end{figure}

Figure \ref{figure:resolution} displays an example of down-sampling an input image dataset \textcolor{black}{along the longitudinal direction of the left ventricle }to various slice thickness of 1 mm, 6 mm, and 10 mm, as well as the corresponding predictions from our method, 2D UNet and 3D UNet, respectively. For low through-plane resolution images, the same linear resampling method was applied as before to interpolate the 3D image volume to the sizes required by the neural network models. As the slice thickness was increased to up to 10 mm, while 2DUNet can generally produce consistent segmentation on 2D slices, \textcolor{black}{it produces uneven 3D geometries due to poor inter-slice consistency.} In contrast, the 3D UNet is able to produce smoother surfaces by accounting for inter-slice information. However, as slice thickness increases, the 3D UNet produces less accurate segmentation, such as incorrectly classifying a part of the RV into the RA \textcolor{black}{and a part of the PA into the aorta, as shown by the arrows} in Figure \ref{figure:resolution}. Our method, however, for all different slice thicknesses, produces consistent reconstructions that closely resembles the ground truth surfaces and are free of any major artifacts. Figure \ref{figure:resolution_line} displays quantitative evaluations of the reconstruction performance on various image resolutions. Regardless of slice thickness values considered, our method out-performed 2D UNet and 3D UNet both in terms of Dice and ASSD. Moreover, as slice thickness increases from 1 mm to 10 mm, in general, we observed increasing improvement of our method compared with 2D UNet or 3D UNet. Furthermore, by taking a 3D image volume as the input, our method and 3D UNet are more robust to additional in-plane resolution changes than the 2D UNet. Both our method and the 3D UNet demonstrated a smaller reduction in accuracy with 4 times reduction of in-plane resolution. 

\subsubsection{Real Low-resolution MR data}
We evaluated the robustness of our method on the challenging cine MR dataset, which significantly differs from our training datasets in terms of the through-plane resolution, imaging plane orientation and coverage of the heart. \q{low_res_mr}{To generate ground truth segmentation and meshes from low-resolution MR data, we re-sampled such 3D image volume and linearly interpolated between the slices to have an isotropic spacing of 1 mm along all three axes. The ground truth segmentations were obtained by manually segmenting the interpolated image data and manually correcting artifacts due to low through-plane resolution based on prior human expert knowledge of the heart to obtain smooth and physiologically plausible geometries that match with the low-resolution image data as much as possible. }Table \ref{table:data_motion_mr} compares the reconstruction accuracy between our method and the baselines. The reconstruction accuracy was evaluated at two time frames, end diastole and end systole, for each patient. Overall, our method demonstrated high reconstruction accuracy and outperformed the other methods for most cardiac structures in terms of average Dice score and ASSD. 

\begin{figure}[H]
\centering
\includegraphics[width=\textwidth]{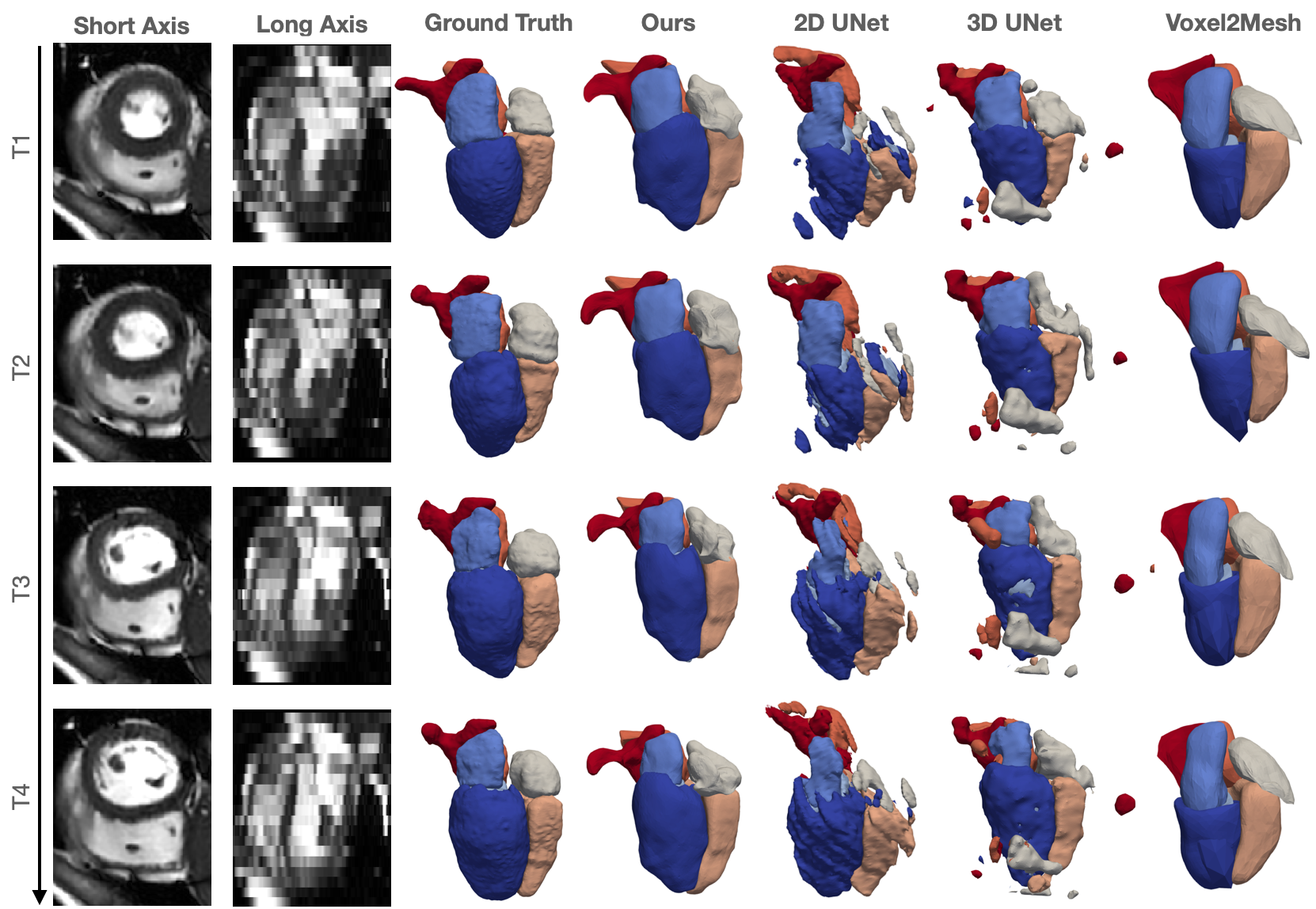}
\caption{Short axis and long axis slices at different time frames for an example cine cardic MR data and the corresponding reconstructed whole heart surfaces from different methods.} 
\label{figure:ucsf}
\end{figure}

Figure \ref{figure:ucsf} compares the whole-heart geometries reconstructed by our method with others for one example of cine cardiac MR images. Our method was able to produce clean surface meshes while at the same capture most of the cardiac structures with reasonable accuracy. In contrast, since these images were acquired on imaging planes that were different from those used in acquiring the training data, 2D UNet produced inaccurate reconstruction and disconnected surfaces. 3D UNet produces more complete reconstruction of the cardiac structures but often produced many disconnected false positive regions. Voxel2Mesh is able to produce clean surface meshes with generally correct topology but the predictions are not accurate. Furthermore, as changes in input images over different time frames are small, our method produced consistent reconstruction over different time phases. However, segentation-based methods, 2D UNet or 3D UNet, often produce inconsistent reconstruction with significant shape or topology changes, despite small changes in input images over different time frames. 
\begin{table}
\centering
\caption{A comparison of prediction accuracy on cine MR dataset from different methods. \textcolor{black}{All accuracy measures are represented by mean $\pm$ standard deviation, which are computed over different patients and  time frames.} }\label{table:data_motion_mr}
\resizebox{\textwidth}{!}{%
\begin{tabular}{llllllllll}
\toprule
          &            &              Epi &               LA &               LV &                RA &               RV &                Ao &                PA &               WH \\
\midrule
\multirow{4}{*}{Dice ($\uparrow$)} & Ours &  \textbf{0.656$\pm$0.169} &  \textbf{0.708$\pm$0.187} &  \textbf{0.822$\pm$0.104} &   \textbf{0.672$\pm$0.114} &  0.643$\pm$0.228 &   \textbf{0.543$\pm$0.255} &   0.445$\pm$0.225 &  \textbf{0.693$\pm$0.112} \\
          & 2D UNet &  0.543$\pm$0.263 &  0.517$\pm$0.283 &  0.734$\pm$0.218 &   0.274$\pm$0.218 &  \textbf{0.644$\pm$0.184} &   0.393$\pm$0.215 &   \textbf{0.487$\pm$0.286} &  0.598$\pm$0.166 \\
          & 3D UNet &  0.546$\pm$0.244 &   0.702$\pm$0.22 &  0.782$\pm$0.134 &   0.598$\pm$0.169 &  0.631$\pm$0.144 &   0.495$\pm$0.175 &   0.285$\pm$0.249 &  0.627$\pm$0.131 \\
          & Voxel2Mesh &  0.438$\pm$0.178 &  0.529$\pm$0.275 &  0.669$\pm$0.135 &    0.54$\pm$0.206 &  0.598$\pm$0.273 &   0.395$\pm$0.246 &   0.223$\pm$0.195 &  0.527$\pm$0.167 \\
\cline{1-10}
\multirow{4}{*}{ASSD (mm) ($\downarrow$)} & Ours &  4.009$\pm$1.118 &  \textbf{4.775$\pm$2.522} &  4.534$\pm$2.195 &   \textbf{5.299$\pm$1.883} &  5.468$\pm$1.856 &   \textbf{6.713$\pm$3.233} &    \textbf{7.463$\pm$3.14} &  \textbf{5.466$\pm$1.613} \\
          & 2D UNet &  4.585$\pm$3.501 &  6.665$\pm$5.147 &    5.204$\pm$3.3 &  10.638$\pm$6.918 &   \textbf{4.12$\pm$2.493} &    8.36$\pm$7.738 &   7.914$\pm$9.257 &  6.784$\pm$3.951 \\
          & 3D UNet &   \textbf{3.498$\pm$2.47} &  4.841$\pm$5.061 &  \textbf{3.228$\pm$2.945} &   8.537$\pm$5.393 &  5.234$\pm$2.466 &  10.022$\pm$6.599 &  11.643$\pm$8.608 &  6.715$\pm$3.091 \\
          & Voxel2Mesh &  5.104$\pm$1.767 &  7.105$\pm$3.082 &  6.763$\pm$2.528 &   6.945$\pm$3.163 &  7.775$\pm$4.613 &   9.181$\pm$4.593 &  12.079$\pm$7.703 &   7.85$\pm$2.881 \\
\bottomrule
\end{tabular}
}
\end{table}

\subsection{Construction of Whole-Heart 4D Models from Motion Image Data}
We further \textcolor{black}{tested} our method on time-series CT datasets. Table \ref{table:data_motion_ct} compares the reconstruction accuracy between our method and the other baseline methods. Similar to above, the reconstruction accuracy was evaluated at two time frames, end diastole and end systole, for each patient. Overall, our method demonstrated high reconstruction accuracy and outperformed the other methods for most cardiac structures in terms of average Dice score and ASSD. 

\begin{figure}[H]
\centering
\includegraphics[width=\textwidth]{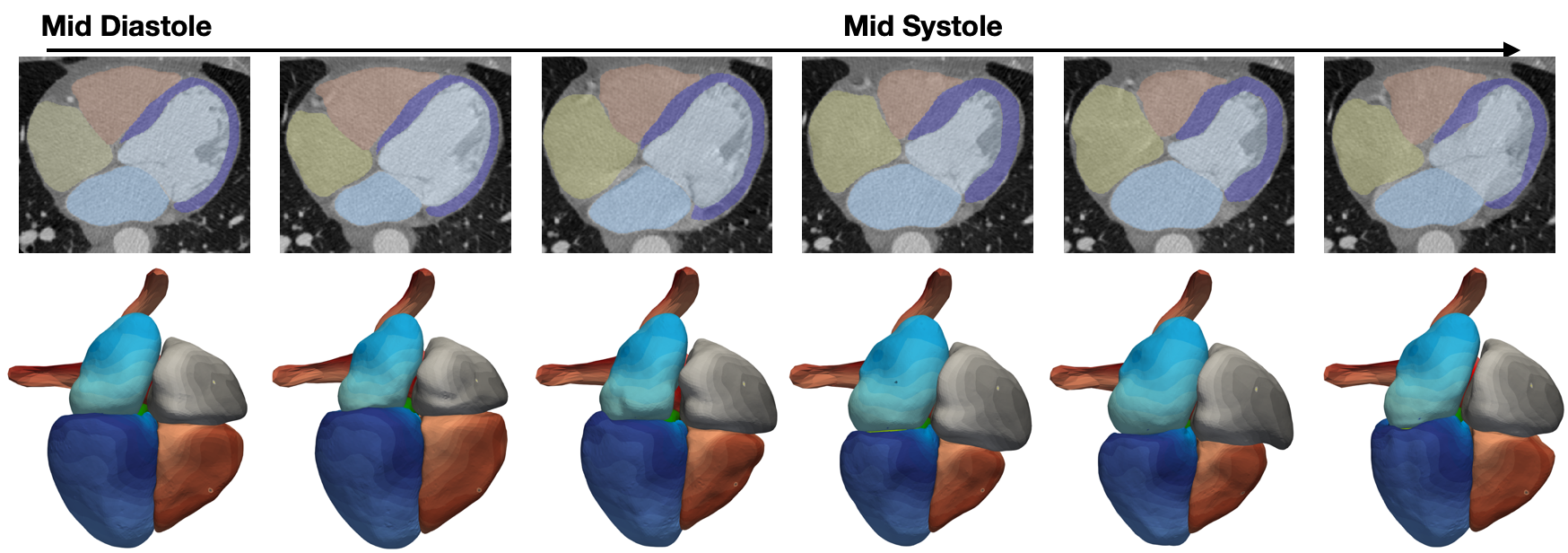}
\caption{Whole-heart reconstruction results for time-series CT data. From left to right, each column displays results at one time frame from middle diastole to early diastole. The top row shows predicted segmentation overlaid with CT images and the bottom row shows the correspondence maps of mesh vertices across reconstructed meshes from different time frames, with same color denoting the same mesh vertices on reconstructed meshes. } 
\label{figure:motion_ct}
\end{figure}

Furthermore, we explore the potential capability of our method to reconstruct dynamic 4D whole-heart models to capture the motion of the heart from time-series image data. Figure \ref{figure:motion_ct} displays example whole-heart reconstruction results of our methods on time-series CT data that consisted of images from 10 time frames over the cardiac cycle for each patient. Although our model predicts mesh reconstructions independently from each time frame, it is able to consistently deform the template meshes such that the same mesh vertices on the template meshes are generally mapped to the same region of the reconstructed geometries across different time frames, as shown by the color maps of vertex IDs in Figure \ref{figure:motion_ct}. Moreover, as demonstrated by the segmentation in Figure \ref{figure:motion_ct}, our method is able to capture the minor changes between time frames. Therefore, our method can potentially be applied to efficiently construct 4D dynamic whole-heart models to capture the motion of a beating heart. 
\begin{table}[H]
\centering
\caption{A comparison of prediction accuracy on time-series CT dataset from different methods. \textcolor{black}{All accuracy measures are represented by mean $\pm$ standard deviation, which are computed over different patients and  time frames.} }\label{table:data_motion_ct}
\resizebox{\textwidth}{!}{%
\begin{tabular}{llllllllll}
\toprule
          &         &              Epi &               LA &               LV &               RA &               RV &               Ao &               PA &               WH \\
\midrule
\multirow{4}{*}{Dice ($\uparrow$)} & Ours &  0.902$\pm$0.035 &   \textbf{0.96$\pm$0.018} &  0.956$\pm$0.033 &  \textbf{0.946$\pm$0.014} &  \textbf{0.944$\pm$0.017} &  \textbf{0.974$\pm$0.006} &  0.798$\pm$0.129 &   \textbf{0.94$\pm$0.012} \\
          & 2D UNet &  \textbf{0.913$\pm$0.028} &  0.958$\pm$0.014 &  \textbf{0.957$\pm$0.023} &  0.927$\pm$0.041 &  0.925$\pm$0.041 &  0.971$\pm$0.009 &  \textbf{0.867$\pm$0.114} &  0.937$\pm$0.022 \\
          & 3D UNet &   0.884$\pm$0.03 &  0.935$\pm$0.012 &   0.946$\pm$0.03 &  0.928$\pm$0.019 &    0.92$\pm$0.02 &   0.955$\pm$0.01 &  0.831$\pm$0.059 &  0.922$\pm$0.014 \\
          & Voxel2Mesh &  0.786$\pm$0.072 &  0.933$\pm$0.019 &  0.928$\pm$0.037 &   0.92$\pm$0.021 &  0.928$\pm$0.019 &  0.924$\pm$0.011 &  0.651$\pm$0.123 &  0.894$\pm$0.014 \\
\cline{1-10}
\multirow{4}{*}{ASSD (mm) ($\downarrow$)} & Ours &  0.697$\pm$0.308 &   \textbf{0.54$\pm$0.205} &  0.574$\pm$0.399 &   \textbf{0.781$\pm$0.21} &  \textbf{0.756$\pm$0.219} &   \textbf{0.28$\pm$0.07}3 &  2.714$\pm$3.079 &  0.906$\pm$0.5 \\
          & 2D UNet &  \textbf{0.634$\pm$0.281} &  0.569$\pm$0.181 &   \textbf{0.538$\pm$0.25} &  1.097$\pm$0.668 &  1.099$\pm$0.737 &  0.281$\pm$0.103 &  \textbf{1.155$\pm$1.019} &  \textbf{0.767$\pm$0.291} \\
          & 3D UNet &   0.811$\pm$0.34 &  0.871$\pm$0.277 &  0.711$\pm$0.381 &  0.993$\pm$0.325 &  1.017$\pm$0.267 &   0.504$\pm$0.19 &  1.598$\pm$1.183 &  0.929$\pm$0.259 \\
          & Voxel2Mesh &  1.297$\pm$0.451 &  0.916$\pm$0.208 &  0.993$\pm$0.423 &  1.194$\pm$0.327 &  1.034$\pm$0.275 &  0.844$\pm$0.124 &  3.788$\pm$2.008 &   1.438$\pm$0.325 \\
\bottomrule
\end{tabular}
}
\end{table}
\textcolor{black}{
\subsection{Impact of Post-Processing on Reconstruction Performance}
}

\q{seg_post}{Post processing techniques have been commonly applied to correct prediction artifacts from segmentation-based deep-learning methods. Therefore, we investigated how the performance of our method compare with that of the 2D UNet and 3D UNet after post-processing. Namely, for each cardiac structure, we applied a median filter with a kernel size of $5\times 5 \times 5$ voxels to fill any small gaps within the segmentation and smooth segmentation boundaries. We then removed any disconnected regions from the segmentation by computing the largest connected component for each cardiac structure.} 
\q{mesh_post}{To correct for gaps between the predicted cardiac structures we leveraged the ability of our method to consistently map the same vertices to the similar regions of the heart. Thus, we can readily identify the vertices on the adjacent surfaces between the cardiac structures from our training data. For test cases, we can then project each of these vertices to the closest vertex on the adjacent surface. }

\begin{figure}[H]
{\color{black}
\centering
\includegraphics[width=0.8\textwidth]{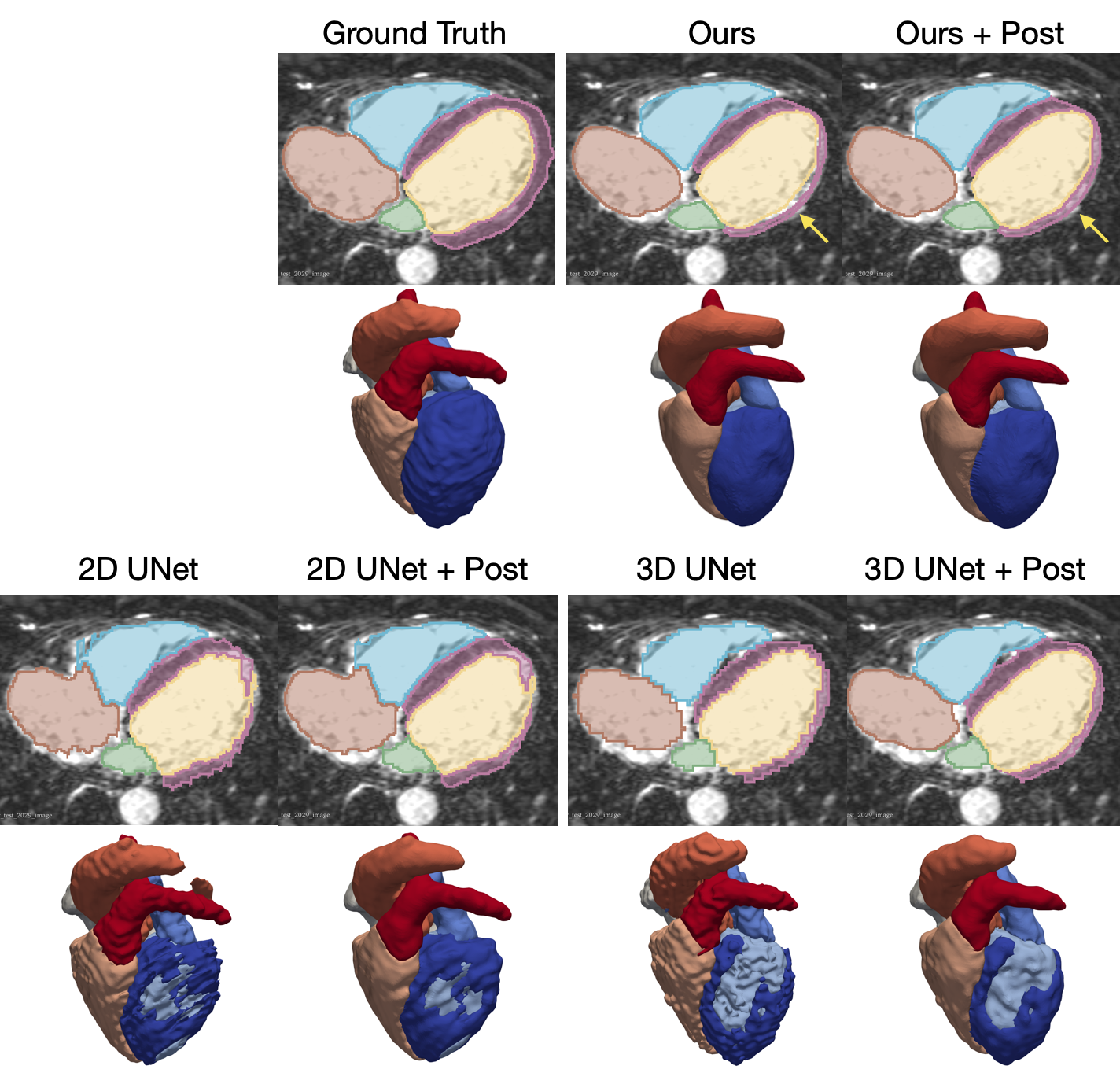}
\caption{Example of whole heart segmentation and surface reconstruction results before and after post-processing.} \label{figure:post}
}
\end{figure}

\q{post_results}{Table \ref{table:post_process} compares the reconstruction accuracy for our method, 2D UNet, and 3D UNet after the above post-processing steps as well as the accuracy differences before and after post-processing for each method. For both CT and MR data, our method consistently outperformed the baselines for all cardiac structures in terms of Dice and Jaccard scores, and for most cardiac structures in terms of ASSD and HD measures, respectively. In general, post-processing techniques did not bring major improvements in Dice, Jaccard or ASSD measures for all the methods. Indeed, these post-processing techniques are designed to correct artifacts small in size and thus do not significantly contribute to the improvements in global accuracy measures. In contrast, for local accuracy measure HD, post-processing techniques brought a major improvement in HD measure for 3D UNet for MR data due to the removal of disconnected regions from the predictions. Figure \ref{figure:post} displays the segmentation and reconstruction results for a challenging MR case before and after post-processing. Segmentation-based approaches, 2D and 3D UNets, predicted topological incorrect LV myocardium geometries with large holes, whereas our template-based method predicted topological-correct geometries. Post-processing techniques were able to reduce, but not fully close these holes. For this MR case, our method produced a small gap between the LV and myocardium as these two structures are represented by individual surfaces. However, our post-processing method on the mesh was able to automatically seal this gap. }

\begin{table}
{\color{black}
\centering
\caption{A comparison post-processed prediction accuracy on MMWHS MR and CT test datasets from different methods. Numbers in parentheses display the accuracy differences (if any) before and after post processing.}\label{table:post_process}
\resizebox{\textwidth}{!}{%
\begin{tabular}{lllllllllll}
\toprule
   &         &             &     Epi &      LA &      LV &      RA &      RV &      Ao &      PA &      WH \\
\midrule
\multirow{12}{*}{CT} & \multirow{3}{*}{Dice ($\uparrow$)} & Ours-Post &   \textbf{0.902} (0.003) &   \textbf{0.933} (0.001) &   \textbf{0.940} &   \textbf{0.892} &   \textbf{0.910} &   \textbf{0.950} &   \textbf{0.856} (0.003) &   \textbf{0.919} (0.001) \\
   &         & 2DUNet-Post &   0.895 (-0.004) &   0.924 (-0.006) &   0.928 (-0.002) &   0.878 (0.001) &   0.904 (-0.001) &   0.926 (-0.008) &    0.831 (-0.001) &   0.908 (-0.002)\\
   &         & 3DUNet-Post &   0.864 (0.001) &   0.903 (0.001) &   0.930 (0.007) &   0.871 (0.003) &   0.877 (0.001) &   0.920 (-0.003) &   0.793 (-0.019) &   0.889 (0.001) \\
\cline{2-11}
   & \multirow{3}{*}{Jaccard ($\uparrow$)} & Ours-Post &   \textbf{0.823} (0.004) &   \textbf{0.876} (0.001) &   \textbf{0.888} &   \textbf{0.809} &   \textbf{0.837} ( -0.001)&   \textbf{0.905} &   \textbf{0.760} (0.005) &   \textbf{0.850} (0.001) \\
   &         & 2DUNet-Post &   0.812 (-0.006) &   0.861 (-0.011)&   0.869 (-0.004)&   0.787 &   0.827 (-0.001)&   0.864 (-0.015)&   0.724 (-0.002) &   0.833 (-0.004)\\
   &         & 3DUNet-Post &   0.763 (0.001) &   0.825 &   0.870 (0.009) &   0.774 (0.005) &   0.785 (0.002)&   0.854 (-0.006)&   0.678 (-0.017) &   0.801 (0.001)\\
\cline{2-11}
   & \multirow{3}{*}{ASSD (mm) ($\downarrow$)} & Ours-Post &   0.874 (-0.461) &   \textbf{1.020} (-0.022) &  \textbf{0.823} (-0.020) &   \textbf{1.549} (-0.034) &   1.139 (-0.037)&   \textbf{0.528} (-0.003)&   1.896 (-0.009)  &   \textbf{1.112} (-0.100) \\
   &         & 2DUNet-Post &   \textbf{0.863} (0.054) &   1.125 (0.0750 &   0.960 (0.056) &   1.681 (-0.038) &   \textbf{1.129} (0.065) &   0.819 (0.174) &   \textbf{1.701} (0.149) &   1.171 (0.083) \\
   &         & 3DUNet-Post &   1.295 (-0.148) &   1.455 (-0.073)  &   0.958 (-0.066) &   1.906 (-0.036) &   1.680 (0.017) &   0.905 (0.090) &   3.135 (0.941) &   1.649 (0.097) \\
\cline{2-11}
   & \multirow{3}{*}{HD (mm) ($\downarrow$)} & Ours-Post &  13.978 (0.415) &   \textbf{7.960} (-2.447) &   \textbf{6.252} (-4.074) &  \textbf{11.735} (-1.904) &  10.958 (-2.401) &   \textbf{9.044} (-0.363) &  26.616 &  28.041 (0.006) \\
   &         & 2DUNet-Post &   \textbf{9.194} (-0.786) &   8.368 (-0.406) &   6.287 (0.189) &  12.243 (-1.381) &   \textbf{9.750} (-0.266) &  10.161 (0.148) &  \textbf{26.100} (-1.734) &  \textbf{26.900} (-1.826) \\
   &         & 3DUNet-Post &  10.250 (-3.386) &   9.828 (-0.986) &   6.618 (-2.961) &  13.251 (-2.779) &  12.614 (-3.020) &  12.500 (-0.826) &  28.700 (1.759) &  30.582 (-0.506) \\
\cline{1-11}
\cline{2-11}
\multirow{12}{*}{MR} & \multirow{3}{*}{Dice ($\uparrow$)} & Ours-Post &   \textbf{0.800} (0.002) &   \textbf{0.879} (-0.002) &   \textbf{0.921} (-0.001) &   \textbf{0.888} &   \textbf{0.892} &   \textbf{0.889} (-0.001) &   \textbf{0.817} &   \textbf{0.881} \\
   &         & 2DUNet-Post &   0.790 (-0.005) &   0.850 (-0.014) &   0.892 (-0.004) &   0.842 (-0.010) &   0.862 (-0.003) &   0.862 (-0.008) &   0.764 (-0.008) &   0.854 (-0.005) \\
   &         & 3DUNet-Post &   0.770 (0.009) &   0.848 (-0.004) &   0.881 (0.002) &   0.868 (0.001) &   0.830 (0.003) &   0.817 (0.076) &   0.761 (-0.003) &   0.844 (0.004)\\
\cline{2-11}
   & \multirow{3}{*}{Jaccard ($\uparrow$)} & Ours-Post &   \textbf{0.674} (0.003) &   \textbf{0.788} (-0.003) &   \textbf{0.856} (-0.002) &  \textbf{ 0.800} (-0.001) &   \textbf{0.812} &   \textbf{0.804} (-0.001) &   \textbf{0.697} &   \textbf{0.790} \\
   &         & 2DUNet-Post &   0.661 (-0.007) &   0.746 (-0.019) &   0.811 (-0.006) &   0.741 (-0.011) &   0.766 (-0.005) &   0.762 (-0.012) &   0.632 (-0.009) &   0.749 (-0.008) \\
   &         & 3DUNet-Post &   0.635 (0.010) &   0.752 (-0.004) &   0.811 (0.009) &   0.768 (0.002) &   0.733 (0.006) &   0.715 (0.065) &   0.633 (-0.007) &   0.737 (0.005) \\
\cline{2-11}
   & \multirow{3}{*}{ASSD (mm) ($\downarrow$)} & Ours-Post &   1.967 (-0.231) &   \textbf{1.373} (-0.028) &   \textbf{1.155} (-0.028) &   \textbf{1.581} (-0.029) &   \textbf{1.310} (-0.023) &   2.650 (0.001) &   2.692 (0.002) &   \textbf{1.713} (-0.061) \\
   &         & 2DUNet-Post &   \textbf{1.805} (-0.013) &   1.699 (0.211) &   1.520 (0.065) &   2.008 (0.288) &   1.523 (0.058) &   2.747 (0.300) &   \textbf{2.151} (0.331) &   1.952 (0.286) \\
   &         & 3DUNet-Post &   2.167 (-0.206) &   2.151 (-0.318) &   1.600 (-0.618) &   1.658 (-0.338) &   2.454 (-0.312) &   \textbf{2.512}(-1.277) &   2.209 (0.265) &   2.042 (-0.073)\\
\cline{2-11}
   & \multirow{3}{*}{HD (mm) ($\downarrow$)} & Ours-Post &  16.516 (-0.406) &   \textbf{9.658} (-2.065) &   \textbf{8.070} (-2.820) &  13.558 (-1.252) &  \textbf{11.025} (-2.438) &  \textbf{22.219} &  19.319 (-0.026) &  27.569 (-0.133) \\
   &         & 2DUNet-Post &  \textbf{13.759} (-5.398) &  11.185 (0.404) &   9.972 (0.014) &  13.825 (-1.005) &  11.544 (-1.556) &  24.912 (2.346) &  17.056 (0.335) &  28.024 (-0.273) \\
   &         & 3DUNet-Post &  17.024 (-11.432) &  11.564 (-12.263) &  11.531 (-11.178) &  \textbf{12.474} (-7.048) &  12.699 (-8.295) &  23.113 (-11.226) &  \textbf{17.021} (0.140) &  \textbf{27.065} (-15.400) \\
\bottomrule
\end{tabular}
}}
\end{table}

\subsection{Impact of Limited Training Data on Reconstruction Performance  }

We investigate how well our method can reconstruct whole-heart geometries using only a small number of training data. In this experiment,  our neural network model was trained using only the training set of MMWHS challenge, which consists of 20 CT images and 20 MR images. 16 out of 20 image volumes from each modality were used for training and the rest were used for validation. We compared our method against the baseline methods for the same MMWHS test set described above. The baseline methods were trained using the same training and validation splits. 

Table \ref{table:data_small} compares the Dice and Jaccard scores, ASSD and HD of the reconstruction results for the methods trained with the reduced training set, as well as the accuracy differences compared with training models using more data, as described above. For CT data, our method consistently outperformed others in terms of Dice and Jaccard scores for the whole heart and individual cardiac structures except for pulmonary arteries. In terms of ASSD and HD, our method outperformed 3D UNet and Voxel2Mesh and was comparable to 2D UNet. For MR data, our method demonstrated better performance than others in terms of whole heart Dice and Jaccard scores, as well as surface HD of whole heart. 2D UNet demonstrated the best whole heart ASSD performance. For individual cardiac structures, our method showed better Dice and Jaccard scores for Epi, LV, RA and RV, smaller ASSD values for Epi, LV, RA and smaller surface HD values for most of the cardiac structures except for LA and Ao. Figure \ref{figure:boxplot-small} shows the distribution of different segmentation accuracy metrics for whole heart and individual cardiac structures among the MMWHS test dataset. 

As shown in Table \ref{table:data_small}, when trained with a smaller training dataset, the methods generally showed reduced Dice or Jaccard scores and increased ASSD and HD values for both whole heart and individual cardiac structures compared with when trained with a larger dataset, as summarized in Table \ref{table:data_large}. Exceptions include the smaller HD values of Epi, LA, LV, RV and PA from our method for CT data and the better LV and aorta segmentation from 3D UNet for MR data in terms of all four metrics. Compared with CT data, all methods generally demonstrated more significant reduction of segmentation accuracy for MR data, in terms of average values of reduction for all four metrics. While performance drops due to reduced size of training data is consistent, the actually amount of performance drop is minor for our method, 2D UNet and 3D UNet. For example, although the number of CT training data was reduced from 87 to 16, we only observed a small average reduction (0.01-0.02) of whole heart Dice scores for 2D UNet, 3D UNet and our method. However, the performance drop for Voxel2Mesh in relation to the number of training data was much more significant, with a 0.27-0.28 reduction of whole-heart Dice scores for CT and MR data. Among all the cardiac structures, our method had the most significant performance reduction of PA reconstruction for both CT and MR data while segmentation based approaches, 2D UNet and 3D UNet, demonstrated a more uniform performance drop across all cardiac structures. Indeed, the shapes of the PA differ significantly from our initial sphere template mesh and therefore accurately capturing the shapes of PA might require more training data for our method. 
\begin{table}[H]
\caption{A comparison of prediction accuracy on MMWHS MR and CT test datasets from different methods trained with images from MMWHS training set. An asterisk $*$ indicates statistically significant accuracy differences, compared with Table \ref{table:data_large}, resulted from training on a smaller datset based on t-tests (p$<$0.05).}\label{table:data_small}
\resizebox{\textwidth}{!}{%
\begin{tabular}{llllllllll|l}
\toprule
   &         &            &     Epi &      LA &      LV &      RA &      RV &      Ao &      PA &      WH \\
\midrule
\multirow{16}{*}{CT} & \multirow{4}{*}{Dice ($\uparrow$)} & Ours &   \textbf{0.880} &   \textbf{0.926} &   \textbf{0.931} &   \textbf{0.868} &   \textbf{0.885}* &  \textbf{0.945} &   0.786* &   \textbf{0.900}* \\
   &         & 2DUNet &   0.877* &   0.916 &   0.926 &   0.855 &   0.876* &   0.916 &   \textbf{0.805} &   0.892* \\
   &         & 3DUNet &   0.816* &   0.916 &   0.914 &   0.848 &   0.878 &   0.923 &   0.793 &   0.877 \\
   &         & Voxel2Mesh &   0.501* &   0.748* &   0.669* &   0.717* &   0.698* &   0.555* &   0.491* &   0.656* \\
\cline{2-11}
   & \multirow{4}{*}{Jaccard ($\uparrow$)} & Ours &   \textbf{0.790} &   \textbf{0.863} &   \textbf{0.874} &   \textbf{0.773} &   \textbf{0.798}* &   \textbf{0.897} &   0.666* &   \textbf{0.819}* \\
   &         & 2DUNet &   0.784* &   0.847 &   0.864 &   0.753 &   0.787* &   0.850 &   \textbf{0.692} &   0.807* \\
   &         & 3DUNet &   0.696* &   0.848 &   0.844 &   0.741 &   0.787 &   0.860 &   0.670 &   0.782 \\
   &         & Voxel2Mesh &   0.337* &   0.600* &   0.510* &   0.570* &   0.543* &   0.397* &   0.337* &   0.491* \\
\cline{2-11}
   & \multirow{4}{*}{ASSD (mm) ($\downarrow$)} & Ours &   1.357 &   \textbf{1.137} &   0.966 &   1.750 &  \textbf{ 1.320} &   \textbf{0.729}* &   2.020 &   1.333* \\
   &         & 2DUNet &   \textbf{1.014}* &   1.141 &   \textbf{0.911} &   \textbf{1.702} &   1.433* &   0.808 &   \textbf{1.754} &   \textbf{1.240}* \\
   &         & 3DUNet &   1.809* &   1.389 &   1.134 &   2.176 &   1.585 &   0.832 &   2.276 &   1.668 \\
   &         & Voxel2Mesh &   3.412* &   3.147* &   4.973* &   3.638* &   4.300* &   4.326* &   5.857* &   4.287* \\
\cline{2-11}
   & \multirow{4}{*}{HD (mm) ($\downarrow$)} & Ours &  13.789 &  10.362 &   9.628 &  \textbf{14.467} &  \textbf{12.766} &  12.740* & \textbf{ 25.362} & \textbf{ 27.567} \\
   &         & 2DUNet &  \textbf{13.582} &  \textbf{10.221} &   \textbf{6.700} &  14.788 &  16.608* &  11.410 &  28.128&  32.514 \\
   &         & 3DUNet &  15.044 &  40.157* &   9.730 &  15.037 &  13.777 &  \textbf{10.821} &  27.467 &  48.731\\
   &         & Voxel2Mesh &  15.526* &  13.683* &  22.146* &  16.834* &  18.390* &  19.419* &  35.322* &  37.065* \\

\cline{1-11}
\cline{2-11}
\multirow{16}{*}{MR} & \multirow{4}{*}{Dice ($\uparrow$)} & Ours &   \textbf{0.773} &   0.826* &   \textbf{0.913} &   \textbf{0.838}* &   \textbf{0.861} &   0.824* &   0.663* &   \textbf{0.846}* \\
   &         & 2DUNet &   0.751 &   \textbf{0.831} &   0.880 &   0.815 &   0.852 &   \textbf{0.838}* &   \textbf{0.747} &   0.834 \\
   &         & 3DUNet &   0.733 &   0.811 &   0.885 &   0.827* &   0.829 &   0.825 &   0.741 &   0.823 \\
   &         & Voxel2Mesh &   0.282* &   0.498* &   0.515* &   0.599* &   0.539* &   0.241* &   0.300* &   0.483* \\
\cline{2-11}
   & \multirow{4}{*}{Jaccard ($\uparrow$)} & Ours &   \textbf{0.639} &   0.712* &   \textbf{0.842} &   \textbf{0.727}* &   \textbf{0.768} &   0.715* &   0.517* &   \textbf{0.737}* \\
   &         & 2DUNet &   0.611* &   \textbf{0.720} &   0.793 &   0.702 &   0.753 &   \textbf{0.726}* &   0.608 &   0.719* \\
   &         & 3DUNet &   0.588 &   0.695* &   0.803 &   0.718* &   0.727 &   0.718 &   \textbf{0.615} &   0.709 \\
   &         & Voxel2Mesh &   0.170* &   0.339* &   0.367* &   0.442* &   0.388* &   0.144* &   0.187* &   0.327* \\
\cline{2-11}
   & \multirow{4}{*}{ASSD (mm) ($\downarrow$)} & Ours &   \textbf{2.385}&   2.166*&   \textbf{1.300} &   \textbf{2.358}* &   1.812 &   3.243 &   3.138 &   2.235* \\
   &         & 2DUNet &   2.692* &   \textbf{1.688} &   1.603 &   3.151* &   \textbf{1.736} &   \textbf{2.920} &   2.281* &   \textbf{1.897} \\
   &         & 3DUNet &   2.713 &   3.866 &   1.551 &   2.475 &   1.931 &   4.049 &   \textbf{2.259} &   2.120 \\
   &         & Voxel2Mesh &   6.886* &   5.987* &   8.679* &   6.173* &   8.192* &   7.877* &   9.200* &   7.419* \\
\cline{2-11}
   & \multirow{4}{*}{HD (mm) ($\downarrow$)} & Ours &  \textbf{16.804} &  15.559* &  \textbf{12.197} &  \textbf{17.286}* &  \textbf{14.480} &  26.012 &  \textbf{19.927} &  \textbf{29.983} \\
   &         & 2DUNet &  23.798 &  \textbf{14.887}* &  14.651 &  22.028* &  22.810* &  \textbf{24.237} &  22.883* &  39.724* \\
   &         & 3DUNet &  20.136 &  32.978 &  13.643 &  23.735 &  22.351 &  31.900 &  21.363 &  43.475 \\
   &         & Voxel2Mesh &  27.272* &  22.748* &  31.327* &  24.456* &  28.987* &  29.381 &  33.637* &  40.072* \\
\bottomrule
\end{tabular}
}
\end{table}

\section{Discussion}

Image-based reconstruction of cardiac anatomy and the concomitant geometric representation using unstructured meshes is important to a number of applications, including visualization of patient-specific heart morphology and computational simulations of cardiac function. Prior deep-learning-based approaches have shown great promise in automatic whole heart segmentation~\citep{ZHUANG2019}, however converting the segmentation results to topologically valid mesh structures requires additional, and often manual, post-processing, and is highly-dependent on the resolution of the image data.  In this work, we present a novel deep-learning-based approach that uses graph convolutional neural networks to directly generate meshes of multiple cardiac structures of the whole heart from volumetric medial image data. Our approach generally demonstrated improved whole heart reconstruction performance compared with the baseline methods in terms of accuracy measures, Dice and Jaccard scores, ASSD and HD. Furthermore, our method demonstrated advantages in generating high-resolution, anatomically and temporally consistent geometries, which are not reflected by the accuracy measures. 

Our method reconstructs cardiac structures by predicting the deformation of mesh vertices from sphere mesh templates. We have demonstrated the advantages of this approach over segmentation-based approaches in terms of both precision and surface quality. Namely, the use of a template mesh can introduce topological constraints so that predicted cardiac structure are \textcolor{black}{homeomorphic} to the template. Thus, our template based approach enables one to eliminate disconnected regions and greatly reduce erroneous topological artifacts often encountered with existing deep-learning-based segmentation methods. While the cardiac structures of interest were \textcolor{black}{homeomorphic} to spheres, the presented method has the potential to be generalized to organs with different topology, by using a different template mesh with the required surface topology. 

When trained on a relatively large dataset with 87 CT and 41 MR images, our method was able to achieve comparable accuracy to manual delineations, which is considered the gold standard. Furthermore, since we explicitly regularized the surface smoothness and normal consistency, our method produced smooth and quality meshes while capturing the detailed features of the cardiac structures. Namely, these factors along with the use of a template enable our method to generate realistic cardiac structures even when image quality was poor and segmentation methods struggled to provide realistic topology. From our observations, the locations on the heart that our neural network models produced high surface errors are consistent with the locations that could suffer from high inter- or intra-observer variations, such as the arbitrary length of aorta and pulmonary arteries, boundaries between atria and ventricles and between the right atrium and the inferior vena cava. Indeed, these boundaries are not distinguishable by voxel intensity differences and are often subject to uncertainties even for human observers. 

Compared with segmentation-based approaches, our method predicts whole heart surfaces directly in the physical space rather than on a voxel grid of the input image. The whole heart geometries are represented using surface meshes rather than a dense voxel grid. Hence, our method is able to generate high-resolution reconstructions (10K mesh vertices for each cardiac structure) efficiently on a limited memory budget and within a shorter or comparable run-time (table \ref{table:s2}). \q{multiple_networks}{Prior 3D segmentation-based approaches have sought to increase the segmentation resolution by training separate neural networks to first locate the region of interest or generate low resolution segmentations, and then generate refined segmentations within the localized region \citep{Payer2018, Isensee2021}. Our method does not require training multiple neural networks and can make predictions directly from the entire down-sampled cardiac image volume. As we used a cascade of three mesh deformation blocks, we observed that the first deformation block can already effectively position and deform the meshes to the correct locations and the subsequent deformation blocks can further refine the predicted mesh vertex locations. }High resolution segmentation may also be obtained by recent methods that represent geometries using implicit surfaces \citep{kirillov2019pointrend}. Namely, for each point in the physical space, this approach predicts the probability of this point belonging to a certain tissue class. Therefore, by sampling a large number of points in the physical space, these methods can also achieve high-resolution reconstruction that are not constrained by the voxel resolution of the input image or GPU memory. However, the inference process for such methods is computationally expensive \citep{NeuralMeshFlow} as it requires prediction on a large number of points. In contrast, our method represents the mesh as a graph (i.e., a sparse matrix) and takes less than a second to predict a high resolution whole heart mesh. 

Compared with prior deep-learning-based mesh reconstruction methods from image data \citep{Pixel2Mesh, Voxel2Mesh}, our method used a shared graph neural network to simultaneously predict surface meshes of multiple cardiac structures. This is made possible by initializing the template meshes at various scales and locations corresponding to individual cardiac structures. We observed that proper template initialization is essential to avoid local minimums due to large mesh deformation at the beginning stage of training. In contrast, prior approaches are designed for predicting a single geometry from image data and require training a separate graph neural network for each anatomical structure and thus do not easily scale to reconstruct multiple cardiac structures at a high-resolution from a single image volume. Furthermore, while prior approaches proposed various up-sampling scheme to construct a dense mesh from a coarse template \citep{Voxel2Mesh, Pixel2Mesh}, we directly deformed a high resolution template. Since the majority of weights is in the image feature encoder to process a dense volumetric input image, more mesh vertices can provide more effective gradient propagation to the image feature encoder. Indeed, using a coarse mesh with 3K mesh vertices for each cardiac structures, we observed a 2\% reduction of whole heart dice score as shown in our supplemental materials (table~\ref{table:s1}). However, our method was still able to outperform Voxel2Mesh by 3\% and 10\% for CT and MR data using a coarse mesh template with a similar amount of mesh vertices. These design choices allowed our method to demonstrate promising generalization capabilities to unseen MR images and maintain good performance when trained with a smaller number of samples. In contrast, Voxel2mesh suffered from a large performance drop when trained on a smaller dataset.

When applied to time-resolved images, our method consistently deformed the template mesh such that mesh vertices were mapped to the similar regions of the heart across different time frames. Learning such semantic correspondence is purely a consequence of our model architecture and did not require any explicit training. This behavior of producing semantic corresponding predictions was also observed in DeepOrganNet, which reconstructed lung shapes from single-view X-ray images by deforming lung templates \citep{DeepOrganNet}. Point-corresponded meshes across different input images are required for numerous applications, such as building statistical shape models, constructing 4D dynamic whole-heart models for motion analysis and deriving boundary conditions for deforming-domain CFD simulations. Current approaches that construct feature corresponding meshes for the heart mostly use surface or image registration methods to deform a reference mesh so that its boundary is consistent with the target surfaces or image segmentation \citep{Ordas2007, Khalafvand2018, KONG2020}. However, registration algorithms are often very computationally expensive to align high-resolution meshes and they often suffer from inaccuracies for complex whole heart geometries due to local minimums during the optimization process. In the case of time-series image data, our method naturally produces point corresponding meshes with high resolution (10K mesh vertices per cardiac structure) across time frames within a couple of seconds, while prior methods could require hours to generate a 4D dynamic whole-heart model at a similar resolution. Although not considered here, it is possible to include another loss function that minimizes the point distances between the vertex locations on the predicted meshes and ground truth landmarks when available to further enhance feature correspondence.

\q{limitation}{Limitations of the proposed method include a lack of diffeomorphic constraints to establish a differentiable mapping from the initial spheres and the predicted surfaces. While we used the Laplacian loss to regularize the smoothness of the meshes, a diffeomorphic constraints may help to further prevent face intersections. Recently, \cite{NeuralMeshFlow} proposed to learn neural ordinary differential equations to predict a diffeomorphic flow that maps a sphere mesh template to the target shapes, thus implicitly preserving the manifoldness of the template mesh without explicit regularizations. This approach could be combined with our image-based whole-heart mesh prediction framework in the future to deform the whole heart geometry while preserving the manifoldness of the meshes so that they could be directly used in applications such as numerical simulations and 3D printing. Furthermore, while our method can simultaneously predict multiple structures from image data, those structures are not coupled to each other. Small intersections or gaps could appear between adjacent cardiac structures. While we have demonstrated that simple projection can generally correct such artifacts, future work could include more explicitly constraining the coupling of cardiac structures within the learning framework. }

\section{Conclusions}
We have developed a deep-learning-based method to directly predict surface mesh reconstructions of the whole heart from volumetric image data. The approach leverages a graph convolutional neural network to predict deformation on mesh vertices from a predefined mesh template to fit multiple anatomical structures in a 3D image volume. The mesh deformation is conditioned on image features extracted by a CNN-based image encoder. The method demonstrated promising performance of generating accurate high-resolution and high-quality whole heart reconstructions and outperformed prior deep-learning-based methods on both CT and MR data. It also demonstrated robust performance when evaluated on MR or CT images from new data sources that differ from our the training datasets. Furthermore, the method produced temporally consistent predictions and feature-corresponding predictions by consistently mapping mesh vertices on the templates to similar structural regions of the heart. Therefore, this method can potentially be applied for efficiently constructing 4D dynamics whole heart model that captures the motion of a beating heart from time-series images data.

\section*{Acknowledgments}
This work was supported by the NSF, Award \#1663747. We thank Drs. Shone Almeida, Amirhossein Arzani and Kashif Shaikh for providing the time-series CT image data, and Dr. Theodore Abraham for providing the time-series MR image data. We thank Dr. Angjoo Kanazawa for her valuable comments concerning this work.

\newpage
\renewcommand{\thesubsection}{\Alph{subsection}}
\counterwithin{figure}{subsection}
\renewcommand{\thefigure}{B\arabic{figure}}
\counterwithin{table}{subsection}
\renewcommand{\thetable}{B\arabic{table}}
\section*{Appendices}
\subsection{Implementation Details}
\subsubsection{Image Pre-Processing}
\label{appendix:pre-processing}
\q{im_pre}{Intensity normalization and resizing were applied to all 3D image volumes to obtain consistent image dimensions and pixel intensity range. We followed the procedures in \cite{KONG2020} to normalize pixel intensity values of each CT or MR image volume such that they ranged from -1 to 1. The 3D image volumes were then resized using linear interpolation to a dimension of 128 $\times$ 128 $\times$ 128, which maintained image resolution with a manageable computational cost. The ground truth meshes were generated by applying the Marching Cube algorithm~\citep{Lorensen1987} on the segmentations, followed by 50 iterations of Laplacian smoothing.}

\subsubsection{Image Augmentation} Data augmentation techniques were applied during training to improve the robustness of the neural network models to the variations of input images. Specifically, we applied random scaling ($-5\%$ to $5\%$), random rotation ($-5^{\circ}$ to $5^{\circ}$), random shearing ($-10^{\circ}$ to $10^{\circ}$) as well as elastic deformations \citep{Simard2003BestPF} on the input images. For elastic deformations, 16 control points were placed along each dimension of the 3D image volume and were randomly perturbed. The input images are then warped according to the displacements of the control points using the B-spline interpolation. 

\subsubsection{Training} The model parameters were computed by minimizing the total loss function using the Adam stochastic gradient descent algorithm \citep{adam}. The initial learning rate was set to be 0.001, while $\beta_1$ and $\beta_2$ for the Adam algorithm were set to 0.9 and 0.999, respectively. Point losses were evaluated on the validation data after each training epoch and the model was saved after one epoch only if the validation point loss had improved. We adopted a learning rate schedule where the learning rate was reduced by 20\% if the validation point losses had not improved for 10 epochs. The minimum learning rate was $5\times10^{-6}$. The network was implemented in TensorFlow and the training was conducted on a Nvidia GeForce GTX 1080 Ti graphics processing unit (GPU) until the validation loss converged.
\textcolor{black}{
\subsubsection{Evaluation Metrics}}
\label{appendix:metrics}
\textcolor{black}{
We used Dice, Jaccard scores as well as average symmetric surface distance (ASSD) and Hausdorff distance (HD) to evaluate the accuracy of our reconstructions. Dice and Jaccard scores are similarity indices that range from 0 to 1 as given by
\begin{equation}
    \text{Dice}(I_P, I_G) = \frac{2 |I_P \cap I_G|}{|I_P|+ |I_G|}
    \label{dice}
\end{equation}
\begin{equation}
    \text{Jaccard}(I_P, I_G) = \frac{|I_P \cap I_G|}{|I_P \cup I_G|}
    \label{jaccard}
\end{equation}
The ASSD and HD measure the average and the largest inconsistency in terms of Euclidean distance between the reconstruction result and the ground truth, respectively. For reconstructed meshes $\mathbf{P}$ and the ground truth meshes $\mathbf{G}$, the ASSD and HD are given by
\begin{equation}
    \text{ASSD}(\mathbf{P}, \mathbf{G}) = \sum_{\mathbf{p} \in \mathbf{P}} \min_{\mathbf{g} \in \mathbf{G}}\frac{  ||\mathbf{p}-\mathbf{g}||_2}{|\mathbf{P}|} + \sum_{\mathbf{g}\in \mathbf{G}} \min_{\mathbf{p}\in \mathbf{P}}\frac{ ||\mathbf{p}-\mathbf{g}||_2 }{|\mathbf{G}|}
\end{equation}
\begin{equation}
    \text{HD}(\mathbf{P}, \mathbf{G}) = \max \left\{\max_{\mathbf{p} \in \mathbf{P}}\min_{\mathbf{g} \in \mathbf{G}} ||\mathbf{p}-\mathbf{g}||_2, \max_{\mathbf{g} \in \mathbf{G}}\min_{\mathbf{p} \in \mathbf{P}} ||\mathbf{p}-\mathbf{g}||_2\right\}
\end{equation}
Normal discrepancy between the reconstruction result and the ground truth was evaluated by an average normal error (ANE). Namely, for $\mathbf{n}_x$, $\mathbf{n}_y$ being the vertex normals at points $\mathbf{x}$ and $\mathbf{y}$, respectively, 
\begin{equation}
    \text{ANE} (\mathbf{P}, \mathbf{G}) = \sum_{\substack{\mathbf{p} \in \mathbf{P}}; \mathbf{g}=\argmin \limits_{\substack{ \mathbf{g} \in \mathbf{G}}}||\mathbf{p}-\mathbf{g}||_2 } \frac{1 - \langle \mathbf{n}_p , \mathbf{n}_g \rangle}{|\mathbf{P}|}
\end{equation}
Surface smoothness was evaluated by the average normalized Laplacian distance (ANLD). ANLD measures the Euclidean distances between the coordinates of mesh vertices and the mean coordinates of their neighbours, normalized by the average edge length between the mesh vertices and their neighbours. Namely, 
\begin{equation}
    \text{ANLD} (\mathbf{P}) = \sum_{\mathbf{p}\in \mathbf{P}} \frac{\left\|\mathbf{p}-\sum_{\mathbf{k_p}\in \mathcal{N}(\mathbf{p})} \frac{1}{|\mathcal{N}(\mathbf{p})|}\mathbf{k_p} \right\|_2}{ \frac{|P|}{|\mathcal{N}(\mathbf{p})|} \sum_{\mathbf{k_p}\in \mathcal{N}(\mathbf{p})}\left\|\mathbf{p} - \mathbf{k_p}\right\|_2}\;.
\end{equation} 
The percentage mesh self-intersection was calculated as the percentage of intersected mesh facets among all mesh facets. The intersected mesh facets were detected by TetGen \cite{Si2015}.
}
\subsection{Supplementary Results}
\subsubsection{Whole Heart Reconstruction for CT and MR images}
\begin{figure}[H]
\centering
\includegraphics[width=\textwidth]{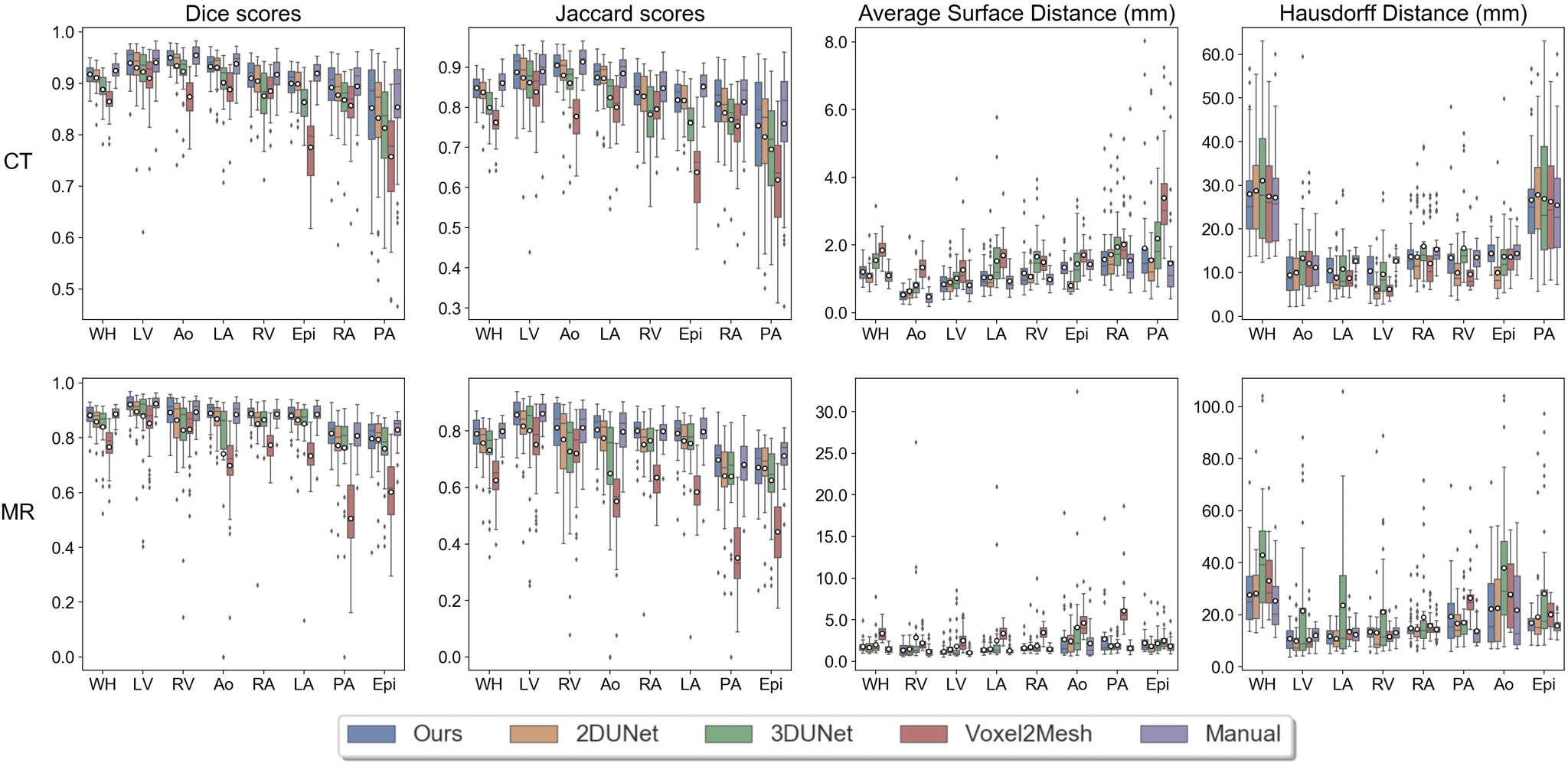}
\caption{Comparison of segmentation accuracy for whole heart and individual cardiac structures from different methods. White circles on the boxes indicate mean values across patients. Cardiac structures are sorted based on the accuracy of our method. } 
\label{figure:boxplot-large}
\end{figure}

\subsubsection{Impact of Limited Training Data on Reconstruction Performance}
\begin{figure}[H]
\centering
\includegraphics[width=\textwidth]{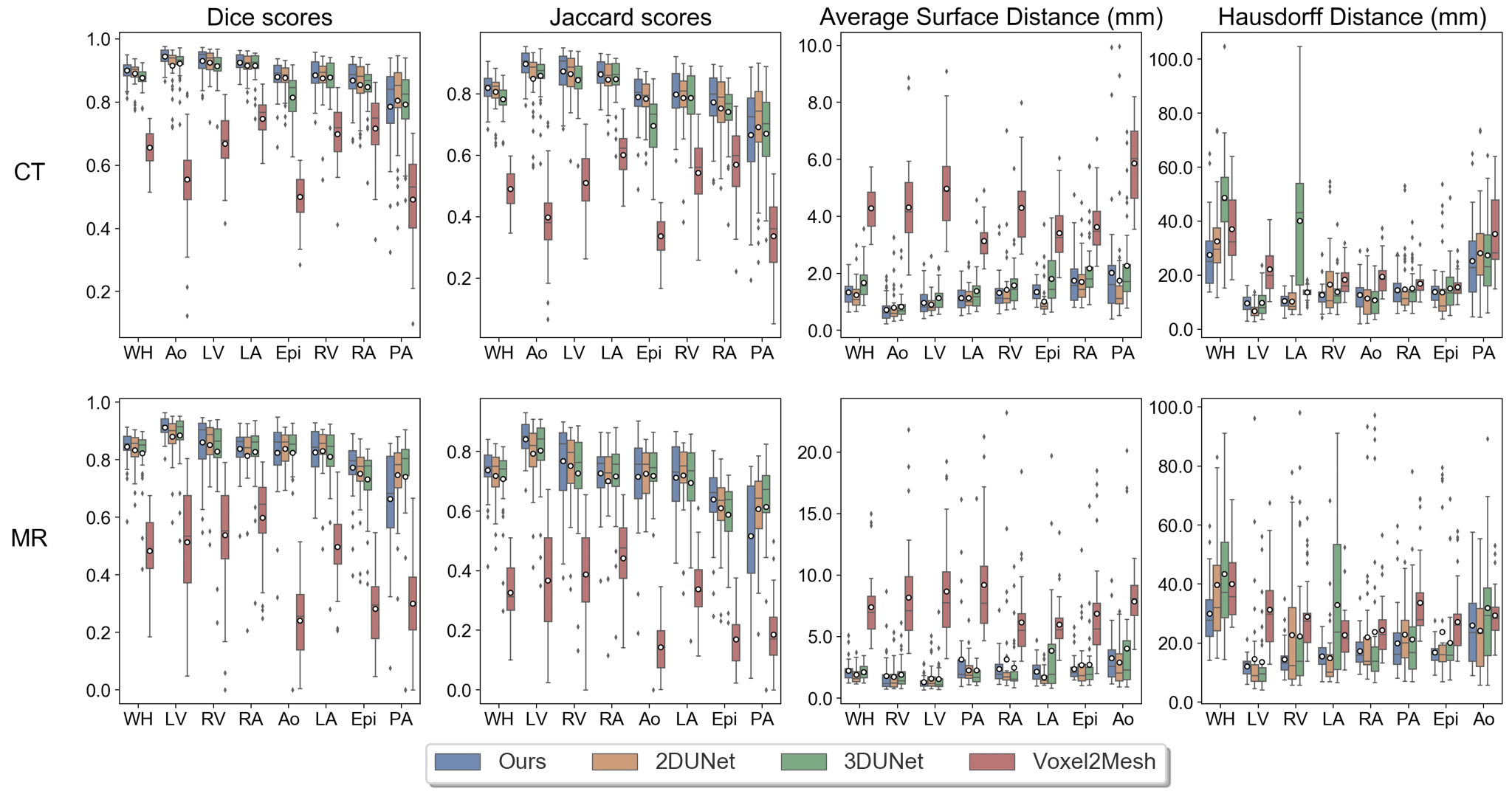}
\caption{Comparison of segmentation accuracy for whole heart and individual cardiac structures from different methods trained using the small MMWHS training dataset. White circles on the boxes indicate mean values across patients. Cardiac structures are sorted based on the accuracy of our method. } 
\label{figure:boxplot-small}
\end{figure}

\subsubsection{Sensitivity Studies}
We compare the effect of design choice changes on the whole heart reconstruction performance of our method. Namely, we trained another three models while, respectively, using reduced number of convolutional filters in the image encoding module, using reduced resolution of template meshes or excluding the elastic deformation from our image augmentation techniques. Specifically, the number of convolutional filters in the last four residual blocks were reduced from 48, 96, 192 and 384 to 32, 64, 128 and 256, respectively. The number of mesh vertices of the template mesh was reduced from 11494 to 3260. As shown in Table \ref{table:s1}, reducing the number of filters or template mesh vertices mildly reduced the reconstruction accuracy of the whole heart or most cardiac structures compared with the our final model.  However, without elastic deformation augmentation, we observed a significant drop in reconstruction performance. Indeed, elastic deformation augmentation may improve model robustness to minor local perturbations of the anatomical structures, thereby facilitating accurate predictions of detailed cardiac structures. 

\begin{table}[H]
    \centering
    \caption{A comparison of prediction accuracy on MMWHS MR and CT test datasets from different variants of our methods. }\label{table:s1}
\resizebox{\textwidth}{!}{%
\begin{tabular}{lllrrrrrrrr}
\toprule
   &         &                                     &     Epi &      LA &      LV &      RA &      RV &      Ao &      PA &      WH \\
\midrule
\multirow{8}{*}{CT} & \multirow{4}{*}{Dice} & Final Model &   0.899 &   0.932 &   0.940 &   0.892 &   0.910 &   0.950 &   0.852 &   0.918 \\
   &         & Reduced Number of Filters &   0.893 &   0.932 &   0.936 &   0.888 &   0.906 &   0.949 &   0.847 &   0.915 \\
   &         & Reduced Number of Vertices &   0.842 &   0.929 &   0.931 &   0.888 &   0.892 &   0.943 &   0.837 &   0.899 \\
   &         & No Elastic Deformation Augmentation &   0.546 &   0.882 &   0.878 &   0.801 &   0.757 &   0.882 &   0.552 &   0.773 \\
\cline{2-11}
   & \multirow{4}{*}{ASSD (mm)} & Final Model &   1.335 &   1.042 &   0.842 &   1.583 &   1.176 &   0.531 &   1.904 &   1.213 \\
   &         & Reduced Number of Filters &   1.404 &   1.063 &   0.892 &   1.523 &   1.122 &   0.536 &   1.744 &   1.195 \\
   &         & Reduced Number of Vertices &   1.768 &   1.126 &   1.029 &   1.664 &   1.425 &   0.583 &   1.768 &   1.378 \\
   &         & No Elastic Deformation Augmentation &   3.055 &   1.742 &   1.731 &   2.479 &   3.188 &   1.117 &   5.823 &   2.920 \\
\cline{1-11}
\multirow{8}{*}{MR} & \multirow{4}{*}{Dice} & Final Model &   0.797 &   0.881 &   0.922 &   0.888 &   0.892 &   0.890 &   0.816 &   0.882 \\
   &         & Reduced Number of Filters &   0.813 &   0.873 &   0.919 &   0.888 &   0.881 &   0.876 &   0.789 &   0.879 \\
   &         & Reduced Number of Vertices &   0.774 &   0.870 &   0.903 &   0.887 &   0.861 &   0.860 &   0.792 &   0.863 \\
   &         & No Elastic Deformation Augmentation &   0.487 &   0.810 &   0.867 &   0.795 &   0.744 &   0.724 &   0.413 &   0.735 \\
\cline{2-11}
   & \multirow{4}{*}{ASSD (mm)} & Final Model &   2.198 &   1.401 &   1.183 &   1.611 &   1.333 &   2.648 &   2.689 &   1.775 \\
   &         & Reduced Number of Filters &   2.053 &   1.556 &   1.238 &   1.488 &   1.429 &   2.143 &   2.205 &   1.645 \\
   &         & Reduced Number of Vertices &   2.405 &   1.615 &   1.516 &   1.561 &   1.651 &   2.390 &   2.222 &   1.845 \\
   &         & No Elastic Deformation Augmentation &   3.794 &   2.151 &   2.003 &   2.976 &   3.575 &   3.700 &   5.166 &   3.348 \\
\bottomrule
\end{tabular}
}
\end{table}

\textcolor{black}{
\subsection{Ablation Studies on Loss components}
}

\textcolor{black}{
We performed an ablation study on the mesh loss functions to evaluate the contribution of individual loss components. Namely, we trained additional models with the normal loss, the edge loss or the Laplacian loss removed by changing $\lambda_2$, $\lambda_3$, or $\lambda_4$ to 0, respectively, in Equation \ref{eq:mesh_loss}. We also evaluated the effect of using $L2$ norm on the edge length loss rather than the $L1$ norm originally used in Equation \ref{eq:edge_loss}. The rest of the hyperparameters were kept the same. As shown in Table \ref{table:s_ablation}, removing the Laplcaian loss caused the most accuracy drop on the MMWHS MR and CT test datasets, followed by the edge length loss, the normal loss and using the L2 edge length loss. }

\begin{table}[H]
{\color{black}
\caption{Impact of using L2 edge length loss or removing edge length loss, Laplacian loss or normal loss on the prediction accuracy on MMWHS MR and CT test datasets} \label{table:s_ablation}
\resizebox{\textwidth}{!}{%
\begin{tabular}{lllrrrrrrrr}
\toprule
   &         &              &     Epi &      LA &      LV &      RA &      RV &      Ao &      PA &      WH \\
\midrule
\multirow{20}{*}{CT} & \multirow{5}{*}{Dice} & Final &   0.899 &   0.932 &   0.940 &   0.892 &   0.910 &   0.950 &   0.852 &   0.918 \\
   &         & L2 edge &   0.871 &   0.926 &   0.932 &   0.886 &   0.904 &   0.938 &   0.840 &   0.907 \\
   &         & No normal &   0.806 &   0.925 &   0.939 &   0.885 &   0.904 &   0.942 &   0.854 &   0.896 \\
   &         & No edge &   0.631 &   0.866 &   0.880 &   0.809 &   0.793 &   0.881 &   0.511 &   0.788 \\
   &         & No Laplacian &   0.439 &   0.870 &   0.803 &   0.799 &   0.760 &   0.870 &   0.546 &   0.746 \\
\cline{2-11}
   & \multirow{5}{*}{Jaccard} & Final &   0.819 &   0.875 &   0.888 &   0.809 &   0.837 &   0.905 &   0.755 &   0.849 \\
   &         & L2 edge &   0.776 &   0.864 &   0.875 &   0.799 &   0.828 &   0.885 &   0.735 &   0.831 \\
   &         & No normal &   0.676 &   0.861 &   0.887 &   0.797 &   0.826 &   0.891 &   0.754 &   0.812 \\
   &         & No edge &   0.464 &   0.766 &   0.792 &   0.684 &   0.663 &   0.790 &   0.349 &   0.652 \\
   &         & No Laplacian &   0.284 &   0.773 &   0.680 &   0.671 &   0.619 &   0.773 &   0.380 &   0.597 \\
\cline{2-11}
   & \multirow{5}{*}{ASSD (mm)} & Final &   1.335 &   1.042 &   0.842 &   1.583 &   1.176 &   0.531 &   1.904 &   1.213 \\
   &         & L2 edge &   1.609 &   1.127 &   0.933 &   1.657 &   1.243 &   0.641 &   1.826 &   1.314 \\
   &         & No normal &   2.039 &   1.154 &   0.838 &   1.701 &   1.200 &   0.726 &   2.147 &   1.469 \\
   &         & No edge &   2.599 &   2.014 &   1.723 &   2.350 &   2.741 &   1.597 &   7.567 &   3.343 \\
   &         & No Laplacian &   3.488 &   1.950 &   3.086 &   2.469 &   3.040 &   1.853 &   6.008 &   3.296 \\
\cline{2-11}
   & \multirow{5}{*}{HD (mm)} & Final &  14.393 &  10.407 &  10.325 &  13.639 &  13.360 &   9.407 &  26.616 &  28.035 \\
   &         & L2 edge &  15.500 &  10.098 &   8.977 &  13.055 &  12.406 &  10.178 &  26.034 &  27.030 \\
   &         & No normal &  14.261 &  11.269 &  10.027 &  13.502 &  11.768 &  12.500 &  27.737 &  29.066 \\
   &         & No edge &  15.000 &  10.304 &   9.668 &  13.104 &  14.236 &  12.677 &  34.336 &  34.852 \\
   &         & No Laplacian &  17.412 &  11.317 &  15.194 &  13.407 &  15.992 &  18.786 &  34.145 &  36.281 \\
\cline{1-11}
\cline{2-11}
\multirow{20}{*}{MR} & \multirow{5}{*}{Dice} & Final &   0.797 &   0.881 &   0.922 &   0.888 &   0.892 &   0.890 &   0.816 &   0.882 \\
   &         & L2 edge &   0.794 &   0.871 &   0.915 &   0.876 &   0.873 &   0.867 &   0.776 &   0.868 \\
   &         & No normal &   0.753 &   0.878 &   0.922 &   0.878 &   0.884 &   0.857 &   0.760 &   0.866 \\
   &         & No edge &   0.505 &   0.745 &   0.853 &   0.818 &   0.789 &   0.783 &   0.498 &   0.743 \\
   &         & No Laplacian &   0.450 &   0.765 &   0.846 &   0.786 &   0.772 &   0.747 &   0.471 &   0.733 \\
\cline{2-11}
   & \multirow{5}{*}{Jaccard} & Final &   0.671 &   0.791 &   0.858 &   0.801 &   0.812 &   0.805 &   0.697 &   0.790 \\
   &         & L2 edge &   0.665 &   0.775 &   0.845 &   0.783 &   0.783 &   0.769 &   0.645 &   0.770 \\
   &         & No normal &   0.609 &   0.787 &   0.857 &   0.786 &   0.798 &   0.755 &   0.629 &   0.765 \\
   &         & No edge &   0.347 &   0.603 &   0.750 &   0.699 &   0.667 &   0.653 &   0.338 &   0.596 \\
   &         & No Laplacian &   0.296 &   0.626 &   0.742 &   0.654 &   0.641 &   0.608 &   0.317 &   0.582 \\
\cline{2-11}
   & \multirow{5}{*}{ASSD (mm)} & Final &   2.198 &   1.401 &   1.183 &   1.611 &   1.333 &   2.648 &   2.689 &   1.775 \\
   &         & L2 edge &   2.224 &   1.563 &   1.288 &   1.738 &   1.599 &   3.017 &   2.345 &   1.909 \\
   &         & No normal &   2.330 &   1.458 &   1.149 &   1.774 &   1.396 &   2.691 &   2.978 &   1.923 \\
   &         & No edge &   4.013 &   3.291 &   2.397 &   2.556 &   3.094 &   3.448 &   6.763 &   3.730 \\
   &         & No Laplacian &   4.117 &   2.658 &   2.362 &   2.827 &   3.007 &   6.512 &   6.047 &   3.850 \\
\cline{2-11}
   & \multirow{5}{*}{HD (mm)} & Final &  16.923 &  11.723 &  10.891 &  14.810 &  13.463 &  22.219 &  19.345 &  27.701 \\
   &         & L2 edge &  18.361 &  10.705 &   8.969 &  14.247 &  13.455 &  22.754 &  17.124 &  29.339 \\
   &         & No normal &  15.460 &  12.190 &  10.354 &  16.143 &  13.493 &  23.968 &  21.291 &  29.490 \\
   &         & No edge &  20.087 &  16.863 &  14.517 &  14.365 &  15.953 &  21.623 &  25.522 &  30.149 \\
   &         & No Laplacian &  21.755 &  13.721 &  12.063 &  15.423 &  16.613 &  35.356 &  25.437 &  38.581 \\
\bottomrule
\end{tabular}
}}
\end{table}

\subsubsection{Comparison of Neural Network Sizes and Run Time}
Table \ref{table:s2} compares the total number of parameters and prediction time among our method, our method with reduced convolutional filters or mesh vertices, 2D UNet, 3D UNet and Voxel2Mesh. The prediction time was measured on a Nvidia GeForce GTX 1080Ti GPU.
\begin{table}[H]
\caption{A comparison of neural network sizes and the average prediction time among our methods, 2D UNet, 3D UNet and Voxel2Mesh. }\label{table:s2}
\resizebox{\textwidth}{!}{%
\begin{tabular}{lllllll}
\hline
                 & Ours       & Ours (Reduced Number of Filters) & Ours (Reduced Number of Vertices) & 2D UNet    & 3D UNet    & Voxel2Mesh \\ \hline
\# of Parameters & 16,765,112 & 8,474,257                    & 16,765,112                    & 31,110,152 & 18,556,552 & 9,124,521  \\ \hline
Time (s)         & 0.425      & 0.378                        & 0.240                         & 1.555      & 0.367      & 3.492      \\ \hline
\end{tabular}
}
\end{table}

\bibliographystyle{model2-names.bst}\biboptions{authoryear}
\bibliography{refs}

\begin{thebibliography}{54}
\expandafter\ifx\csname natexlab\endcsname\relax\def\natexlab#1{#1}\fi
\providecommand{\url}[1]{\texttt{#1}}
\providecommand{\href}[2]{#2}
\providecommand{\path}[1]{#1}
\providecommand{\DOIprefix}{doi:}
\providecommand{\ArXivprefix}{arXiv:}
\providecommand{\URLprefix}{URL: }
\providecommand{\Pubmedprefix}{pmid:}
\providecommand{\doi}[1]{\href{http://dx.doi.org/#1}{\path{#1}}}
\providecommand{\Pubmed}[1]{\href{pmid:#1}{\path{#1}}}
\providecommand{\bibinfo}[2]{#2}
\ifx\xfnm\relax \def\xfnm[#1]{\unskip,\space#1}\fi
\bibitem[{Augustin et~al.(2016)Augustin, Neic, Liebmann, Prassl, Niederer,
  Haase and Plank}]{Augustin2016EM}
\bibinfo{author}{Augustin, C.M.}, \bibinfo{author}{Neic, A.},
  \bibinfo{author}{Liebmann, M.}, \bibinfo{author}{Prassl, A.J.},
  \bibinfo{author}{Niederer, S.A.}, \bibinfo{author}{Haase, G.},
  \bibinfo{author}{Plank, G.}, \bibinfo{year}{2016}.
\newblock \bibinfo{title}{{Anatomically accurate high resolution modeling of
  human whole heart electromechanics: A strongly scalable algebraic multigrid
  solver method for nonlinear deformation}}.
\newblock \bibinfo{journal}{Journal of computational physics}
  \bibinfo{volume}{305}, \bibinfo{pages}{622--646}.
\newblock \URLprefix \url{https://pubmed.ncbi.nlm.nih.gov/26819483
  https://www.ncbi.nlm.nih.gov/pmc/articles/PMC4724941/},
  \DOIprefix\doi{10.1016/j.jcp.2015.10.045}.
\bibitem[{Bai et~al.(2015)Bai, Shi, Ledig and Rueckert}]{BAI201598}
\bibinfo{author}{Bai, W.}, \bibinfo{author}{Shi, W.}, \bibinfo{author}{Ledig,
  C.}, \bibinfo{author}{Rueckert, D.}, \bibinfo{year}{2015}.
\newblock \bibinfo{title}{Multi-atlas segmentation with augmented features for
  cardiac mr images}.
\newblock \bibinfo{journal}{Medical Image Analysis} \bibinfo{volume}{19},
  \bibinfo{pages}{98 -- 109}.
\newblock \URLprefix
  \url{http://www.sciencedirect.com/science/article/pii/S136184151400142X},
  \DOIprefix\doi{https://doi.org/10.1016/j.media.2014.09.005}.
\bibitem[{{Bronstein} et~al.(2017){Bronstein}, {Bruna}, {LeCun}, {Szlam} and
  {Vandergheynst}}]{Bronstein2017}
\bibinfo{author}{{Bronstein}, M.M.}, \bibinfo{author}{{Bruna}, J.},
  \bibinfo{author}{{LeCun}, Y.}, \bibinfo{author}{{Szlam}, A.},
  \bibinfo{author}{{Vandergheynst}, P.}, \bibinfo{year}{2017}.
\newblock \bibinfo{title}{Geometric deep learning: Going beyond euclidean
  data}.
\newblock \bibinfo{journal}{IEEE Signal Processing Magazine}
  \bibinfo{volume}{34}, \bibinfo{pages}{18--42}.
\newblock \DOIprefix\doi{10.1109/MSP.2017.2693418}.
\bibitem[{{Bucioli} et~al.(2017){Bucioli}, {Cyrino}, {Lima}, {Peres},
  {Cardoso}, {Lamounier}, {Neto} and {Botelho}}]{Bucioli2017}
\bibinfo{author}{{Bucioli}, A.A.B.}, \bibinfo{author}{{Cyrino}, G.F.},
  \bibinfo{author}{{Lima}, G.F.M.}, \bibinfo{author}{{Peres}, I.C.S.},
  \bibinfo{author}{{Cardoso}, A.}, \bibinfo{author}{{Lamounier}, E.A.},
  \bibinfo{author}{{Neto}, M.M.}, \bibinfo{author}{{Botelho}, R.V.},
  \bibinfo{year}{2017}.
\newblock \bibinfo{title}{Holographic real time 3d heart visualization from
  coronary tomography for multi-place medical diagnostics}, in:
  \bibinfo{booktitle}{2017 IEEE 15th Intl Conf on Dependable, Autonomic and
  Secure Computing, 15th Intl Conf on Pervasive Intelligence and Computing, 3rd
  Intl Conf on Big Data Intelligence and Computing and Cyber Science and
  Technology Congress(DASC/PiCom/DataCom/CyberSciTech)}, pp.
  \bibinfo{pages}{239--244}.
\newblock \DOIprefix\doi{10.1109/DASC-PICom-DataCom-CyberSciTec.2017.51}.
\bibitem[{Bui et~al.(2020)Bui, Hsu, Shanbhag, Tran, Bandettini, Chang and
  Chen}]{BUI2020104019}
\bibinfo{author}{Bui, V.}, \bibinfo{author}{Hsu, L.Y.},
  \bibinfo{author}{Shanbhag, S.M.}, \bibinfo{author}{Tran, L.},
  \bibinfo{author}{Bandettini, W.P.}, \bibinfo{author}{Chang, L.C.},
  \bibinfo{author}{Chen, M.Y.}, \bibinfo{year}{2020}.
\newblock \bibinfo{title}{Improving multi-atlas cardiac structure segmentation
  of computed tomography angiography: A performance evaluation based on a
  heterogeneous dataset}.
\newblock \bibinfo{journal}{Computers in Biology and Medicine}
  \bibinfo{volume}{125}, \bibinfo{pages}{104019}.
\newblock \URLprefix
  \url{http://www.sciencedirect.com/science/article/pii/S0010482520303504},
  \DOIprefix\doi{https://doi.org/10.1016/j.compbiomed.2020.104019}.
\bibitem[{{Bui} et~al.(2020){Bui}, {Shanbhag}, {Levine}, {Jacobs},
  {Bandettini}, {Chang}, {Chen} and {Hsu}}]{Bui2020}
\bibinfo{author}{{Bui}, V.}, \bibinfo{author}{{Shanbhag}, S.M.},
  \bibinfo{author}{{Levine}, O.}, \bibinfo{author}{{Jacobs}, M.},
  \bibinfo{author}{{Bandettini}, W.P.}, \bibinfo{author}{{Chang}, L.C.},
  \bibinfo{author}{{Chen}, M.Y.}, \bibinfo{author}{{Hsu}, L.Y.},
  \bibinfo{year}{2020}.
\newblock \bibinfo{title}{Simultaneous multi-structure segmentation of the
  heart and peripheral tissues in contrast enhanced cardiac computed tomography
  angiography}.
\newblock \bibinfo{journal}{IEEE Access} \bibinfo{volume}{8},
  \bibinfo{pages}{16187--16202}.
\newblock \DOIprefix\doi{10.1109/ACCESS.2020.2966985}.
\bibitem[{Chennupati et~al.(2019)Chennupati, Sistu, Yogamani and
  A~Rawashdeh}]{Chennupati_2019_CVPR_Workshops}
\bibinfo{author}{Chennupati, S.}, \bibinfo{author}{Sistu, G.},
  \bibinfo{author}{Yogamani, S.}, \bibinfo{author}{A~Rawashdeh, S.},
  \bibinfo{year}{2019}.
\newblock \bibinfo{title}{Multinet++: Multi-stream feature aggregation and
  geometric loss strategy for multi-task learning}, in:
  \bibinfo{booktitle}{Proceedings of the IEEE/CVF Conference on Computer Vision
  and Pattern Recognition (CVPR) Workshops}.
\bibitem[{Chnafa et~al.(2016)Chnafa, Mendez and Franck}]{Chnafa2016}
\bibinfo{author}{Chnafa, C.}, \bibinfo{author}{Mendez, S.},
  \bibinfo{author}{Franck, N.}, \bibinfo{year}{2016}.
\newblock \bibinfo{title}{Image-based simulations show important flow
  fluctuations in a normal left ventricle: What could be the implications?}
\newblock \bibinfo{journal}{Annals of biomedical engineering}
  \bibinfo{volume}{44}.
\newblock \DOIprefix\doi{10.1007/s10439-016-1614-6}.
\bibitem[{{\c{C}}i{\c{c}}ek et~al.(2016){\c{C}}i{\c{c}}ek, Abdulkadir,
  Lienkamp, Brox and Ronneberger}]{cicek3DUNET}
\bibinfo{author}{{\c{C}}i{\c{c}}ek, {\"O}.}, \bibinfo{author}{Abdulkadir, A.},
  \bibinfo{author}{Lienkamp, S.S.}, \bibinfo{author}{Brox, T.},
  \bibinfo{author}{Ronneberger, O.}, \bibinfo{year}{2016}.
\newblock \bibinfo{title}{3d u-net: Learning dense volumetric segmentation from
  sparse annotation}, in: \bibinfo{editor}{Ourselin, S.},
  \bibinfo{editor}{Joskowicz, L.}, \bibinfo{editor}{Sabuncu, M.R.},
  \bibinfo{editor}{Unal, G.}, \bibinfo{editor}{Wells, W.} (Eds.),
  \bibinfo{booktitle}{Medical Image Computing and Computer-Assisted
  Intervention -- MICCAI 2016}, \bibinfo{publisher}{Springer International
  Publishing}, \bibinfo{address}{Cham}. pp. \bibinfo{pages}{424--432}.
\bibitem[{Defferrard et~al.(2016)Defferrard, Bresson and
  Vandergheynst}]{Defferrard2016}
\bibinfo{author}{Defferrard, M.}, \bibinfo{author}{Bresson, X.},
  \bibinfo{author}{Vandergheynst, P.}, \bibinfo{year}{2016}.
\newblock \bibinfo{title}{Convolutional neural networks on graphs with fast
  localized spectral filtering}, in: \bibinfo{editor}{Lee, D.},
  \bibinfo{editor}{Sugiyama, M.}, \bibinfo{editor}{Luxburg, U.},
  \bibinfo{editor}{Guyon, I.}, \bibinfo{editor}{Garnett, R.} (Eds.),
  \bibinfo{booktitle}{Advances in Neural Information Processing Systems},
  \bibinfo{publisher}{Curran Associates, Inc.}. pp.
  \bibinfo{pages}{3844--3852}.
\newblock \URLprefix
  \url{https://proceedings.neurips.cc/paper/2016/file/04df4d434d481c5bb723be1b6df1ee65-Paper.pdf}.
\bibitem[{{Ecabert} et~al.(2008){Ecabert}, {Peters}, {Schramm}, {Lorenz}, {von
  Berg}, {Walker}, {Vembar}, {Olszewski}, {Subramanyan}, {Lavi} and
  {Weese}}]{Ecabert2008}
\bibinfo{author}{{Ecabert}, O.}, \bibinfo{author}{{Peters}, J.},
  \bibinfo{author}{{Schramm}, H.}, \bibinfo{author}{{Lorenz}, C.},
  \bibinfo{author}{{von Berg}, J.}, \bibinfo{author}{{Walker}, M.J.},
  \bibinfo{author}{{Vembar}, M.}, \bibinfo{author}{{Olszewski}, M.E.},
  \bibinfo{author}{{Subramanyan}, K.}, \bibinfo{author}{{Lavi}, G.},
  \bibinfo{author}{{Weese}, J.}, \bibinfo{year}{2008}.
\newblock \bibinfo{title}{Automatic model-based segmentation of the heart in ct
  images}.
\newblock \bibinfo{journal}{IEEE Transactions on Medical Imaging}
  \bibinfo{volume}{27}, \bibinfo{pages}{1189--1201}.
\newblock \DOIprefix\doi{10.1109/TMI.2008.918330}.
\bibitem[{Ecabert et~al.(2011)Ecabert, Peters, Walker, Ivanc, Lorenz, {von
  Berg}, Lessick, Vembar and Weese}]{ECABERT2011863}
\bibinfo{author}{Ecabert, O.}, \bibinfo{author}{Peters, J.},
  \bibinfo{author}{Walker, M.J.}, \bibinfo{author}{Ivanc, T.},
  \bibinfo{author}{Lorenz, C.}, \bibinfo{author}{{von Berg}, J.},
  \bibinfo{author}{Lessick, J.}, \bibinfo{author}{Vembar, M.},
  \bibinfo{author}{Weese, J.}, \bibinfo{year}{2011}.
\newblock \bibinfo{title}{Segmentation of the heart and great vessels in ct
  images using a model-based adaptation framework}.
\newblock \bibinfo{journal}{Medical Image Analysis} \bibinfo{volume}{15},
  \bibinfo{pages}{863 -- 876}.
\newblock \URLprefix
  \url{http://www.sciencedirect.com/science/article/pii/S1361841511000910},
  \DOIprefix\doi{https://doi.org/10.1016/j.media.2011.06.004}.
\bibitem[{{Gonz{\'{a}}lez Izard} et~al.(2020){Gonz{\'{a}}lez Izard},
  {S{\'{a}}nchez Torres}, {Alonso Plaza}, {Juanes M{\'{e}}ndez} and
  Garc{\'{i}}a-Pe{\~{n}}alvo}]{GonzalezIzard2020}
\bibinfo{author}{{Gonz{\'{a}}lez Izard}, S.}, \bibinfo{author}{{S{\'{a}}nchez
  Torres}, R.}, \bibinfo{author}{{Alonso Plaza}, {\'{O}}.},
  \bibinfo{author}{{Juanes M{\'{e}}ndez}, J.A.},
  \bibinfo{author}{Garc{\'{i}}a-Pe{\~{n}}alvo, F.J.}, \bibinfo{year}{2020}.
\newblock \bibinfo{title}{{Nextmed: Automatic Imaging Segmentation, 3D
  Reconstruction, and 3D Model Visualization Platform Using Augmented and
  Virtual Reality}}.
\newblock \bibinfo{journal}{Sensors (Basel, Switzerland)} \bibinfo{volume}{20},
  \bibinfo{pages}{2962}.
\newblock \URLprefix \url{https://pubmed.ncbi.nlm.nih.gov/32456194
  https://www.ncbi.nlm.nih.gov/pmc/articles/PMC7288297/},
  \DOIprefix\doi{10.3390/s20102962}.
\bibitem[{Gupta and Chandraker(2020)}]{NeuralMeshFlow}
\bibinfo{author}{Gupta, K.}, \bibinfo{author}{Chandraker, M.},
  \bibinfo{year}{2020}.
\newblock \bibinfo{title}{Neural mesh flow: 3d manifold mesh generationvia
  diffeomorphic flows}.
\newblock \href{http://arxiv.org/abs/2007.10973}{\tt arXiv:2007.10973}.
\bibitem[{Habijan et~al.(2020)Habijan, Babin, Gali{\'{c}}, Leventi{\'{c}},
  Romi{\'{c}}, Velicki and Pi{\v{z}}urica}]{Habijan2020}
\bibinfo{author}{Habijan, M.}, \bibinfo{author}{Babin, D.},
  \bibinfo{author}{Gali{\'{c}}, I.}, \bibinfo{author}{Leventi{\'{c}}, H.},
  \bibinfo{author}{Romi{\'{c}}, K.}, \bibinfo{author}{Velicki, L.},
  \bibinfo{author}{Pi{\v{z}}urica, A.}, \bibinfo{year}{2020}.
\newblock \bibinfo{title}{{Overview of the Whole Heart and Heart Chamber
  Segmentation Methods}}.
\newblock \bibinfo{journal}{Cardiovascular Engineering and Technology}
  \URLprefix \url{https://doi.org/10.1007/s13239-020-00494-8},
  \DOIprefix\doi{10.1007/s13239-020-00494-8}.
\bibitem[{He et~al.(2016)He, Zhang, Ren and Sun}]{He2016}
\bibinfo{author}{He, K.}, \bibinfo{author}{Zhang, X.}, \bibinfo{author}{Ren,
  S.}, \bibinfo{author}{Sun, J.}, \bibinfo{year}{2016}.
\newblock \bibinfo{title}{Identity mappings in deep residual networks}, in:
  \bibinfo{editor}{Leibe, B.}, \bibinfo{editor}{Matas, J.},
  \bibinfo{editor}{Sebe, N.}, \bibinfo{editor}{Welling, M.} (Eds.),
  \bibinfo{booktitle}{Computer Vision -- ECCV 2016},
  \bibinfo{publisher}{Springer International Publishing},
  \bibinfo{address}{Cham}. pp. \bibinfo{pages}{630--645}.
\bibitem[{Heller et~al.(2021)Heller, Isensee, Maier-Hein, Hou, Xie, Li, Nan,
  Mu, Lin, Han, Yao, Gao, Zhang, Wang, Hou, Yang, Xiong, Tian, Zhong, Ma,
  Rickman, Dean, Stai, Tejpaul, Oestreich, Blake, Kaluzniak, Raza, Rosenberg,
  Moore, Walczak, Rengel, Edgerton, Vasdev, Peterson, McSweeney, Peterson,
  Kalapara, Sathianathen, Papanikolopoulos and Weight}]{HELLER2021101821}
\bibinfo{author}{Heller, N.}, \bibinfo{author}{Isensee, F.},
  \bibinfo{author}{Maier-Hein, K.H.}, \bibinfo{author}{Hou, X.},
  \bibinfo{author}{Xie, C.}, \bibinfo{author}{Li, F.}, \bibinfo{author}{Nan,
  Y.}, \bibinfo{author}{Mu, G.}, \bibinfo{author}{Lin, Z.},
  \bibinfo{author}{Han, M.}, \bibinfo{author}{Yao, G.}, \bibinfo{author}{Gao,
  Y.}, \bibinfo{author}{Zhang, Y.}, \bibinfo{author}{Wang, Y.},
  \bibinfo{author}{Hou, F.}, \bibinfo{author}{Yang, J.},
  \bibinfo{author}{Xiong, G.}, \bibinfo{author}{Tian, J.},
  \bibinfo{author}{Zhong, C.}, \bibinfo{author}{Ma, J.},
  \bibinfo{author}{Rickman, J.}, \bibinfo{author}{Dean, J.},
  \bibinfo{author}{Stai, B.}, \bibinfo{author}{Tejpaul, R.},
  \bibinfo{author}{Oestreich, M.}, \bibinfo{author}{Blake, P.},
  \bibinfo{author}{Kaluzniak, H.}, \bibinfo{author}{Raza, S.},
  \bibinfo{author}{Rosenberg, J.}, \bibinfo{author}{Moore, K.},
  \bibinfo{author}{Walczak, E.}, \bibinfo{author}{Rengel, Z.},
  \bibinfo{author}{Edgerton, Z.}, \bibinfo{author}{Vasdev, R.},
  \bibinfo{author}{Peterson, M.}, \bibinfo{author}{McSweeney, S.},
  \bibinfo{author}{Peterson, S.}, \bibinfo{author}{Kalapara, A.},
  \bibinfo{author}{Sathianathen, N.}, \bibinfo{author}{Papanikolopoulos, N.},
  \bibinfo{author}{Weight, C.}, \bibinfo{year}{2021}.
\newblock \bibinfo{title}{The state of the art in kidney and kidney tumor
  segmentation in contrast-enhanced ct imaging: Results of the kits19
  challenge}.
\newblock \bibinfo{journal}{Medical Image Analysis} \bibinfo{volume}{67},
  \bibinfo{pages}{101821}.
\newblock \URLprefix
  \url{https://www.sciencedirect.com/science/article/pii/S1361841520301857},
  \DOIprefix\doi{https://doi.org/10.1016/j.media.2020.101821}.
\bibitem[{Isensee et~al.(2021)Isensee, Jaeger, Kohl, Petersen and
  Maier-Hein}]{Isensee2021}
\bibinfo{author}{Isensee, F.}, \bibinfo{author}{Jaeger, P.F.},
  \bibinfo{author}{Kohl, S.A.A.}, \bibinfo{author}{Petersen, J.},
  \bibinfo{author}{Maier-Hein, K.H.}, \bibinfo{year}{2021}.
\newblock \bibinfo{title}{nnu-net: a self-configuring method for deep
  learning-based biomedical image segmentation}.
\newblock \bibinfo{journal}{Nature Methods} \bibinfo{volume}{18},
  \bibinfo{pages}{203--211}.
\newblock \URLprefix \url{https://doi.org/10.1038/s41592-020-01008-z},
  \DOIprefix\doi{10.1038/s41592-020-01008-z}.
\bibitem[{Isensee et~al.(2018)Isensee, Kickingereder, Wick, Bendszus and
  Maier-Hein}]{Isensee2017}
\bibinfo{author}{Isensee, F.}, \bibinfo{author}{Kickingereder, P.},
  \bibinfo{author}{Wick, W.}, \bibinfo{author}{Bendszus, M.},
  \bibinfo{author}{Maier-Hein, K.H.}, \bibinfo{year}{2018}.
\newblock \bibinfo{title}{Brain tumor segmentation and radiomics survival
  prediction: Contribution to the brats 2017 challenge}, in:
  \bibinfo{editor}{Crimi, A.}, \bibinfo{editor}{Bakas, S.},
  \bibinfo{editor}{Kuijf, H.}, \bibinfo{editor}{Menze, B.},
  \bibinfo{editor}{Reyes, M.} (Eds.), \bibinfo{booktitle}{Brainlesion: Glioma,
  Multiple Sclerosis, Stroke and Traumatic Brain Injuries},
  \bibinfo{publisher}{Springer International Publishing},
  \bibinfo{address}{Cham}. pp. \bibinfo{pages}{287--297}.
\bibitem[{Isensee and Maier-Hein(2019)}]{Isensee2019AnAA}
\bibinfo{author}{Isensee, F.}, \bibinfo{author}{Maier-Hein, K.},
  \bibinfo{year}{2019}.
\newblock \bibinfo{title}{An attempt at beating the 3d u-net}.
\newblock \bibinfo{journal}{ArXiv} \bibinfo{volume}{abs/1908.02182}.
\bibitem[{Karim et~al.(2018)Karim, Blake, Inoue, Tao, Jia, Housden, Bhagirath,
  Duval, Varela, Behar, Cadour, {van der Geest}, Cochet, Drangova, Sermesant,
  Razavi, Aslanidi, Rajani and Rhode}]{SLAWT}
\bibinfo{author}{Karim, R.}, \bibinfo{author}{Blake, L.E.},
  \bibinfo{author}{Inoue, J.}, \bibinfo{author}{Tao, Q.}, \bibinfo{author}{Jia,
  S.}, \bibinfo{author}{Housden, R.J.}, \bibinfo{author}{Bhagirath, P.},
  \bibinfo{author}{Duval, J.L.}, \bibinfo{author}{Varela, M.},
  \bibinfo{author}{Behar, J.M.}, \bibinfo{author}{Cadour, L.},
  \bibinfo{author}{{van der Geest}, R.J.}, \bibinfo{author}{Cochet, H.},
  \bibinfo{author}{Drangova, M.}, \bibinfo{author}{Sermesant, M.},
  \bibinfo{author}{Razavi, R.}, \bibinfo{author}{Aslanidi, O.},
  \bibinfo{author}{Rajani, R.}, \bibinfo{author}{Rhode, K.},
  \bibinfo{year}{2018}.
\newblock \bibinfo{title}{Algorithms for left atrial wall segmentation and
  thickness – evaluation on an open-source ct and mri image database}.
\newblock \bibinfo{journal}{Medical Image Analysis} \bibinfo{volume}{50},
  \bibinfo{pages}{36 -- 53}.
\newblock \URLprefix
  \url{http://www.sciencedirect.com/science/article/pii/S1361841518306431},
  \DOIprefix\doi{https://doi.org/10.1016/j.media.2018.08.004}.
\bibitem[{Khalafvand et~al.(2012)Khalafvand, Ng, Zhong and
  Hung}]{Khalafvand2012}
\bibinfo{author}{Khalafvand, S.S.}, \bibinfo{author}{Ng, E.},
  \bibinfo{author}{Zhong, L.}, \bibinfo{author}{Hung, T.K.},
  \bibinfo{year}{2012}.
\newblock \bibinfo{title}{Fluid-dynamics modelling of the human left ventricle
  with dynamic mesh for normal and myocardial infarction: Preliminary study}.
\newblock \bibinfo{journal}{Computers in biology and medicine}
  \bibinfo{volume}{42}, \bibinfo{pages}{863--70}.
\newblock \DOIprefix\doi{10.1016/j.compbiomed.2012.06.010}.
\bibitem[{Khalafvand et~al.(2018)Khalafvand, Voorneveld, Muralidharan, Gijsen,
  Bosch, Walsum, Haak, Jong and Kenjeres}]{Khalafvand2018}
\bibinfo{author}{Khalafvand, S.S.}, \bibinfo{author}{Voorneveld, J.},
  \bibinfo{author}{Muralidharan, A.}, \bibinfo{author}{Gijsen, F.},
  \bibinfo{author}{Bosch, J.G.}, \bibinfo{author}{Walsum, T.},
  \bibinfo{author}{Haak, A.}, \bibinfo{author}{Jong, N.},
  \bibinfo{author}{Kenjeres, S.}, \bibinfo{year}{2018}.
\newblock \bibinfo{title}{Assessment of human left ventricle flow using
  statistical shape modelling and computational fluid dynamics}.
\newblock \bibinfo{journal}{Journal of Biomechanics} \bibinfo{volume}{74}.
\newblock \DOIprefix\doi{10.1016/j.jbiomech.2018.04.030}.
\bibitem[{Kingma and Ba(2014)}]{adam}
\bibinfo{author}{Kingma, D.}, \bibinfo{author}{Ba, J.}, \bibinfo{year}{2014}.
\newblock \bibinfo{title}{Adam: A method for stochastic optimization}.
\newblock \bibinfo{journal}{International Conference on Learning
  Representations} .
\bibitem[{Kirillov et~al.(2019)Kirillov, Wu, He and
  Girshick}]{kirillov2019pointrend}
\bibinfo{author}{Kirillov, A.}, \bibinfo{author}{Wu, Y.}, \bibinfo{author}{He,
  K.}, \bibinfo{author}{Girshick, R.}, \bibinfo{year}{2019}.
\newblock \bibinfo{title}{{PointRend}: Image segmentation as rendering}.
\bibitem[{Kong and Shadden(2020)}]{KONG2020}
\bibinfo{author}{Kong, F.}, \bibinfo{author}{Shadden, S.C.},
  \bibinfo{year}{2020}.
\newblock \bibinfo{title}{{Automating Model Generation for Image-Based Cardiac
  Flow Simulation}}.
\newblock \bibinfo{journal}{Journal of Biomechanical Engineering}
  \bibinfo{volume}{142}.
\newblock \URLprefix \url{https://doi.org/10.1115/1.4048032},
  \DOIprefix\doi{10.1115/1.4048032},
  \href{http://arxiv.org/abs/https://asmedigitalcollection.asme.org/biomechanical/article-pdf/142/11/111011/6565541/bio\_142\_11\_111011.pdf}{\tt
  arXiv:https://asmedigitalcollection.asme.org/biomechanical/article-pdf/142/11/111011/6565541/bio\_142\_11\_111011.pdf}.
  \bibinfo{note}{111011}.
\bibitem[{Lorensen and Cline(1987)}]{Lorensen1987}
\bibinfo{author}{Lorensen, W.E.}, \bibinfo{author}{Cline, H.E.},
  \bibinfo{year}{1987}.
\newblock \bibinfo{title}{Marching cubes: A high resolution 3d surface
  construction algorithm}.
\newblock \bibinfo{journal}{SIGGRAPH Comput. Graph.} \bibinfo{volume}{21},
  \bibinfo{pages}{163–169}.
\newblock \URLprefix \url{https://doi.org/10.1145/37402.37422},
  \DOIprefix\doi{10.1145/37402.37422}.
\bibitem[{Maher et~al.(2019)Maher, Wilson and Marsden}]{Maher2019}
\bibinfo{author}{Maher, G.}, \bibinfo{author}{Wilson, N.},
  \bibinfo{author}{Marsden, A.}, \bibinfo{year}{2019}.
\newblock \bibinfo{title}{Accelerating cardiovascular model building with
  convolutional neural networks}.
\newblock \bibinfo{journal}{Medical \& Biological Engineering \& Computing}
  \bibinfo{volume}{57}, \bibinfo{pages}{2319--2335}.
\bibitem[{Mittal et~al.(2015)Mittal, Seo, Vedula, Choi, Liu, Huang, Jain,
  Younes, Abraham and George}]{Mittal}
\bibinfo{author}{Mittal, R.}, \bibinfo{author}{Seo, J.H.},
  \bibinfo{author}{Vedula, V.}, \bibinfo{author}{Choi, Y.},
  \bibinfo{author}{Liu, H.}, \bibinfo{author}{Huang, H.},
  \bibinfo{author}{Jain, S.}, \bibinfo{author}{Younes, L.},
  \bibinfo{author}{Abraham, T.}, \bibinfo{author}{George, R.},
  \bibinfo{year}{2015}.
\newblock \bibinfo{title}{Computational modeling of cardiac hemodynamics:
  Current status and future outlook}.
\newblock \bibinfo{journal}{Journal of Computational Physics}
  \bibinfo{volume}{305}.
\newblock \DOIprefix\doi{10.1016/j.jcp.2015.11.022}.
\bibitem[{Ordas et~al.(2007)Ordas, Oubel, Leta, Carreras and
  Frangi}]{Ordas2007}
\bibinfo{author}{Ordas, S.}, \bibinfo{author}{Oubel, E.},
  \bibinfo{author}{Leta, R.}, \bibinfo{author}{Carreras, F.},
  \bibinfo{author}{Frangi, A.}, \bibinfo{year}{2007}.
\newblock \bibinfo{title}{A statistical shape model of the heart and its
  application to model-based segmentation}.
\newblock \bibinfo{journal}{Proceedings of SPIE - The International Society for
  Optical Engineering} \bibinfo{volume}{6511}.
\newblock \DOIprefix\doi{10.1117/12.708879}.
\bibitem[{Payer et~al.(2018)Payer, Štern, Bischof and Urschler}]{Payer2018}
\bibinfo{author}{Payer, C.}, \bibinfo{author}{Štern, D.},
  \bibinfo{author}{Bischof, H.}, \bibinfo{author}{Urschler, M.},
  \bibinfo{year}{2018}.
\newblock \bibinfo{title}{Multi-label whole heart segmentation using cnns and
  anatomical label configurations}, in: \bibinfo{booktitle}{Statistical Atlases
  and Computational Models of the Heart. ACDC and MMWHS Challenges},
  \bibinfo{publisher}{Springer}, \bibinfo{address}{Cham}. pp.
  \bibinfo{pages}{190--198}.
\bibitem[{Peng et~al.(2016)Peng, Lekadir, Gooya, Shao, Petersen and
  Frangi}]{Peng2016}
\bibinfo{author}{Peng, P.}, \bibinfo{author}{Lekadir, K.},
  \bibinfo{author}{Gooya, A.}, \bibinfo{author}{Shao, L.},
  \bibinfo{author}{Petersen, S.E.}, \bibinfo{author}{Frangi, A.F.},
  \bibinfo{year}{2016}.
\newblock \bibinfo{title}{{A review of heart chamber segmentation for
  structural and functional analysis using cardiac magnetic resonance
  imaging}}.
\newblock \bibinfo{journal}{Magnetic Resonance Materials in Physics, Biology
  and Medicine} \bibinfo{volume}{29}, \bibinfo{pages}{155--195}.
\newblock \URLprefix \url{https://doi.org/10.1007/s10334-015-0521-4},
  \DOIprefix\doi{10.1007/s10334-015-0521-4}.
\bibitem[{Peters et~al.(2010)Peters, Ecabert, Meyer, Kneser and
  Weese}]{PETERS201070}
\bibinfo{author}{Peters, J.}, \bibinfo{author}{Ecabert, O.},
  \bibinfo{author}{Meyer, C.}, \bibinfo{author}{Kneser, R.},
  \bibinfo{author}{Weese, J.}, \bibinfo{year}{2010}.
\newblock \bibinfo{title}{Optimizing boundary detection via simulated search
  with applications to multi-modal heart segmentation}.
\newblock \bibinfo{journal}{Medical Image Analysis} \bibinfo{volume}{14},
  \bibinfo{pages}{70 -- 84}.
\newblock \URLprefix
  \url{http://www.sciencedirect.com/science/article/pii/S1361841509001194},
  \DOIprefix\doi{https://doi.org/10.1016/j.media.2009.10.004}.
\bibitem[{Pontes et~al.(2019)Pontes, Kong, Sridharan, Lucey, Eriksson and
  Fookes}]{Image2mesh}
\bibinfo{author}{Pontes, J.K.}, \bibinfo{author}{Kong, C.},
  \bibinfo{author}{Sridharan, S.}, \bibinfo{author}{Lucey, S.},
  \bibinfo{author}{Eriksson, A.}, \bibinfo{author}{Fookes, C.},
  \bibinfo{year}{2019}.
\newblock \bibinfo{title}{Image2mesh: A learning framework for single image 3d
  reconstruction}, in: \bibinfo{editor}{Jawahar, C.V.}, \bibinfo{editor}{Li,
  H.}, \bibinfo{editor}{Mori, G.}, \bibinfo{editor}{Schindler, K.} (Eds.),
  \bibinfo{booktitle}{Computer Vision -- ACCV 2018},
  \bibinfo{publisher}{Springer International Publishing},
  \bibinfo{address}{Cham}. pp. \bibinfo{pages}{365--381}.
\bibitem[{Prakosa et~al.(2018)Prakosa, Arevalo, Deng, Boyle, Nikolov, Ashikaga,
  Blauer, Ghafoori, Park, Blake, Han, MacLeod, Halperin, Callans, Ranjan,
  Chrispin, Nazarian and Trayanova}]{Prakosa2018}
\bibinfo{author}{Prakosa, A.}, \bibinfo{author}{Arevalo, H.J.},
  \bibinfo{author}{Deng, D.}, \bibinfo{author}{Boyle, P.M.},
  \bibinfo{author}{Nikolov, P.P.}, \bibinfo{author}{Ashikaga, H.},
  \bibinfo{author}{Blauer, J.J.E.}, \bibinfo{author}{Ghafoori, E.},
  \bibinfo{author}{Park, C.J.}, \bibinfo{author}{Blake, R.C.},
  \bibinfo{author}{Han, F.T.}, \bibinfo{author}{MacLeod, R.S.},
  \bibinfo{author}{Halperin, H.R.}, \bibinfo{author}{Callans, D.J.},
  \bibinfo{author}{Ranjan, R.}, \bibinfo{author}{Chrispin, J.},
  \bibinfo{author}{Nazarian, S.}, \bibinfo{author}{Trayanova, N.A.},
  \bibinfo{year}{2018}.
\newblock \bibinfo{title}{Personalized virtual-heart technology for guiding the
  ablation of infarct-related ventricular tachycardia}.
\newblock \bibinfo{journal}{Nature Biomedical Engineering} \bibinfo{volume}{2},
  \bibinfo{pages}{732--740}.
\newblock \URLprefix \url{https://doi.org/10.1038/s41551-018-0282-2},
  \DOIprefix\doi{10.1038/s41551-018-0282-2}.
\bibitem[{Ronneberger et~al.(2015)Ronneberger, Fischer and
  Brox}]{Ronneberger2015}
\bibinfo{author}{Ronneberger, O.}, \bibinfo{author}{Fischer, P.},
  \bibinfo{author}{Brox, T.}, \bibinfo{year}{2015}.
\newblock \bibinfo{title}{U-net: Convolutional networks for biomedical image
  segmentation}, in: \bibinfo{editor}{Navab, N.}, \bibinfo{editor}{Hornegger,
  J.}, \bibinfo{editor}{Wells, W.M.}, \bibinfo{editor}{Frangi, A.F.} (Eds.),
  \bibinfo{booktitle}{Medical Image Computing and Computer-Assisted
  Intervention -- MICCAI 2015}, \bibinfo{publisher}{Springer International
  Publishing}, \bibinfo{address}{Cham}. pp. \bibinfo{pages}{234--241}.
\bibitem[{Si(2015)}]{Si2015}
\bibinfo{author}{Si, H.}, \bibinfo{year}{2015}.
\newblock \bibinfo{title}{Tetgen, a delaunay-based quality tetrahedral mesh
  generator}.
\newblock \bibinfo{journal}{ACM Trans. Math. Softw.} \bibinfo{volume}{41}.
\newblock \URLprefix \url{https://doi.org/10.1145/2629697},
  \DOIprefix\doi{10.1145/2629697}.
\bibitem[{Simard et~al.(2003)Simard, Steinkraus and Platt}]{Simard2003BestPF}
\bibinfo{author}{Simard, P.}, \bibinfo{author}{Steinkraus, D.},
  \bibinfo{author}{Platt, J.C.}, \bibinfo{year}{2003}.
\newblock \bibinfo{title}{Best practices for convolutional neural networks
  applied to visual document analysis}.
\newblock \bibinfo{journal}{Seventh International Conference on Document
  Analysis and Recognition, 2003. Proceedings.} , \bibinfo{pages}{958--963}.
\bibitem[{{Tobon-Gomez} et~al.(2015){Tobon-Gomez}, {Geers}, {Peters}, {Weese},
  {Pinto}, {Karim}, {Ammar}, {Daoudi}, {Margeta}, {Sandoval}, {Stender},
  {Zheng}, {Zuluaga}, {Betancur}, {Ayache}, {Chikh}, {Dillenseger}, {Kelm},
  {Mahmoudi}, {Ourselin}, {Schlaefer}, {Schaeffter}, {Razavi} and
  {Rhode}}]{LASC}
\bibinfo{author}{{Tobon-Gomez}, C.}, \bibinfo{author}{{Geers}, A.J.},
  \bibinfo{author}{{Peters}, J.}, \bibinfo{author}{{Weese}, J.},
  \bibinfo{author}{{Pinto}, K.}, \bibinfo{author}{{Karim}, R.},
  \bibinfo{author}{{Ammar}, M.}, \bibinfo{author}{{Daoudi}, A.},
  \bibinfo{author}{{Margeta}, J.}, \bibinfo{author}{{Sandoval}, Z.},
  \bibinfo{author}{{Stender}, B.}, \bibinfo{author}{{Zheng}, Y.},
  \bibinfo{author}{{Zuluaga}, M.A.}, \bibinfo{author}{{Betancur}, J.},
  \bibinfo{author}{{Ayache}, N.}, \bibinfo{author}{{Chikh}, M.A.},
  \bibinfo{author}{{Dillenseger}, J.}, \bibinfo{author}{{Kelm}, B.M.},
  \bibinfo{author}{{Mahmoudi}, S.}, \bibinfo{author}{{Ourselin}, S.},
  \bibinfo{author}{{Schlaefer}, A.}, \bibinfo{author}{{Schaeffter}, T.},
  \bibinfo{author}{{Razavi}, R.}, \bibinfo{author}{{Rhode}, K.S.},
  \bibinfo{year}{2015}.
\newblock \bibinfo{title}{Benchmark for algorithms segmenting the left atrium
  from 3d ct and mri datasets}.
\newblock \bibinfo{journal}{IEEE Transactions on Medical Imaging}
  \bibinfo{volume}{34}, \bibinfo{pages}{1460--1473}.
\newblock \DOIprefix\doi{10.1109/TMI.2015.2398818}.
\bibitem[{Updegrove et~al.(2016)Updegrove, Wilson and
  Shadden}]{UPDEGROVE201616}
\bibinfo{author}{Updegrove, A.}, \bibinfo{author}{Wilson, N.M.},
  \bibinfo{author}{Shadden, S.C.}, \bibinfo{year}{2016}.
\newblock \bibinfo{title}{Boolean and smoothing of discrete polygonal
  surfaces}.
\newblock \bibinfo{journal}{Advances in Engineering Software}
  \bibinfo{volume}{95}, \bibinfo{pages}{16 -- 27}.
\newblock \URLprefix
  \url{http://www.sciencedirect.com/science/article/pii/S0965997816300230},
  \DOIprefix\doi{https://doi.org/10.1016/j.advengsoft.2016.01.015}.
\bibitem[{{Wang} et~al.(2020a){Wang}, {Zhang}, {Li}, {Fu}, {Yu}, {Liu}, {Xue}
  and {Jiang}}]{Pixel2Mesh}
\bibinfo{author}{{Wang}, N.}, \bibinfo{author}{{Zhang}, Y.},
  \bibinfo{author}{{Li}, Z.}, \bibinfo{author}{{Fu}, Y.},
  \bibinfo{author}{{Yu}, H.}, \bibinfo{author}{{Liu}, W.},
  \bibinfo{author}{{Xue}, X.}, \bibinfo{author}{{Jiang}, Y.},
  \bibinfo{year}{2020}a.
\newblock \bibinfo{title}{Pixel2mesh: 3d mesh model generation via image guided
  deformation}.
\newblock \bibinfo{journal}{IEEE Transactions on Pattern Analysis and Machine
  Intelligence} , \bibinfo{pages}{1--1}.
\bibitem[{{Wang} et~al.(2020b){Wang}, {Zhong} and {Hua}}]{DeepOrganNet}
\bibinfo{author}{{Wang}, Y.}, \bibinfo{author}{{Zhong}, Z.},
  \bibinfo{author}{{Hua}, J.}, \bibinfo{year}{2020}b.
\newblock \bibinfo{title}{Deeporgannet: On-the-fly reconstruction and
  visualization of 3d / 4d lung models from single-view projections by deep
  deformation network}.
\newblock \bibinfo{journal}{IEEE Transactions on Visualization and Computer
  Graphics} \bibinfo{volume}{26}, \bibinfo{pages}{960--970}.
\newblock \DOIprefix\doi{10.1109/TVCG.2019.2934369}.
\bibitem[{Wei et~al.(2013)Wei, Shen, Qin, Wu, Wong and Heng}]{WEI20131223}
\bibinfo{author}{Wei, M.}, \bibinfo{author}{Shen, W.}, \bibinfo{author}{Qin,
  J.}, \bibinfo{author}{Wu, J.}, \bibinfo{author}{Wong, T.T.},
  \bibinfo{author}{Heng, P.A.}, \bibinfo{year}{2013}.
\newblock \bibinfo{title}{Feature-preserving optimization for noisy mesh using
  joint bilateral filter and constrained laplacian smoothing}.
\newblock \bibinfo{journal}{Optics and Lasers in Engineering}
  \bibinfo{volume}{51}, \bibinfo{pages}{1223 -- 1234}.
\newblock \URLprefix
  \url{http://www.sciencedirect.com/science/article/pii/S0143816613001395},
  \DOIprefix\doi{https://doi.org/10.1016/j.optlaseng.2013.04.018}.
\bibitem[{Wei et~al.(2018)Wei, Wang, Guo, Wu, Xie, Wang and Qin}]{Wei2018}
\bibinfo{author}{Wei, M.}, \bibinfo{author}{Wang, J.}, \bibinfo{author}{Guo,
  X.}, \bibinfo{author}{Wu, H.}, \bibinfo{author}{Xie, H.},
  \bibinfo{author}{Wang, F.L.}, \bibinfo{author}{Qin, J.},
  \bibinfo{year}{2018}.
\newblock \bibinfo{title}{{Learning-based 3D surface optimization from medical
  image reconstruction}}.
\newblock \bibinfo{journal}{Optics and Lasers in Engineering}
  \bibinfo{volume}{103}, \bibinfo{pages}{110--118}.
\newblock \URLprefix \url{https://doi.org/10.1016/j.optlaseng.2017.11.014},
  \DOIprefix\doi{10.1016/j.optlaseng.2017.11.014}.
\bibitem[{Wen et~al.(2019)Wen, Zhang, Li and Fu}]{Wen2019Pixel2MeshM3}
\bibinfo{author}{Wen, C.}, \bibinfo{author}{Zhang, Y.}, \bibinfo{author}{Li,
  Z.}, \bibinfo{author}{Fu, Y.}, \bibinfo{year}{2019}.
\newblock \bibinfo{title}{Pixel2mesh++: Multi-view 3d mesh generation via
  deformation}.
\newblock \bibinfo{journal}{2019 IEEE/CVF International Conference on Computer
  Vision (ICCV)} , \bibinfo{pages}{1042--1051}.
\bibitem[{Wickramasinghe et~al.(2020)Wickramasinghe, Remelli, Knott and
  Fua}]{Voxel2Mesh}
\bibinfo{author}{Wickramasinghe, U.}, \bibinfo{author}{Remelli, E.},
  \bibinfo{author}{Knott, G.}, \bibinfo{author}{Fua, P.}, \bibinfo{year}{2020}.
\newblock \bibinfo{title}{Voxel2mesh: 3d mesh model generation from volumetric
  data}, in: \bibinfo{editor}{Martel, A.L.}, \bibinfo{editor}{Abolmaesumi, P.},
  \bibinfo{editor}{Stoyanov, D.}, \bibinfo{editor}{Mateus, D.},
  \bibinfo{editor}{Zuluaga, M.A.}, \bibinfo{editor}{Zhou, S.K.},
  \bibinfo{editor}{Racoceanu, D.}, \bibinfo{editor}{Joskowicz, L.} (Eds.),
  \bibinfo{booktitle}{Medical Image Computing and Computer Assisted
  Intervention -- MICCAI 2020}, \bibinfo{publisher}{Springer International
  Publishing}, \bibinfo{address}{Cham}. pp. \bibinfo{pages}{299--308}.
\bibitem[{Wolterink et~al.(2016)Wolterink, Leiner, de~Vos, Coatrieux, Kelm,
  Kondo, Salgado, Shahzad, Shu, Snoeren, Takx, van Vliet, van Walsum, Willems,
  Yang, Zheng, Viergever and Išgum}]{orCaScore}
\bibinfo{author}{Wolterink, J.M.}, \bibinfo{author}{Leiner, T.},
  \bibinfo{author}{de~Vos, B.D.}, \bibinfo{author}{Coatrieux, J.L.},
  \bibinfo{author}{Kelm, B.M.}, \bibinfo{author}{Kondo, S.},
  \bibinfo{author}{Salgado, R.A.}, \bibinfo{author}{Shahzad, R.},
  \bibinfo{author}{Shu, H.}, \bibinfo{author}{Snoeren, M.},
  \bibinfo{author}{Takx, R.A.P.}, \bibinfo{author}{van Vliet, L.J.},
  \bibinfo{author}{van Walsum, T.}, \bibinfo{author}{Willems, T.P.},
  \bibinfo{author}{Yang, G.}, \bibinfo{author}{Zheng, Y.},
  \bibinfo{author}{Viergever, M.A.}, \bibinfo{author}{Išgum, I.},
  \bibinfo{year}{2016}.
\newblock \bibinfo{title}{An evaluation of automatic coronary artery calcium
  scoring methods with cardiac ct using the orcascore framework}.
\newblock \bibinfo{journal}{Medical Physics} \bibinfo{volume}{43},
  \bibinfo{pages}{2361--2373}.
\newblock \URLprefix
  \url{https://aapm.onlinelibrary.wiley.com/doi/abs/10.1118/1.4945696},
  \DOIprefix\doi{https://doi.org/10.1118/1.4945696},
  \href{http://arxiv.org/abs/https://aapm.onlinelibrary.wiley.com/doi/pdf/10.1118/1.4945696}{\tt
  arXiv:https://aapm.onlinelibrary.wiley.com/doi/pdf/10.1118/1.4945696}.
\bibitem[{{Ye} et~al.(2019){Ye}, {Wang}, {Zhang} and {Wang}}]{Ye2019}
\bibinfo{author}{{Ye}, C.}, \bibinfo{author}{{Wang}, W.},
  \bibinfo{author}{{Zhang}, S.}, \bibinfo{author}{{Wang}, K.},
  \bibinfo{year}{2019}.
\newblock \bibinfo{title}{Multi-depth fusion network for whole-heart ct image
  segmentation}.
\newblock \bibinfo{journal}{IEEE Access} \bibinfo{volume}{7},
  \bibinfo{pages}{23421--23429}.
\newblock \DOIprefix\doi{10.1109/ACCESS.2019.2899635}.
\bibitem[{Ye et~al.(2021)Ye, Huang, Yang, Wu, Yi, Axel and
  Metaxas}]{YeMeng2020}
\bibinfo{author}{Ye, M.}, \bibinfo{author}{Huang, Q.}, \bibinfo{author}{Yang,
  D.}, \bibinfo{author}{Wu, P.}, \bibinfo{author}{Yi, J.},
  \bibinfo{author}{Axel, L.}, \bibinfo{author}{Metaxas, D.},
  \bibinfo{year}{2021}.
\newblock \bibinfo{title}{Pc-u net: Learning to jointly reconstruct and segment
  the cardiac walls in 3d from ct data}, in: \bibinfo{editor}{Puyol~Anton, E.},
  \bibinfo{editor}{Pop, M.}, \bibinfo{editor}{Sermesant, M.},
  \bibinfo{editor}{Campello, V.}, \bibinfo{editor}{Lalande, A.},
  \bibinfo{editor}{Lekadir, K.}, \bibinfo{editor}{Suinesiaputra, A.},
  \bibinfo{editor}{Camara, O.}, \bibinfo{editor}{Young, A.} (Eds.),
  \bibinfo{booktitle}{Statistical Atlases and Computational Models of the
  Heart. M{\&}Ms and EMIDEC Challenges}, \bibinfo{publisher}{Springer
  International Publishing}, \bibinfo{address}{Cham}. pp.
  \bibinfo{pages}{117--126}.
\bibitem[{Zhang et~al.(2020)Zhang, Dong and Li}]{ZhangMo2020}
\bibinfo{author}{Zhang, M.}, \bibinfo{author}{Dong, B.}, \bibinfo{author}{Li,
  Q.}, \bibinfo{year}{2020}.
\newblock \bibinfo{title}{Deep active contour network for medical image
  segmentation}, in: \bibinfo{editor}{Martel, A.L.},
  \bibinfo{editor}{Abolmaesumi, P.}, \bibinfo{editor}{Stoyanov, D.},
  \bibinfo{editor}{Mateus, D.}, \bibinfo{editor}{Zuluaga, M.A.},
  \bibinfo{editor}{Zhou, S.K.}, \bibinfo{editor}{Racoceanu, D.},
  \bibinfo{editor}{Joskowicz, L.} (Eds.), \bibinfo{booktitle}{Medical Image
  Computing and Computer Assisted Intervention -- MICCAI 2020},
  \bibinfo{publisher}{Springer International Publishing},
  \bibinfo{address}{Cham}. pp. \bibinfo{pages}{321--331}.
\bibitem[{Zhuang(2013)}]{Zhuang2013}
\bibinfo{author}{Zhuang, X.}, \bibinfo{year}{2013}.
\newblock \bibinfo{title}{{Challenges and Methodologies of Fully Automatic
  Whole Heart Segmentation: A Review}}.
\newblock \bibinfo{journal}{Journal of Healthcare Engineering}
  \bibinfo{volume}{4}, \bibinfo{pages}{981729}.
\newblock \URLprefix \url{https://doi.org/10.1260/2040-2295.4.3.371},
  \DOIprefix\doi{10.1260/2040-2295.4.3.371}.
\bibitem[{Zhuang et~al.(2015)Zhuang, Bai, Song, Zhan, Qian, Shi, Lian and
  Rueckert}]{Zhuang2015MAS}
\bibinfo{author}{Zhuang, X.}, \bibinfo{author}{Bai, W.}, \bibinfo{author}{Song,
  J.}, \bibinfo{author}{Zhan, S.}, \bibinfo{author}{Qian, X.},
  \bibinfo{author}{Shi, W.}, \bibinfo{author}{Lian, Y.},
  \bibinfo{author}{Rueckert, D.}, \bibinfo{year}{2015}.
\newblock \bibinfo{title}{{{M}ultiatlas whole heart segmentation of {C}{T} data
  using conditional entropy for atlas ranking and selection}}.
\newblock \bibinfo{journal}{Med Phys} \bibinfo{volume}{42},
  \bibinfo{pages}{3822--3833}.
\bibitem[{Zhuang et~al.(2019)Zhuang, Li, Payer, Štern, Urschler, Heinrich,
  Oster, Wang, Örjan Smedby, Bian, Yang, Heng, Mortazi, Bagci, Yang, Sun,
  Galisot, Ramel, Brouard, Tong, Si, Liao, Zeng, Shi, Zheng, Wang,
  MacGillivray, Newby, Rhode, Ourselin, Mohiaddin, Keegan, Firmin and
  Yang}]{ZHUANG2019}
\bibinfo{author}{Zhuang, X.}, \bibinfo{author}{Li, L.}, \bibinfo{author}{Payer,
  C.}, \bibinfo{author}{Štern, D.}, \bibinfo{author}{Urschler, M.},
  \bibinfo{author}{Heinrich, M.P.}, \bibinfo{author}{Oster, J.},
  \bibinfo{author}{Wang, C.}, \bibinfo{author}{Örjan Smedby},
  \bibinfo{author}{Bian, C.}, \bibinfo{author}{Yang, X.},
  \bibinfo{author}{Heng, P.A.}, \bibinfo{author}{Mortazi, A.},
  \bibinfo{author}{Bagci, U.}, \bibinfo{author}{Yang, G.},
  \bibinfo{author}{Sun, C.}, \bibinfo{author}{Galisot, G.},
  \bibinfo{author}{Ramel, J.Y.}, \bibinfo{author}{Brouard, T.},
  \bibinfo{author}{Tong, Q.}, \bibinfo{author}{Si, W.}, \bibinfo{author}{Liao,
  X.}, \bibinfo{author}{Zeng, G.}, \bibinfo{author}{Shi, Z.},
  \bibinfo{author}{Zheng, G.}, \bibinfo{author}{Wang, C.},
  \bibinfo{author}{MacGillivray, T.}, \bibinfo{author}{Newby, D.},
  \bibinfo{author}{Rhode, K.}, \bibinfo{author}{Ourselin, S.},
  \bibinfo{author}{Mohiaddin, R.}, \bibinfo{author}{Keegan, J.},
  \bibinfo{author}{Firmin, D.}, \bibinfo{author}{Yang, G.},
  \bibinfo{year}{2019}.
\newblock \bibinfo{title}{Evaluation of algorithms for multi-modality whole
  heart segmentation: An open-access grand challenge}.
\newblock \bibinfo{journal}{Medical Image Analysis} \bibinfo{volume}{58},
  \bibinfo{pages}{101537}.
\newblock \URLprefix
  \url{http://www.sciencedirect.com/science/article/pii/S1361841519300751},
  \DOIprefix\doi{https://doi.org/10.1016/j.media.2019.101537}.
\bibitem[{Zhuang and Shen(2016)}]{ZHUANG2016MAS}
\bibinfo{author}{Zhuang, X.}, \bibinfo{author}{Shen, J.}, \bibinfo{year}{2016}.
\newblock \bibinfo{title}{Multi-scale patch and multi-modality atlases for
  whole heart segmentation of mri}.
\newblock \bibinfo{journal}{Medical Image Analysis} \bibinfo{volume}{31},
  \bibinfo{pages}{77 -- 87}.
\newblock \URLprefix
  \url{http://www.sciencedirect.com/science/article/pii/S1361841516000219},
  \DOIprefix\doi{https://doi.org/10.1016/j.media.2016.02.006}.

\end{thebibliography}
\end{document}